\documentclass[11pt]{article}

\usepackage{amsmath,amsfonts,amssymb,amsthm,bbm,stmaryrd}
\usepackage[dvips]{graphicx}
\usepackage{latexsym}
\usepackage{psfrag}
\usepackage[latin1]{inputenc}
\usepackage{hyperref}
\usepackage{color}
\usepackage{tcolorbox}
\numberwithin{equation}{section}
\usepackage{float}
\usepackage{multirow}
\usepackage[bb=stix]{mathalpha}
\usepackage{upgreek}

\newcommand{\be}{\begin{equation}}
\newcommand{\ee}{\end{equation}}
\newcommand{\barray}{\begin{array}}
\newcommand{\earray}{\end{array}}
\newcommand{\bea}{\begin{eqnarray}}
\newcommand{\eea}{\end{eqnarray}}
\newcommand{\bs}{\begin{subequations}}
\newcommand{\es}{\end{subequations}}

\newcommand{\bit}{\begin{itemize}}
\newcommand{\eit}{\end{itemize}}
\newcommand{\bd}{\begin{description}}
\newcommand{\ed}{\end{description}}

\def\nn{\nonumber}

\newcommand{\re}{\mathrm{Re}}
\newcommand{\im}{\mathrm{Im}}

\def\la{\langle}
\def\ra{\rangle}

\newcommand{\mat} [4] {\left ( \begin{array}{cc}{#1}&{#2}\\{#3}&{#4} \end{array} \right ) }

\def\w{\wedge}
\def\cw{\curlywedge}
\newcommand{\p}{\partial}
\newcommand{\na}{\nabla}

\newcommand{\R}{\mathbb{R}}

\newcommand{\tr}{{\rm Tr}}
\newcommand{\f}{\frac}
\newcommand{\tl}{\tilde}

\renewcommand{\a}{\alpha} \renewcommand{\b}{\beta} \newcommand{\g}{\gamma}  
\renewcommand{\d}{\delta}  \newcommand{\eps}{\epsilon} 
 \newcommand{\veps}{\varepsilon}
 \renewcommand{\th}{\theta}  \newcommand{\vth}{\vartheta} 
  \renewcommand{\k}{\kappa}  \renewcommand{\l}{\lambda}
\let\m=\mu    \let\n=\nu   \let\r=\rho \let\om=\omega
 \newcommand{\s}{\sigma}      
\let\G=\Gamma \let\D=\Delta  \let\Th=\Theta \let\L=\Lambda 
\let\Si=\Sigma \let\Om=\Omega

\newcommand{\SL}{\mathrm{SL}(2,\mathbb{C})}

\def\cL{{\cal L}}

\def\cN{{\cal N}}
\newcommand{\cR}{{\mathcal R}}

\newcommand{\norm}[1]{|\!|#1|\!|}

\usepackage{slashed}
\newcommand{\sd}{\slashed{\delta}}

\newcommand{\eqonS}{\stackrel{\scri}=}
\newcommand{\eqons}{\,\hat{=}\,}
\newcommand{\eqon}[1]{\stackrel{{#1}}=}
\newcommand{\eqonN}{\,\eqon{\sscr\cN}\,}

\usepackage{pgf}
\makeatletter
\newcommand*{\pgfunderleftarrow}{%
  \@ifstar
    {\let\ifpgf@depth\iftrue\mathpalette\@pgfunderleftarrow}
    {\let\ifpgf@depth\iffalse\mathpalette\@pgfunderleftarrow}%
}
\newcommand*{\@pgfunderleftarrow}[2]{%
  #2%
  \edef\pgf@math@fam{\the\fam}%
  \pgfpicture
    \pgfsetbaseline{0pt}
    \pgf@relevantforpicturesizefalse      
    \pgfsetroundcap                       
    \pgfsetarrowsend{to}
    \pgfutil@tempdima=0.28pt%
    \advance\pgfutil@tempdima by.8\pgflinewidth%
    \pgfutil@tempdima-4\pgfutil@tempdima
    \sbox\pgfutil@tempboxa{$\m@th\fam\pgf@math@fam#1#2$}%
    \advance\pgfutil@tempdima-\dp\pgfutil@tempboxa
    \pgfutil@tempdimb\wd\pgfutil@tempboxa
    \pgfpathmoveto{\pgfqpoint{0pt}{\pgfutil@tempdima}}%
    \pgfpathlineto{\pgfqpoint{-\pgfutil@tempdimb}{\pgfutil@tempdima}}%
    \pgfusepath{stroke}
    \ifpgf@depth
      \pgf@relevantforpicturesizetrue
      \pgfpathmoveto{\pgfqpoint{0pt}{-\pgfutil@tempdimb}}%
      \pgfusepath{use as bounding box}%
    \fi
  \endpgfpicture
}
\makeatother
\newcommand{\pbi}[1]{\pgfunderleftarrow{#1}}

\newcommand{\sscr}{\scriptscriptstyle\rm}

\usepackage{manfnt}
\reversemarginpar

\newcommand{\qB}{{\bar q}}

\newcommand{\KB}{{\bar K}}

\newcommand{\Diff}{{\rm{Diff}}}

\newcommand{\note}[1]{{\begin{quotation} {\noindent\footnotesize{#1}} \end{quotation}}}
\usepackage{mathrsfs}
\newcommand{\Dd}{{\mathscr{D}}}
\newcommand{\scri}{{\mathscr{I}}} 
\newcommand{\exb}{{\cal B}} 
\newcommand{\bb}{{\upbeta}}
\newcommand{\Dr}{{\cal{D}}}
\newcommand{\nG}{{\mathbb n}}

\newcommand{\os}[1]{\overset{\scriptscriptstyle\circ}{#1}{}}

\usepackage{tikz}
\usetikzlibrary{arrows, decorations.pathmorphing, positioning,calc}
\newcommand{\ala}[1]{{\stackrel{\leftrightarrow}{#1}}{}}

\begin{document}

\title{\bf GGI lectures on boundary and asymptotic symmetries }

\author{\Large{Simone Speziale}
\smallskip \\ 
\small{\it{Aix Marseille Univ., Univ. de Toulon, CNRS, CPT, UMR 7332, 13288 Marseille, France}} }
\date{June 30, 2026}

\maketitle

\begin{abstract}
\noindent Support material for lectures at the May '25 Galileo Galilei Institute school on asymptotic symmetries and flat holography.
Contains an introduction to Noether theorem for gauge theories and gravity, covariant phase space formalism, boundary and asymptotic symmetries, 
flux-balance laws on null hypersurfaces, future null infinity in Bondi-Sachs coordinates and with Penrose's conformal compactification,  BMS symmetries and their charges and fluxes. Includes an original and pedagogical derivation of the BMS group using only Minkowski, and an original derivation of an integral Hamiltonian generator for diffeomorphisms of a scalar field on a null hypersurface.
\end{abstract}

\tableofcontents

\section{Introduction}

Gauge symmetries 
map solutions to physically equivalent solutions: the same electric field described by a different potential, or the same spacetime geometry described in different coordinates. They manifest a redundancy in the field equations, a degeneracy in the symplectic structure, and don't usually lead to conservation laws or other useful insights about the dynamics.
The situation can however change in the presence of boundaries. Boundaries, and more in particular the boundary conditions one chooses, can make gauge transformations physically 
distinguishable. 
Residual gauge transformation at the boundary of a spacetime region are the \emph{boundary symmetries} object of these lectures. And if the boundary is at infinity, one speaks about asymptotic symmetries. 
Examples of asymptotic symmetries are the ADM and BMS symmetries of general relativity, and the celestial symmetry of electromagnetism.
Examples at finite distance include the symmetries behind the mechanical laws of horizons, isolated horizons and general null hypersurfaces in general relativity, and the edge modes of Chern-Simons theory.

When talking about symmetries, a prominent role is taken by Noether's theorem, which identifies conserved currents that can be used to study conservation laws or flux-balance laws. This theorem is particularly useful to understand gauge symmetries, however its application to this case requires care. 
There are two reasons for this. The first, is that Noether currents are only defined up to exact forms. But in gauge theories, the Noether current is itself an exact form, on-shell. Therefore, \emph{ the whole Noether current is completely ambiguous}. Second, one cannot always fix such ambiguities looking at the canonical generator, because in the presence of radiation, \emph{some symmetry generators correspond to non-Hamiltonian vector fields}, hence don't admit a canonical generator in the standard sense.
For these reasons for instance,  charges for the gravitational BMS asymptotic symmetries where identified first using physical arguments \cite{Geroch:1977jn,Ashtekar:1981hw,Ashtekar:1981bq,Dray:1984rfa,Dray:1984gz}, and only later it was shown how to derive them in a consistent and unambiguous way from Noether's theorem \cite{Wald:1999wa} (see also discussion in \cite{Ashtekar:2024stm}). 
This achievement sets the ground for what we call the `generalized Wald-Zoupas prescription', an approach to resolving the ambiguities by focusing on the physical 
applications of the Noether current, 
and the imposition of general covariance at the boundaries. This approach allows one to derive many useful results such as the BMS flux-balance laws and the first and second laws of horizons and isolated horizons in an elegant way, to understand them all under the unifying principle of Noether's theorem, and to set the ground for new applications.

In the last few years the interest in boundary and asymptotic symmetries has increased enormously. We remark the connection between BMS symmetries and soft theorems in perturbative quantum gravity pioneered by Strominger, the relation between  corner symmetries and entanglement,
the experimental and theoretical work around memory effects, the relation between asymptotic symmetries and perturbation theory,
the new explorations proposed by celestial holography, flat holography and Carollian holography. The current ongoing research has motivated 
the program of the Galileo Galilei Institute (GGI) \href{https://www.ggi.infn.it/showevent.pl?id=510}{workshop} \emph{From Asymptotic Symmetries to Flat Holography: Theoretical Aspects and Observable Consequences}, as well as the series of lectures that we have proposed for its introductory \href{https://www.ggi.infn.it/showevent.pl?id=511}{\emph{GGI school}} \emph{on asymptotic symmetries and flat holography}: introduction to asymptotic symmetries, to celestial holography, to twistor methods for amplitudes, to amplitude methods for gravitational waves, to Carollian holography. I have the pleasure to propose you the first in this series.

\medskip

\emph{Disclaimer:} these lecture notes cover only a small part of the large amount of interesting work that has been done in this topic. They furthermore present a rather personal viewpoint, built on my own perspective and work, and limited by it. I have added in the end a more extended bibliography. Any feedback, corrections and comments are welcome.

\section{Boundaries and the covariant phase space}

In the context of gauge theories and gravity, we will talk about boundary symmetries in the sense of residual gauge transformations allowed by the boundary conditions. 
Boundary gauge transformations acquire a special status if they are not degenerate directions of the symplectic 2-form, suggesting that they may not be a redundancy of the description. Rather, they could change the way the boundary data influence the physical interpretation of the solution.
The simplest, and possibly oldest application of this idea, is an asymptotic diffeomorphism at spatial infinity. Assuming fall-off conditions to a flat metric leaves as residual diffeomorphisms the isometries of the flat metric, namely Poincar\'e transformations, and their interpretation is to describe the same physical spacetime as it would look like from the perspective of observers that can be translated, rotated or boosted with respect to one another.
This idea can be applied also to null infinity, to finite boundaries, and to other gauge theories than gravity. In all cases, one needs first a study of boundary conditions to identify the residual gauge transformations, and then an analysis of Noether's theorem and canonical generators in order to determine the dynamical properties of the symmetry charges.

\begin{figure}[H] 
\begin{center}  \includegraphics[width=12cm]{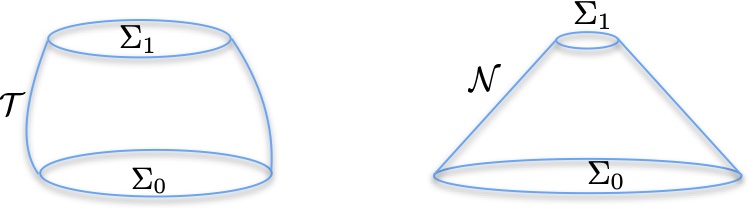} \end{center}
\caption{\label{FigB}    {\small {\emph{Two space-like hypersurfaces $\Si_1$ and $\Si_2$ joined by a time-like boundary (left panel, $\cal T$) or a null boundary (right panel, $\cal N$).}} }}
\end{figure}

A useful setup to have in mind when thinking about boundary conditions is a finite region of spacetime, bounded by two space-like hypersurfaces, as in Fig.~\ref{FigB}.
The codimension-1 boundary connecting the two hypersurfaces could be time-like, or null. If the spatial hypersurfaces extend all the way to infinity and data on them captures all the solutions of the physical theory under consideration, we refer to them as Cauchy hypersurfaces. Otherwise we will generically refer to them as partial Cauchy hypersurfaces. 

\subsection{Hamiltonian phase space}
 
 The covariant phase space is a very convenient tool to discuss boundary symmetries. Before introducing it, let us briefly recall the more conventional construction of phase space through the canonical formalism. For a simple mechanical system with second order equations of motion, one intersects the space of trajectories with an `initial time surface', whose position and velocity can be taken as initial conditions identifying each trajectory. 
The space of such initial conditions, $\cal P,$ can be equipped with a symplectic structure induced from an action principle in Hamiltonian form:
\be
S=\int dt \left(p\dot q-H\right)\quad \Rightarrow\quad \th:=pdq, \quad \om:=d \th = dp\w dq.
\ee
We call $\cal P$ so equipped the \emph{phase space}, with $\th$ the symplectic potential, and $\om$ the symplectic 2-form,
which is closed, non-degenerate, and conserved on solutions: 
\be\label{dotom0}
d\om=0, \qquad \det\om\neq 0,\qquad\dot \om\eqons 0.
\ee
Here and in the following the short-hand notation $\eqons$ means on-shell of the equations of motion.
To prove the last statement, we define the vector field 
\be
\p_t=\dot q\p_q+\dot p\p_p, \qquad \dot\om=\pounds_{\p_t}\om =di_{\p_t}\om=d( -\dot pdq+\dot qdp)\eqons d^2H = 0,
\ee
where in the last step we used the identity $d^2=0$.

More in general, it is useful to recall the definition of Hamiltonian vector field.
A generic vector field on $T{\cal P}$ can be written as $v=v^q\p_q+v^p\p_p$. 
We will call it Hamiltonian if it preserves the symplectic structure, namely if $\pounds_v\om=0$. This guarantees (assuming trivial cohomology, so that every positive-degree closed form is also exact) that its flow is generated by a scalar function in the phase space, denoted $h_v$ and called Hamiltonian of the vector field:
\be
\pounds_v \om=i_vd\om+di_v\om=di_v\om=0 \quad \Rightarrow \quad -i_v\om=d h_v.
\ee
This is a kinematical property of $v$, independent of the dynamics. The reason for the name Hamiltonian comes from the example of the energy in canonical variables and the evolution flow. 
Other notable examples are the momentum and angular momentum, which are the Hamiltonians of translations and rotations, respectively. The map between Hamiltonian vector fields and scalar functions is also known as moment map in the more mathematical literature. 
Conversely, any phase space function $F$ defines an Hamiltonian vector field $\hat F$ via $-i_{\hat F}\om=dF$, thanks to the non-degeneracy of the symplectic 2-form.
We will equivalently use the notation $v\leftrightarrow h_v$ or $\hat F\leftrightarrow F$ for the moment map, depending on whether we want to stress the vector field or Hamiltonian side of the map. A flow that preserves the symplectic 2-form is a canonical transformation, also called a \emph{symplectomorphism}.

Invertibility of the symplectic form guarantees its equivalence to the Poisson bracket, defined as 
\be
\{F,G\}:=\p_qF\p_pG-\p_pF\p_qG, \qquad \{q,p\}=1,
\ee
and satisfying
\be
\{F,G\}= i_{\hat F}i_{\hat G}\om = \om(\hat G,\hat F).
\ee
From this perspective, the special property that makes  $\hat F$ a Hamiltonian vector field is that its flow can be written using Poisson brackets, 
\be
\pounds_{\hat F}= \{\cdot,F\}.
\ee

An important property of the symplectic and Poisson structures is that they provide a representation of the Lie bracket of Hamiltonian vector fields:
\be\label{PBalgebra}
i_{v}i_{w}\om = \{h_v,h_w\}= \pounds_w h_v = h_{[w,v]} + c,
\ee
where $c$ is a closed 0-form: $dc=0$ everywhere in the phase space. 
The proof follows from standard exterior calculus and the property of Hamiltonian vector fields:
\be
di_v i_w\om = \pounds_v i_w \om = [\pounds_v, i_w] \om = i_{[v,w]} \om = -dh_{[v,w]}.
\ee
The property \eqref{PBalgebra} is particularly useful when the vector fields generate a symmetry algebra, say $[v_i,v_j]=c_{ij}{}^kv_k$ where $c_{ij}{}^k$ are the structure constants, because it guarantees that the canonical generators satisfy the same algebra under Poisson brackets: $\{h_i,h_j\}=c_{ij}{}^kh_k$, again up to a closed 0-form $c$.
With our simplifying assumption of trivial topology, all closed 0-forms are constants. Hence $c$ is a constant in phase space, and Poisson-commutes with every phase space function. This is known as \emph{central extension}, using language from group theory. The value of $c$ cannot be determined from the general analysis and must be computed on a case by case basis. In classical mechanics it is most commonly zero, but there are notable examples such as Brown-Henneaux \cite{Brown:1986nw} and Virasoro algebras \cite{Virasoro:1969zu}.

\subsection{Covariant phase space}
The starting steps of the canonical formalism just reviewed are: a choice of time, of initial value surface, and of momenta identified by the chosen time. These steps hide covariance, a drawback that becomes more significative in relativistic field theory, where the initial data are associated with a choice of Cauchy slice, and even more so in general relativistic field theories, where there is no preferred simultaneity surface to be chosen. The idea of the covariant phase space (whose germ actually goes back to Lagrange himself and pre-dates the canonical formalism) is to associate a symplectic structure to the trajectories themselves, as opposed to the initial data identifying them. Such a construction does not require any choice of time or momenta, and manifestly preserves covariance. To realize this idea, we first define the \emph{field space} as the ensemble of all functions $q(t)$ (not necessarily solutions of equations of motion). We can think of this functional space as an uncountable infinite-dimensional space, for which we can take coordinates $q(t)$ that are labelled by a continuous index $t$, and define the functional derivative $\f{\d q(t)}{\d q(t')}=\d(t-t')$. We view the infinitesimal variations $\d q(t)$ as coordinate differentials, namely $\d$ now denotes the exterior derivative for differential forms on the field space.\footnote{
Promoting functional variations to exterior derivatives is always possible if we introduce a suitable notion of differentiability in field space, typically Fr\'echet.} We denote a generic 1-form $F=\int dt F[q(t)]\d q(t)$, and the wedge product $\cw$. Notice that a 2-form like $\d q(t)\cw \d \dot q(t)$ is not zero since $q(t)$ and $\dot q(t)$ are different functions, just like $dx^\m\w dx^\n$ is not zero as long as $\m\neq \n$. Vector fields have functionals as components, and can be represented  in the coordinate basis as $V=\int dt V[q(t)]\f{\d}{\d q(t)}$. 
If these notions feel initially too abstract, it is useful to ground them by thinking in terms of a standard vector space whose coordinate label has been made continuous,
namely $\m=1,2,\ldots$ becomes $t\in\R$, and we have the following associations:
\be\label{contindex}
x^\m\to q(t), \qquad \f{\p x^\m}{\p x^\n}=\d^\m_\n \to \f{\d q(t)}{\d q(t')}=\d(t-t'), \qquad \p_\m\to \f{\d}{\d q(t)}, \qquad dx^\m = \d q(t).
\ee

Next, we define the interior product $I_V$ pairing forms and vectors, 
\be
I_V F = \int dt dt' V[q(t')] F[q(t)]\f{\d q(t)}{\d q(t')} = \int dt V[q(t)] F[q(t)],
\ee
just like $i_v \a=v^\m \a_\m$ in the finite-dimensional case, 
and the field-space Lie derivative 
\be
\d_V=I_V\d +\d I_V
\ee satisfying Cartan's formula, just like $\pounds_v=i_v d+d i_v$.
Notice that when acting on scalar functionals of $q(t)$, like $q(t)$ itself or the Lagrangian, the field-space Lie derivative has a single term: $\d_V=I_V\d$. It then acquires the connotation of a variation specialized to the direction identified by the vector field $V$. The field-space Lie derivative is thus the way to write specific functional variations in the covariant phase space.\footnote{And it is the reason why I tend to prefer the notation $\d_V$ for the Lie derivative, as opposed to some variants of $L$ that are often found in the literature: it reproduces the standard symbol for functional directional variations when acting on scalars.}

The construction scales up immediately to field theories:
We simply replace $q(t)$ by $\phi(x^\m)$, where $\phi$ is the dynamical field under consideration, and instead of a single continuous label $t$, we have $n$ continuous labels $x^0, \ldots x^{n-1}$ representing $n$-dimensional spacetime. From this perspective, the finite-dimensional case can be thought of as a special case of field theory in $0+1$ dimensions.
Table~\eqref{TableCPS} below summarizes the notation for the exterior calculus in spacetime and in field space.
\begin{table}[H]\begin{center}
\begin{tcolorbox}[title=Exterior calculus, colback=gray!10, colframe=gray, width=0.7\textwidth,
    boxsep=0mm,     
    left=1mm, right=1mm, top=1mm, bottom=4mm,  
    ]
\begin{center}\begin{tabular}{ccc}
{\bf spacetime} & & {\bf field space} \\
$x^\m$ & coordinates & $\phi(x)$ \\  $v=v^\m\p_\m$ & vector field & $V=\int dx V[\phi]\f{\d}{\d\phi}$ \\
$d$ & exterior derivative & $\d$ \\ $i_v$ & interior product & $I_{V}$ \\
$\pounds_v$ & Lie derivative & $\d_{V}$ \\ $\w$ & wedge product & $\cw$ 
\end{tabular}\end{center}\end{tcolorbox} 
\end{center}
\caption{\label{TableCPS} \small{\emph{Notation for the exterior calculus in spacetime and field space.}}}
\end{table}
To keep track of the two sets of operations, it is convenient to introduce a grading for each object, say $(p,P)$ for a quantity that is a $p$-form in spacetime and a $P$-form in field space, and define the bi-graded commutator $[F^{ (p,P)},G^{ (q,Q)}]=FG-(-1)^{pq+PQ}GF$. With these conventions, the two sets of operations are basically `decoupled', and the two differentials commute:
\be
[d,\d]=d\d-\d d=0.
\ee 
Some useful basic commutators are 
\begin{center}\begin{tabular}{lll}
$[d,i_v]=\pounds_v$  & $[d, I_W] =0$ & $[\d,I_V]=\d_V$ \\  $[\pounds_v, d] = 0$ & $[\d,i_v]=0$ & $[\d_V, \d] = 0$ \\
$[\pounds_v, i_w] = i_{[v,w]}$ & $[i_v, I_W] = 0$ & $ [{\d_V}, I_W] = -I_{[V,W]}$ 
\\ $[\pounds_v,\pounds_w] = \pounds_{[v,w]}$ & $[\d_V, \pounds_w] =0$ & $[{\d_V}, \d_W] = -\d_{[V,W] }$
\end{tabular}\end{center}
Notice the opposite signs in the last two lines between spacetime and field-space commutators. This is a direct consequence of our convention of commuting $d$ and $\d$.

The formal construction we just described can be made mathematically precise using the formalism of \emph{jet bundles}, which extends the tangent and cotangent bundles used in the canonical formalism.
In a jet bundle the base is the argument of the field, and the fibres are the field and all its derivatives seen as independent elements  (each a different \emph{jet}), up to the desired (finite) order, called degree of the bundle.
Just like a vector field is a section of the tangent bundle, a field and its derivatives seen as functions of the coordinates are a section of the jet bundle. 
We also remark that the two differential structures summarized in Table~\ref{TableCPS} can be unified into a single \emph{variational bi-complex} \cite{Anderson,Barnich:2000zw,Barnich:2001jy}, with total differential $d+\d$, 
and a common wedge product.\footnote{This requires that $d$ and $\d$ \emph{anti-commute}, in order for the total differential to square to zero. 
Notice that components still commute, namely $\d\p_\m\phi=\p_\m \d\phi$: the minus sign in the anti-commutation $\d d\phi=-d \d\phi$ comes instead from swapping the differential forms. We also remark that changing this convention flips the global sign of the field equations and symplectic 2-form, with the commuting convention picking $\om=\d\pi\cw \d\phi$ and the anti-commuting one picking $\om=\d\phi\cw \d\pi$.}
We will not use the theory of jet bundles and the variational bi-complex in the following, but they are useful to keep in mind as a precise mathematical framework for our formal manipulations.

We are now ready to define the covariant phase space (CPS).
To equip the field space with a symplectic structure, we look at the variational principle, and the boundary term induced when we derive the Euler-Lagrange equations
\be\label{dLdth}
\d L=E+d\th\eqons d\th, \qquad \om:=\d\th.
\ee
The field-space 2-form $\om$ so defined is closed and conserved on-shell,
\be\label{dom0}
\d \om =0, \qquad d\om\eqons 0.
\ee
The first property follows by construction since $\om$ is field-space exact, and the second from
\be
d\om=\d d\th = \d E+ \d^2L =\d E.
\ee
 
In this formalism, it is convenient to think of the Lagrangian as a top-form in spacetime, rather than as  a scalar. In other words, we define
\be
S = \int L, \qquad L=\cL \eps,
\ee
where $\cL$ is the Lagrangian scalar, and $\eps=\sqrt{-g}d^nx$ the volume form in $n$ spacetime dimensions. In background-independent theories $g$ is a dynamical variable, it is then also convenient to introduce the Lagrangian density $\tl\cL=\sqrt{-g}\cL$ so that $\d L=\d\tl\cL \, d^nx$.
Having done so, the short-hand notation $E$ for the Euler-Lagrange equations used in \eqref{dLdth} represents a 1-form in field space and $n$-form in spacetime,
\be
E = \f{\d L}{\d\phi}\d\phi = \left(\f{\p L}{\p\phi}-\p_\m\f{\p L}{\p\p_\m\phi}\right)\d\phi,
\ee
with the second equality restricted to the case when the Lagrangian is first order in derivatives.

The fact that the boundary term in \eqref{dLdth} is a good definition of symplectic potential should be clear by comparison with the canonical formalism and the Legendre transform, but can be also verified explicitly. For instance for a non-relativistic point particle in a conservative potential, 
\be\label{L0d}
L=\left(\f12m\dot q^2-V(q)\right)dt, \qquad \d L = (m\dot q\d\dot q-\p_q V\d q) dt= -(m\ddot q +\p_qV)\d q\, dt +d(m\dot q\d q),
\ee
hence
\be
\th = m\dot q\d q, \qquad \om = m\d \dot q\cw \d q.
\ee
We can  check the equivalence to the canonical formulation if we introduce a constant time slice $t=t_0$, and project the trajectories there. The functions become their values at $t_0$, the variations become standard variations of the function's values at that point, and we recover  the canonical formulation:
\be\label{CPStoCan}
q(t)|_{t_0}=q, \quad \d q(t)|_{t_0}=dq, \quad m \dot q(t)|_{t_0}=p, \quad  \d p(t)|_{t_0}=dp, \quad \th|_{t_0}=pdq, \quad \om|_{t_0}=dp\w dq.
\ee

The space of fields equipped with the symplectic structure \eqref{dLdth} is the \emph{covariant phase space}.
From the viewpoint of the variational bi-complex, $\th$ has grading $(n-1,1)$ and $\om$ has $(n-1,2)$. Namely, they are both co-dimension 1 forms in the base manifold, and respectively a 1-form and a 2-form in field space. In the finite-dimensional case, $n=1$, and $\om$ is directly the symplectic 2-form, as \eqref{CPStoCan} shows. In field theory, this is the symplectic 2-form \emph{current}. The actual symplectic structure is its integral over a Cauchy hypersurface $\Si$. See Table~\ref{TableCPS2} for a quick reference. 
By Cauchy hypersurface we mean in the canonical sense that knowledge of initial data on it determines the solutions everywhere. One can also consider `smaller' hypersurfaces that contain only part of the full data, and we will see examples below. In this case one can talk of a \emph{partial} Cauchy slice, and partial phase space associated with it. Given a symplectic 2-form, one can define the analogue of Poisson brackets via $ \{F,G\}_\Si:= I_{\hat F}I_{\hat G}\Om_\Si$, and a bracket density $\{F,G\}:= I_{\hat F}I_{\hat G}\om$. More precisely, these are \emph{Peierls brackets} \cite{Forger:2002ak,Khavkine:2014kya}. 

\begin{table}[H]\begin{center}
\begin{tcolorbox}[title=CPS symplectic structure, colback=gray!10, colframe=gray, width=0.8\textwidth,
    boxsep=0mm,     
    left=1mm, right=1mm, top=1mm, bottom=4mm,  
    ]
\begin{center}
\begin{tabular}{ll}
 $\th$ \quad symplectic potential current; & $\Th_\Si=\int_\Si\th$ \quad symplectic potential  \\
$\om$\quad symplectic 2-form current;  &  $\Om_\Si=\int_\Si\om$ \quad symplectic 2-form\\
\end{tabular}\end{center}\end{tcolorbox} \end{center}
\caption{\label{TableCPS2} \small{\emph{Components of the symplectic structure of the covariant phase space. Here $\Si$ can be a complete or partial Cauchy slice,
and it can be space-like, or null.}} } 
\end{table}

In these lectures we will restrict attention to $n=4$, so the currents are 3-forms. Their Hodge dual is a vector, and we will use the following conventions:
\be\label{Hodge}
\th_{\m\n\r}=\th^\a\eps_{\a\m\n\r}, \qquad \th^\m:=-\f1{3!}\eps^{\m\n\r\s}\th_{\n\r\s}, \qquad d\th=\p_\m\tl\th^\m d^4x, \qquad \tl\th^\m=\sqrt{-g}\th^\m.
\ee

There are two important caveats to discuss. First, in order to be a symplectic structure, $\om$ needs to be non-degenerate. This is typically the case when it is constructed starting from a Lagrangian whose fields are all independent physical degrees of freedom. But it is not the case in the presence of gauge symmetries: as we will see, gauge transformations are degenerate directions of $\om$. 
It is then more accurate to refer to $\om$ as \emph{pre}-symplectic; the actual symplectic 2-form is obtained on the reduced phase space with the degenerate directions removed. 

Second, there are two ambiguities in the construction of the covariant phase space, as was pointed out in \cite{Iyer:1994ys,Jacobson:1993vj}:
\bit
\item[$(b)$] Adding a boundary term to the Lagrangian. This changes $\th$ but not $\om$:
\be
L'=L+d\ell, \qquad \th'=\th+\d\ell, \qquad \om'=\om.
\ee
\item[$(c)$] Adding an exact form to the symplectic potential. This changes both $\th$ and $\om$:
\be
L'=L, \qquad \th'=\th-d\vth, \qquad \om'=\om-\d d\vth.
\ee
\eit
We will refer to these as \emph{boundary} and \emph{corner} ambiguities, respectively. The boundary ambiguity arises because adding a boundary term to the Lagrangian does not change the field equations. The corner ambiguity because the variation of the Lagrangian only defines $d\th$, therefore $\th$ is defined only up to a closed 3-form, or equivalently, $\th^\m$ up to a vector with vanishing divergence. 
Given an $n$-dimensional manifold $M$, a $p$-form functional (with $0<p<n$) that is closed everywhere in field space is also necessarily exact, regardless of the topology of $M$ \cite{Wald:1990mme}.\footnote{The theorem fortunately fails for $p=n$, otherwise every Lagrangian, being a top form hence closed, would be a boundary term!}
It follows that $\th$ is defined up to an exact 3-form, as used in $(c)$, or equivalently, $\th^\m$ up to a total derivative.
We will refer to the choice of $\th$ corresponding to simply removing $d$ as the `reference' choice. Other names such as `standard', or `bare', can also be found in the literature. 
These ambiguities and how to fix them play an important role in the realization of boundary symmetries, and we will discuss them in more details below.
Before doing that, we finish the review of the covariant phase space with the discussion of Hamiltonian and non-Hamiltonian vector fields.

\subsection{Hamiltonian vector fields}

A vector field $V$ in the covariant phase space is Hamiltonian if it Lie-drags the symplectic 2-form current. This guarantees that $I_V\om$ is field-space closed, hence exact if we assume that the field space has trivial topology:
\be
\d_V \om=0 \quad \Rightarrow \quad -I_V\om=\d h_V.
\ee
Conversely, non-degeneracy of the symplectic 2-form guarantees that any field-space functional $F$ defines a Hamiltonian vector field via $-I_{\hat F}\om=\d F$.
The potential $h_V$ is the Hamiltonian current, or Hamiltonian \emph{aspect}, the actual Hamiltonian occurring after integration over $\Si$. 
As before, Hamiltonian vector fields have their flow generated by Poisson brackets, via
\be
\d_{\hat F}= \{\cdot,F\},
\ee
and the Hamiltonian currents provide a realization of the Lie bracket of Hamiltonian vector fields:
\be\label{PBalgebraCPS}
 \{h_V,h_W\}= I_{V}I_{W}\om =\d_W h_V= h_{[V,W]} + c,
\ee
where $c$ is closed in field space: $\d c=0$, and the difference in the final minus sign with respect to \eqref{PBalgebra} stems from the commuting conventions for the bi-complex as explained earlier. The proof of \eqref{PBalgebraCPS} is identical to the canonical case, and follows from
\be
- \d \d_W h_V = \d I_V I_W\om = \d_V I_W \om = [\d_V, I_W] \om = -I_{[V,W]} \om = \d h_{[V,W]}.
\ee
A remark about $c$: this  central extension can be also referred to as a \emph{field-independent 2-cocycle}, since closed-ness of a scalar in field space can equally be stated as field-independence, and it is a functional of two field variations that satisfies the Jacobi identity, hence a 2-cocycle.

\subsection{Non-Hamiltonian vector fields and the Barnich-Troessaert bracket}

Consider now the case of a non-Hamiltonian vector field. Generically, we can write the inner product of any vector with the symplectic 2-form as
\be
-I_V\om =\d I_V\th - \d_V\th = \d h_V - {\cal F}_V, 
\ee
where
\be
\d_V\om = \d I_V\om = \d {\cal F}_V\neq 0
\ee
for a non-Hamiltonian vector field. 
The non-closed term ${\cal F}_V$ is often referred to as `non-integrable'. Notice that the split between `integrable' and `non-integrable' contributions is ambiguous: we can add and subtract an arbitrary integrable quantity $\d X_V$ and distribute it to the two terms, so that
\be\label{Xshift}
h_V=I_V\th+X_V, \qquad {\cal F}_V=\d_V\th+\d X_V.
\ee
The split ambiguity can also be parametrized using the ambiguity $(b)$ of the symplectic potential, with $X_V=\d_V\ell$.
In the following, we will be interested in a specific form of non-integrability: one that is associated with dissipation, namely the non-conservation in time of the symplectic 2-form. In this case, we will address the ambiguity above requiring that \emph{on the subset of the phase space in which dissipation is absent, ${\cal F}_V$  should vanish for all vector fields of interest.}

To see what replaces \eqref{PBalgebraCPS} for non-Hamiltonian vector fields, we compute
\begin{align}
-\d I_W I_V\om &= -\d_W I_V\om +I_W\d I_V\om  =\d\d_Wh_V-\d I_W{\cal F}_V \\\nn
&\hspace{-.7cm}= I_{[W,V]}\om -I_V\d{\cal F}_W +I_W\d{\cal F}_V 
 = -\d h_{[W,V]}+{\cal F}_{[W,V]} +\d(I_V{\cal F}_W-I_W{\cal F}_V) -\d_V{\cal F}_W+\d_W{\cal F}_V,
\end{align}
where in the second line we used the commutator $[\d_W,I_V]$. 
From this we read
\begin{align}
-\d( I_W I_V\om +I_V{\cal F}_W-I_W{\cal F}_V) =\d(\d_Wh_V- I_V{\cal F}_W) 
= \d h_{[V,W]}+\d_W{\cal F}_V -\d_V{\cal F}_W+{\cal F}_{[W,V]}. 
\end{align}
Suppose now that the 2-cocycle at the end of the right-hand side is field-space exact, namely that there exists a functional 
 $ K_{(V,W)}$ such that
\be\label{cocycle}
\d_W{\cal F}_V -\d_V{\cal F}_W+{\cal F}_{[W,V]} = \d K_{(V,W)}.
\ee
We can then define the (generalization of the) Barnich-Troessaert bracket \cite{Barnich:2011mi} (see also \cite{Speranza:2017gxd,Chandrasekaran:2020wwn,Freidel:2021cjp,Chandrasekaran:2021vyu,Rignon-Bret:2024wlu}
\begin{align}\label{BTbracketgen}
\{h_V,h_W\}_*&:=- I_W I_V\om +I_W{\cal F}_V -I_V{\cal F}_W = \d_Wh_V- I_V{\cal F}_W = h_{[V,W]}+ K_{(V,W)}+c,
\end{align}
with $ \d c=0$. Notice that this definition depends on the split between integrable and non-integrable, and changes as $\d_VX_W-\d_WX_V$ under \eqref{Xshift}.
The last equality shows that the Barnich-Troessaert bracket can provide a realization of the algebra, but can be spoiled by the 
\emph{field-dependent 2-cocycle} $K_{(V,W)}$.
Indeed the literature abounds with field-dependent 2-cocycles for symmetry vector fields, including the original paper \cite{Barnich:2011mi}. These typically arise because of a wrong choice of integrable/non-integrable split, and should be removed if possible. It was proved in \cite{Rignon-Bret:2024wlu} that  field-dependent 2-cocycles can be removed if one can implement the generalized Wald-Zoupas prescription of \cite{Odak:2022ndm}. We will come back to this below.

\subsection{The equivalence class of symplectic structures}

As anticipated above, the symplectic potential and 2-form used to construct the covariant phase space of a given Lagrangian theory are not unique.
Instead, the field equations only determine the  equivalence class 
\be\label{defth'}
L'=L+d\ell, \qquad \th\sim\th'=\th+\d\ell-d\vth,\qquad \om\sim\om'=\om-d\d\vth,
\ee
spanned by the freedom to add terms that are exact in field-space or spacetime, associated respectively with boundary terms of the Lagrangian, and corner terms of the symplectic potential. These ambiguities were denoted $(b)$ and $(c)$ in the previous Section, and we now discuss in some details their origin, meaning, and how they can be fixed.

\subsubsection*{Polarizations and boundary Lagrangians} 

In Hamiltonian mechanics, a \emph{polarization} of the phase space is a choice of configuration variables, namely a split of the phase space that identifies a maximal subspace of Poisson-commuting coordinates (also called a Lagrangian submanifold). For instance in the case of $T^*\R^n$,  
standard choices include:
\begin{center}\begin{tabular}{|l|c|l|}\hline
\emph{polarization} & \emph{configuration variable} &  \emph{symplectic potential} \\ \hline
position & $q^i$ & $\th =p_idq^i$ \\ momentum & $p_i$ & $\th =-q^idp_i$ \\ holomorphic & $z^i$ & $\th =\f i4\bar z_idz^i+cc$\\ \hline
\end{tabular}\end{center}
A similar classification can be done also in the covariant phase space, without necessarily referring to positions and momenta, but rather to different jets of the bundle: we can interpret the symplectic potential obtained through  \eqref{dLdth} as $\th=`p\d q$', namely identify the fields appearing under the variation as the `configuration variables' in the jet bundle.
In this way we can attribute a notion of polarization to a given $\th$.
From this perspective, the freedom to add a boundary term to the Lagrangian is akin to the freedom to change polarization, like $\th'=p\d q - \d(pq) = -q\d p$.
In more general systems, not every boundary Lagrangian, or `integration by parts in field space', corresponds to an actual choice of polarization. Nonetheless, it is useful to keep in mind the reference to polarizations in order to guide our intuition.

The boundary term and thus the polarization can in turn be related to specific boundary conditions. Recall in fact that
that the boundary term $\th$ in \eqref{dLdth} has also another application, independent of the covariant phase space: it indicates which boundary conditions make the variational principle well defined. For a given $`p\d q$' form, we have a well-defined variational principle if we hold $q$ fixed at the boundary, or if $q$ varies but $p$ vanishes. Adding a boundary Lagrangian changes the 
 conditions relevant to the variational principle. 
For instance
in our basic example \eqref{L0d}, we can take
\be
\ell = -mq\dot q , \qquad \th'=\th+\d\ell=-mq\d\dot q,
\ee
which changes the boundary conditions from Dirichlet ($\d q=0$, fixed position) to Neumann ($\d\dot q=0$, fixed velocity).
This suggests that we can use physical considerations about the boundary conditions in order to control this ambiguity, and this is what we will do below.

\subsubsection*{Corner terms and the symplectic form}

The second source of ambiguity in $\th$ is that without further prescriptions,  the Lagrangian defines the symplectic potential only up to an exact form. If this is not field-space exact, it will change also the symplectic 2-form by an exact form:
\be\label{om'}
L'=L, \qquad \th'=\th -d\vth, \qquad \om'=\om-d\d\vth.
\ee
We will refer to $\vth$ as to a \emph{corner term} modification to the symplectic potential. It is distinct from adding a corner term to the boundary Lagrangian, which would necessarily produce a field-space exact corner potential, and have no effect on the symplectic 2-form:
\be
\ell'=\ell+dc, \qquad  \th'=\th +\d\ell+\d dc, \qquad \om'=\om.
\ee 

The modification \eqref{om'} of the symplectic structure by a corner term occurs
naturally in different formulations of the same theory. For instance, the Einstein-Hilbert symplectic potential \eqref{thEH} differs by such an exact form from both the ADM one \cite{Burnett:1990tww} and the the tetrad one \cite{DePaoli:2018erh}. It can also occur within the same formulation if one derives the symplectic structure not from \eqref{dLdth} but  using homotopy methods as in \cite{Barnich:2000zw,Barnich:2001jy}, see e.g. discussion in \cite{Oliveri:2019gvm}. 

In principle, one could expect to control this freedom analysing the role played by corner degrees of freedom with respect to the boundary conditions given, see e.g. \cite{Reisenberger:2007ku,Reisenberger:2012zq,Reisenberger:2018xkn,Chandrasekaran:2021vyu,Chandrasekaran:2023vzb,Barnich:2024aln} for discussions in this direction. 
In the meantime, one can also take a more pragmatic attitude, and fix this ambiguity \emph{a posteriori}, for instance from requirements on the realization of the symmetry algebras. This is what we will do in the following.

The importance of corner terms on general grounds was brought to the foreground by \cite{Harlow:2019yfa}, which prompted a more systematic analysis (see e.g. \cite{Freidel:2020xyx,Margalef-Bentabol:2020teu,Freidel:2021cjp,Ciambelli:2021nmv,Chandrasekaran:2021vyu,Odak:2022ndm}).
They play a very important role in the recent developments of asymptotic symmetries. 
Among the applications of corner terms that will be most relevant to us: 
 they allow to remove divergences in the case of asymptotic symmetries (a procedure sometimes called `symplectic renormalization') \cite{Compere:2008us,Compere:2018ylh,Freidel:2021yqe,McNees:2024iyu}, and to achieve covariance and select the right phase space realization of the asymptotic symmetries \cite{Campiglia:2020qvc,Rignon-Bret:2024gcx}.

\subsection{Choosing a representative: conservative vs. dissipative boundary conditions}\label{SecDissOm}

The key question is how to handle the ambiguities \eqref{defth'}. 
Restricting attention to invariant quantities only would be too restrictive and not be particularly useful: for instance, we would not be able to talk about flux-balance laws for boundary symmetries as Noether currents.
Instead, we look for a prescription to select a specific representative of the equivalence class.  A possibility is to use 
a \emph{mathematical} prescription, for instance one could choose a specific boundary Lagrangian and a unique symplectic potential associated to it via Anderson's homotopy operator \cite{Iyer:1994ys,Anderson:1996sc,Barnich:2000zw,Freidel:2020xyx}. We will follow a different approach, 
and use instead a more \emph{physical} prescription, aimed at selecting a representative on a case by case basis, adequate to the problem under consideration.
Our approach to the problem is similar to the perspective used in thermodynamics, where one does not have a universal choice of state functions, but the most suitable ones are chosen only after one specifies the physical system and its boundary conditions.
In this perspective, the initial $\th$ may well be taken with a mathematical prescription, for instance the reference or homotopy one, but it does not matter very much: in the end, only the preferred $\th'$ selected on physical grounds matters.
In other words, we do not look at the ambiguities as a problem, with the associated burden of looking only at ambiguity-independent quantities; but rather as a richness, which allows us the make choices appropriate to different contexts. Pretty much like in the original spirit of Noether's theorem, where one can select the symmetric Maxwell energy-momentum tensor as opposed to the reference one in order to work with gauge-invariant expressions, or in thermodynamics, where one selects the free energy rather than the internal energy for different boundary conditions.

In many cases, and specifically in the context of boundary symmetries, we are only interested in the symplectic potential evaluated on a specific hypersurface, for instance the boundary $\cal B$, or the initial data surface $\Si$,
and in the (possibly partial) phase space there defined by the pull-back $\pbi{\th}$. The ambiguities then show up in the possibility of rearranging the pull-back $\pbi{\th}$ as follows,
\be\label{pbth'}
\pbi{\th}=\th_{\sscr\cal B}-\d\ell+d\vth, \qquad \om_{\sscr\cal B}=\d\th_{\sscr \cal B} = \pbi\om -d\d\vth.
\ee
Any $\th_{\sscr\cal B}$ so defined is a good symplectic potential for the phase space at $\cal B$. Notice that this  is a more generic situation than \eqref{defth'}, which requires $\ell$ and $\vth$ to be defined everywhere. In the special case \eqref{pbth'}, all new quantities may be defined only at $\cal B$, and not on the whole spacetime. This occurs specifically if the pull-back involves extra fields defined only at $\cal B$. Extra care is then needed over whether the extra fields should or should not affect the dynamics, and we will talk about this below.

To select a representative $\th_{\sscr\cal B}$, we will be guided by the boundary conditions and their physical interpretation. These can be broadly classified in two categories: conservative and not. 
Recall that the symplectic current satisfies $d\om\eqons 0$, which is the field theory equivalent of $ \dot\om\eqons 0$. While at first sight similar, the field theory equation does not immediately imply that the symplectic 2-form is conserved in time.
To understand why, let us integrate it over a bounded region of spacetime, as in Fig~\ref{FigB}.
By Stokes' theorem, 
\be
\Om_{\Si_1}\eqons \Om_{\Si_2}+\Om_{\cal B},
\ee
where the boundary $\cal B$ can be for instance time-like or null.
If the fields vanish at the boundary $\cal B$, then the symplectic 2-form is the same at the two space-like hypersurfaces: we can then conclude that it is constant in time.
This is the situation we are most familiar with in field theory, where it is typically implemented  taking the boundary at infinite distance, and the fields falling off sufficiently fast.

More in general, it is the boundary conditions at $\cal B$ that determine whether $\Om_{\cal B}$ vanishes or not, be it at finite or infinite distance.
The two broad categories are:
Conservative boundary conditions, for which $\Om_{\cal B}=0$, and not conservative, or open, boundary conditions, for which 
$\Om_{\cal B}\neq 0$. In the second case there is `symplectic flux' through the boundary, and the data specified on $\Si_1$ are not sufficient to reconstruct the data on $\Si_2$.
Intuitively, one can think of information being lost or added through the radiation outgoing $\cal B$ or incoming.
The open boundary conditions are also known as radiative, leaky, or dissipative/absorbing. In the following we will be mostly interested in the dissipative case, hence we will use this term.

A finer characterization of the conservative/dissipative distinction can be obtained if we look at the symplectic potential on the boundary. This will generically be of the form $\th_{\sscr \cal B}=p\d q$ for some set of $q$'s and $p$'s, where $q$ here represents a complete set of independent configuration variables, in other words $p\d q$ should be an admissible choice of polarization.  We will characterise  conservative boundary conditions as $\d q\stackrel{\sscr \cal B}=0$: this guarantees that $\om_{\sscr \cal B}=0$. 
For dissipative boundary conditions, it is crucial to first be able to identify the degrees of freedom responsible for the dissipation, and
a special class of solutions for which these degrees of freedom are not excited, and dissipation does not occur. 
We will refer to the special solutions using the generic term `stationary', having in mind the time-independence of the symplectic form.\footnote{This is weaker than the use of the term in general relativity, where stationarity has the technical meaning of a solution admitting a time-translational Killing vector, and there can be solutions of Einstein's equations with no radiation (hence `stationary') but also no time-translational Killing vector (hence not stationary).} 
Depending on the context, these special solutions can be non-dissipative, non-radiative, equilibrium configurations, etc. Having identified a relevant class of `stationary' solutions, we seek a polarization of the symplectic potential such that $p$ vanishes on them.
The power of this criterion is that we can study dynamics in dissipative situation with the solid benchmark that conservation automatically occurs when the systems undergoes a non-dissipative epoch.

\begin{table}[H]\begin{center}\begin{tabular}{|c|c|l|}\hline
\emph{boundary conditions} & \emph{definition} & \emph{chosen representative} \\ \hline
conservative & $\om_{\sscr\cal B}=0$ &  $\th_{\sscr\cal B} = p\d q$ with $\d q=0$ on every solution \\
dissipative & $\om_{\sscr\cal B}\neq 0$ & ${\th}_{\sscr\cal B} = p\d q$ with $p=0$  on `stationary' solutions \\ \hline
\end{tabular}\end{center}\caption{\label{TableTwoCases} \emph{\small{Generic criteria to define conservative and dissipative boundary conditions, and to select a representative in the equivalence class \eqref{pbth'} of symplectic potentials.}} }\end{table}

Notice that while in the conservative case we should impose $\d q=0$ for every variable, in the dissipative case we can have mixed situations as well. For instance, the identification of a notion of stationarity may require some partial restrictions on the field variations at the boundary. In this case some of the variables would still satisfy `conservative' boundary conditions, and only the remaining ones the `stationarity' one. This can occur in gauge theories and gravity, with the variables fixed corresponding to gauge-fixing conditions. 

The two categories are summarized in Table~\ref{TableTwoCases}. This discussion is a generalization and a formalization of the Wald-Zoupas prescription \cite{Wald:1999wa} (where 
the conservative and dissipative case are referred to as I and II, respectively) that we proposed in \cite{Odak:2022ndm}. 
Similar ideas have also been put forward in \cite{Chandrasekaran:2018aop,Chandrasekaran:2021vyu}. 
The original Wald-Zoupas prescription is extended in various directions: in the physical sense, by stressing the flexibility of the notions of `conservative' and `stationarity' so that they can be applied to different contexts, by relating them to choices of boundary Lagrangians and boundary conditions, 
and in the technical construction, with the crucial inclusion of corner terms following \cite{Harlow:2019yfa}, and of anomalies and field-dependent transformations following \cite{Freidel:2021cjp}, which allows one to include renormalization procedures and extend the applicability of the prescription.

This \emph{generalized Wald-Zoupas prescription} needs however a more precise characterization than the boundary conditions of \eqref{TableTwoCases}: if the selected symplectic potential depends on non-dynamical boundary fields, a prescription on how to handle them has to be given as well. Specifically in the case of general relativity, one should make sure that the chosen polarization is \emph{general covariant} and background-independent, and this imposes restrictions on the available choices. In many cases of interest, general covariance can be guaranteed by a weaker condition, that we referred to as `Wald-Zoupas covariance' in  \cite{Odak:2022ndm}: background-independence of the action of the boundary symmetry group. 
This is particularly relevant in the construction of Noether charges for boundary symmetries, as we will see below.

Summarizing, we take a physical perspective on the equivalence class \eqref{pbth'}: identify a representative that satisfies the requirements dictated by the physical problem of interest, schematically divided in two categories as in Table~\ref{TableTwoCases}. If the identification is possible using only the $\ell$ freedom, then the change is akin to a change of polarization. If a corner term is needed as well, then the change is more subtle and means that corner degrees of freedom play a role. As we will see below, the physical considerations are typically enough to select a unique potential is unique. Then the problem of ambiguities is completely eliminated. If on the other hand it is not unique, further analysis would be needed. 

\subsection{Examples}
Let us work out the standard CPS symplectic structure for a few field theories of interest.
\bit
\item Klein-Gordon scalar field in Minkowski, with $g_{\m\n}=\eta_{\m\n}$, and $\eps$ the flat volume form:
\be\label{Lphi}
L=\left(-\f12\p_\m\phi\p^\m\phi -V(\phi)\right)\eps, \qquad \th^\m=-\p^\m \phi \d\phi.
\ee
\note{Proof:
\begin{align}\nn
& \d L = \left(-\p_\m\d\phi\p^\m\phi -\p_\phi V\d\phi\right)\eps = (\square \phi - \p_\phi V) \d\phi \, \eps - \p_\m(\d\phi\p^\m\phi)\eps. 
\end{align}
}
Projecting on a space-like slice we recover the usual canonical formalism,
\be
\th^t=\pi \d\phi, \qquad \pi=\dot\phi.
\ee
The reference symplectic potential corresponds to Dirichlet conservative boundary conditions, and $\dot\phi=0$ stationarity condition.
A boundary term to switch from Dirichlet to Neumann conservative boundary conditions on a time-like boundary with unit normal $n_\m$ is $\ell = -\phi\pounds_n\phi\, i_n\eps$.

\item Maxwell and Yang-Mills fields in Minkowski:
\be
L=-\f14\tr(F_{\m\n}F^{\m\n})\eps, \qquad \th^\m=-\tr(F^{\m\n}  \d A_\n).
\ee
Projecting on a space-like slice we recover the usual canonical formalism with the (non-abelian) electric field as the conjugated momentum,
\be
\th^t=\tr(\pi^\m \d A_\m), \qquad \pi^\m_i=-F^{0\m}_i = \dot A^\m_i - \p^\m A^0_i - c_i{}^{jk}A^0_jA^\m_k=: E^\m_i.
\ee
\note{Proof: using $\d F=d_A\d A$, we find
\begin{align}\nn
& \d L =-\f12\tr(\d F_{\m\n}F^{\m\n})\eps =- \tr(D_\m\d A_\n F^{\m\n})\eps =\tr(D_\m F^{\m\n}\d A_\n)\eps - \p_\m\tr(\d A_\n F^{\m\n})\eps.
\end{align}
}
This polarization corresponds to Dirichlet conservative boundary conditions $\d A_a=0$, and a notion of stationarity as solutions with vanishing (non-abelian) electric field. A boundary term to switch from Dirichlet to Neumann conservative boundary conditions is $\ell = -n_\m F^{\m\n}A_\n i_n\eps$, the effect being to hold fixed the electric field as opposed to the magnetic vector potential.

\item Chern-Simons theory:
\begin{align}
& L=\f{k}{4\pi}\tr (A\w dA+\f23 A\w A\w A), \qquad \th = -\f{k}{4\pi}\tr(A\w \d A)  
\end{align}
\note{Proof:
\begin{align}\nn
& \f{4\pi}k\d L =\tr(\d A\w dA+A\w d\d A+2\d A\w A\w A) = -d\tr(A\d A) + 2\tr(\d A\w F).
\end{align}
}
Notice that the symplectic potential is not gauge invariant, a fact that captures the well-known fact that the Chern-Simons Lagrangian is only gauge-invariant up to a boundary term.

\item General Relativity, with the Einstein-Hilbert Lagrangian
\be
L = \f1{16\pi}(R-2\L)\eps, \qquad \th^\m = \f1{8\pi} g^{\r[\s} \d \G^{\m]}_{\r\s} = \f1{8\pi} g^{\m[\r}g^{\n]\s} \na_\n\d g_{\r\s}. \label{thEH}
\ee
\note{Proof: follows from the Palatini identity $g^{\m\n}\d R_{\m\n}= 2g^{\r[\s}\na_\m \d \G^{\m]}_{\r\s}$.
}
The presence of normal derivatives of the metric in the boundary term suggests that this polarization corresponds to Neumann conservative boundary conditions.
We will confirm this below with a dedicated analysis, and discuss alternative boundary terms and polarizations.

\item General Relativity in tetrad variables, with the Einstein-Cartan Lagrangian:
\be\label{LEC}
L  = 
\f1{32\pi}\eps_{IJKL}~e^I\w e^J \w \Big(F^{KL}(\om) - \f\L6 ~ e^K\w e^L\Big), \qquad \th^\m= \f1{8\pi}e^{[\m}_I e^{\n]}_J\d\om_\n^{IJ},
\ee
where $\om=\om(e)$ is the Levi-Civita spin connection, satisfying metricity and torsion-freeness: $d_\om\eta^{IJ}=d_\om e^I=0$.
\note{Proof: using the identity $\d F=d_\om\d\om$, and the torsion-free condition, we have
\begin{align}
& \d L =-\f1{8\pi} \d e^I\w E_I +d\th, \\
& E^\m_I = \f1{4}\eps_{IJKL} \eps^{\m\n\r\s} e^J_\n \left(F^{KL}_{\r\s}-\f23\L e^K_\r  e^L_\s \right) = (G_{\m\n}+\L g_{\m\n}) e^\m_I, \\
& \th= \f1{32\pi}\eps_{IJKL}~e^I\w e^J\w \d\om^{KL}. \label{thEC1}
\end{align}
The expression for $\th^\m$ follows from the Hodge dual convention \eqref{Hodge}.}

The polarization corresponds to holding the spin connection fixed at the boundary, which is the analogue of Neumann boundary conditions for the first order system.
The tetrad formulation offers a simple example of how `reference' symplectic potentials for the same theory in different variables can differ by an exact 3-form.
In fact, even though \eqref{thEH} and \eqref{LEC} give equivalent field equations, their reference symplectic potentials differ:
\be\label{thda}
\th_{\sscr EH} = \th_{\sscr EC} + d\a,
\ee
where
\be \label{DPS}
\a = \star (e_{I}\w \d e^{I}) = -\frac{1}{2}\eps_{IJKL}~e^I\w e^J \left(e^{\r K}\d e^L_\r\right)
\ee
 is the 2-form introduced in \cite{DePaoli:2018erh}. Another example of exact-3-form difference occurs with the ADM formulation \cite{Burnett:1990tww,Freidel:2020xyx}.

\item General relativity in the first-order formulation: same  Lagrangian as \eqref{thEH} or \eqref{LEC}, but now the connection and metric, or spin connection and tetrads, are taken as independent variables. This offers the possibility of adding to the Lagrangian the parity-odd term
\be\label{Holst}
e_I\w e_J \w F^{KL}(\om) = -\tl\eps^{\m\n\r\s}R_{\m\n\r\s}(\G) \, d^4x.
\ee
Often referred to as Holst term, it is the starting point of Loop Quantum Gravity \cite{Holst:1995pc}, with coupling constant being (the inverse of) the Barbero-Immirzi parameter $\g$. 
Since it contains derivatives, it gives a contribution to the symplectic potential. In vacuum or with matter Lagrangians not involving the connection, the only solution of the connection  equations is Levi-Civita. We thus have equivalence with the second-order formulation, and \eqref{Holst} vanishes. But if we derive the symplectic potentials off-shell, \eqref{Holst} leads to different reference choices in a way similar to the difference between \eqref{LEC} and \eqref{thEH}. The result is now \eqref{thda} with $\a$ given by \eqref{DPS} plus the extra term
$\a_\g = (1/\g)e_{I}\w \d e^{I}.$ 
This effect has been studied in the context of \emph{dual} gravitational charges \cite{Godazgar:2018qpq,Godazgar:2020kqd,Oliveri:2020xls}.
If the matter Lagrangian involves the connection, the field equations introduce torsion. In this case the metric and tetrad theories not only differ from the second order formulation,  they also have different field equations, and the reference symplectic potentials differ by a non-exact 3-form \cite{Oliveri:2019gvm}.

\eit

\section{Noether's theorem for gauge symmetries and gravity}

Noether exposed a profound relation between conservation laws and differentiable symmetries (continuous and connected to the identity), and which is at the heart of our lectures.

\medskip

\begin{tcolorbox}[title=Definition: differentiable symmetry, colback=gray!10, colframe=gray, width=\textwidth,
    boxsep=0mm,     
    left=1mm, right=1mm, top=1mm, bottom=4mm,  
    ]
Given a Lagrangian $L(\phi)$, an infinitesimal transformation $\d_\veps\phi$ with continuous parameter $\veps$ is a symmetry if it leaves the field equations invariant, namely the variation of the Lagrangian is at most a boundary term:
\be\label{infsym}
\d_\veps L = dY_\veps.
\ee
\end{tcolorbox}

\noindent Here the field-space Lie derivative has a single term $\d_\veps=I_\veps\d$, since the Lagrangian is a field-space scalar, and $I_\veps:=I_{V_\veps}$ is a short-hand notation for the internal product with the symmetry vector field
\be
V_\eps:=\int \d_\veps\phi\f\d{\d\phi}.
\ee

\begin{tcolorbox}[title=Noether theorem, colback=gray!10, colframe=gray, width=\textwidth,
    boxsep=0mm,     
    left=1mm, right=1mm, top=1mm, bottom=4mm,  
    ]
For every differentiable symmetry of the Lagrangian there exists a current conserved on-shell, given by
\be\label{defjN}
j_\veps:=I_\veps\th-Y_\veps, \qquad dj_\veps = -I_\veps E\eqons 0.
\ee 
Furthermore, if the symmetry transformation depends on derivatives of the symmetry parameters, then the field equations are not independent, and the conserved current is on-shell exact: 
\be\label{NT2}
j_\veps  \eqons dq_\veps.
\ee
\end{tcolorbox}

The 3-form $j_\veps$, or its Hodge dual vector via \eqref{Hodge}, is the Noether current of the symmetry $\veps$.
Its integral $Q_\veps[\Si]:=\int_\Si j_\veps$ is the Noether charge, and we will refer to $q_\veps$ as  charge \emph{aspect}. 
The proof of the first statement follows immediately from $\d_\veps L = I_\veps E + dI_\veps \th= dY_\veps$. The proof of the second is only slightly longer, and we give it below. The two statements are also known separately as first and second Noether's theorems. 
Notice the power and elegance of the covariant phase space methods: compact and transparent formulas, and straightforward proofs.

\note{{\bf Proof of Noether's second theorem.} 
Suppose that the infinitesimal transformation of the fields contains \emph{derivatives} of the symmetry parameters, schematically $\d_\veps\phi = \veps \phi +\phi \p \veps$ (the terms in $\phi$ are just bookkeepers, these could in principle contain functions of the fields and their derivatives), as it happens in gauge transformations and diffeomorphisms, see \eqref{Agt} and \eqref{gdiffeo} below. In this case we can write
\be
I_\veps E = \f{\d L}{\d\phi}\d_\veps\phi = \f{\d L}{\d\phi}(\veps \phi +\phi \p \veps) = \veps\left(\f{\d L}{\d\phi} \phi - \p (\f{\d L}{\d\phi} \phi)\right) +d \left( \f{\d L}{\d\phi}\veps\phi \right).
\ee
Then from \eqref{defjN}
\be
d(j_\veps + \f{\d L}{\d\phi}\veps\phi ) =  \veps\left(\p (\f{\d L}{\d\phi} \phi) - \f{\d L}{\d\phi} \phi \right).
\ee
The round bracket on the RHS must vanish in the bulk, because $\eps$ is an arbitrary parameter, and the LHS is a boundary term only. By continuity, it has to vanish on the boundary as well. From this, we can conclude two things. First, that the round bracket on the RHS vanishes, and this is an off-shell identity (these are called Noether identities, or generalized Bianchi identities). Second, that the round bracket on the LHS is a closed form everywhere in field space, hence exact. Denoting the exact 3-form $dq_\veps$, we can write
\be\label{j2}
j_\veps = dq_\veps - \f{\d L}{\d\phi}\veps\phi \eqons dq_\veps.
\ee
Dependence on derivatives of the symmetry parameters is precisely what happens in gauge theories and gravity. 
After pull-back on a chosen hypersurface, one gets those components of the field equations that are identified as constraints in the canonical analysis. 

 }

The second statement has the consequence that gauge symmetries have trivial Noether charges if there are no boundaries.\footnote{In the following, we will loosely refer to symmetries whose transformation laws depend on derivatives of the parameters as `gauge symmetries', including in this term both gauge transformations of connection theories, and diffeomorphisms of gravitational theories, and refer generically to the connections and metric as `gauge fields'. We will be more specific if the difference is relevant.}
If there are  boundaries, and \emph{if the boundary conditions allow all or some gauge transformations at the boundary}, we obtain a non-trivial Noether charge, given by a surface integral.
Boundary conditions thus play a crucial role in determining if, and which, gauge transformations can become symmetries. For the rest of the discussion, it is sufficient to assume that some  gauge transformations are allowed at the boundary.

Ambiguities: an important point is that \emph{the Noether current is only defined up to cohomology}, namely up to a closed 3-form, or exact with our assumption of trivial cohomology.
The ambiguity can be understood as the possibility of adding to both sides of \eqref{NT2} a current $ds_\veps$ that is trivially conserved, i.e. without using the field equations. The new Noether currents and charge aspects are $j'_\veps=j_\veps+ds_\veps$ and $q'_\veps=q_\veps+s_\veps$.
In fact, both $\th$ and $Y$ are only defined modulo exact forms, so we need to choose specific \emph{representatives} in order to write the formula \eqref{defjN}.\footnote{The boundary Lagrangian ambiguity $(b)$ on the other hand does not affect the Noether current: $I_\veps\th'=I_\veps\th+\d_\veps\ell$, $Y'_\veps=Y_\veps+\d_\veps\ell$ and $j'_\veps=j_\veps$. Notice furthermore that there is no guarantee that $\d_\veps\ell$ be closed, so generically this would not be an  allowed modification of the Noether current.}
The ambiguity of the Noether current is well-known already in the case of global symmetries, albeit more often stated in terms of adding total divergences to the current seen as a vector, rather that exact forms. It becomes more  consequential in the gauge case, where the current itself is exact on-shell: the whole current and its aspect $q_\veps$ are a priori  \emph{completely ambiguous}. For instance, one can use this freedom to choose a Noether current for gauge symmetries that is exactly zero on-shell, aka `weakly vanishing' \cite{Barnich:2001jy}. 
We will instead fix the ambiguity choosing the `reference' $\th$ and $Y$ (namely the ones obtained removing the $d$).
With either choice, one should keep in mind the existence of ambiguities, and investigate how they affect any result obtained.
This analysis is simpler if there is no $Y_\veps$ term, as it happens for instance in Maxwell and Yang-Mills gauge transformations. A non-trivial $Y_\veps$ occurs for instance in Chern-Simons and gravity. Fixing the cohomology ambiguity of $j_\veps$ still leaves a cohomology ambiguity in $q_\veps$, but this will play no role if we restrict attention to charges integrated over compact surfaces. The freedom of adding closed 2-forms to the charge aspects will be neglected throughout the rest of the discussion.

Let us explore the consequences of Noether's first and second statements, assuming that we have fixed the ambiguity with the reference choice, for instance. By Stokes theorem,
\be\label{Qcons}
Q_\veps[\Si_1]\eqons Q_\veps[\Si_2]+Q_\veps[\cal B].
\ee
If the boundary conditions at $\cal B$ make $Q_\veps[\cal B]$ vanish, then the Noether charges are conserved between one hypersurface and the next, namely they are constant in time.
Observe the difference between a mechanical system and a field theory: in the first case the Noether charges are automatically conserved in time on solutions, whereas in the latter this requires specific boundary conditions. 

On top of this codimension-1  laws, in gauge theories and gravity we also have codimension-2  laws relating the Noether current on $\Si$ to the boundary of $\Si$  
via \eqref{NT2}. Upon integration of this equation, we find
\be\label{chargeflux}
Q_\veps[\Si]=\int_\Si j_\veps \eqons \oint_{\p\Si} q_\veps = Q_\veps[\p\Si].
\ee
Here we denoted $\p\Si$ the boundary of $\Si$, which can have disconnected components, and for simplicity we assume it to be closed.
We refer to the term on the right as \emph{surface charges}, because the support of the integral is on surfaces (or codimension-2 space in general dimensions).
We thus find an important property of gauge symmetries:  their Noether charges are surface integrals, a result that follows immediately from the second statement of Noether's theorem. Another important property that we will see below is that they correspond to degenerate directions of the symplectic 2-form.

\begin{figure}[H] 
\begin{center}  \includegraphics[width=12cm]{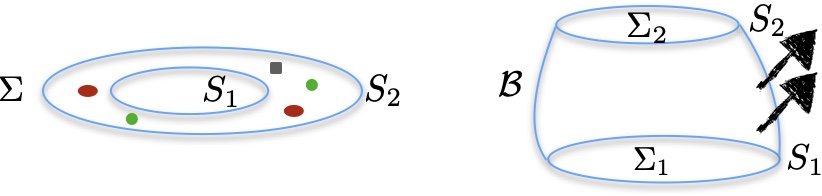} \end{center}
\caption{\label{FigU1}    {\small {\emph{Different applications of the co-dimension 2 flux-balance laws. 
} Left panel: \emph{On a single space-like hypersurface $\Si$ with two boundaries, the surface charge difference is determined by the matter content in between.} Right panel: \emph{Between different times, with dissipative boundary conditions -- allowing residual gauge transformations at the boundary, to which $\veps$ must belong -- the surface charge difference is determined by the flux, here drawn as purely outgoing for example.}} }}
\end{figure}

The Noether relation \eqref{chargeflux} can have different important applications, depending on the nature of $\Si$ and of its boundary. 
If $\Si$ has a single boundary, this equation determines the value of the Noether surface charge as a function of the bulk field configuration, which can be typically related to the matter sources via the field equations. If $\Si$ has two disconnected boundaries, the integral of the Noether current is a \emph{difference} of surface charges. This can also be called \emph{flux}, 
in the sense that it captures the variation of the surface charges associated with the two boundaries. The Noether relation \eqref{chargeflux} then provides a \emph{flux-balance law} between the variation of the surface charges and the field behaviour in the bulk. This flux-balance law can be `instantaneous', of $\Si$ is space-like, or dynamical, if we integrate instead over a time-like or null boundary $\cal B$. See Fig.~\ref{FigU1}. In other words, the Noether charge of a gauge symmetry defined as the integral of the Noether current  can be either a surface charge, or a flux, namely a variation of surface charges, depending on the integration surface chosen. 
To highlight this versatility, we will often refer to the 3-form equation \eqref{NT2} as a flux-balance law, in a more abstract sense before specializing to a given hypersurface and the allowed gauge transformations on it.

The simplest example of such co-dimension-2 conservation law is Gauss's theorem relating the total electric charge in a region of space to the flux of the electric field, and we will cover this example in details shortly. In this case, the caveat about language becomes even more important, because the surface charge aspect is the electric flux, so in the case of $\Si$ with two boundaries, the flux between surface charges is a difference of electric fluxes.
You will notice that depending on context, what one calls charge and flux can easily be swapped and confused. It is important to keep your eyes open and not just your ears, to avoid misunderstandings.

Finally, but also very important, there is also an addendum to the second statement  in Noether's theorem.
When a gauge theory is coupled to matter, there is a special case that can occur: gauge transformations that leave the gauge fields invariant, but affect the matter. These are usually referred to as \emph{global gauge transformations}, because in electromagnetism this occurs for constant gauge transformations;
in the non-abelian case these are covariantly-constant gauge parameters, also referred to as rigid, or residual, gauge transformations, or as isotropies;
in the gravitational case these are Killing vectors, or isometries.
They are still `global' in the sense that they are fixed everywhere in the region of interest starting from a reference value, and not independently attributed at each point. 
The Noether current for global gauge transformations is exact on-shell as for any other value of the gauge parameter, so the statement of the theorem does not change. 
The main difference is that since global gauge symmetries leave the gauge field invariant, it is possible to consider them also as symmetries of a theory with only matter, and no dynamical gauge fields (they could be set to zero, or to any value, and be treated as a non-dynamical, background field). 
We can then think of the global gauge symmetry as a standard physical symmetry for the matter fields alone. 
Removing the gauge fields from the dynamics clearly changes the physics of the problem however, so some properties will be different, notably concerning boundaries.
We will see this explicitly in the examples below.

\subsection{Noether charges as canonical generators}\label{SecHamGen}

Another important property of Noether charges is that they provide canonical generators for the flow of the symmetry vector fields. 
This can be easily proved using covariant phase space methods. We will see that it works only if the symmetry vector fields are Hamiltonian, and what obstruction arises if they are not.
We start by contracting $\om=\d\th$ with a symmetry vector field, obtaining
\be\label{Iepsom}
-I_\veps\om = -\d_\veps\th+\d I_\veps\th =-\d_\veps\th+ \d j_\veps +\d Y_\veps.
\ee
Here we have taken as before the short-hand notation of representing the symmetry vector field with its symmetry parameter,
and the second equality follows from the definition of the Noether current \eqref{defjN}.
In order to write these formulas, we are making an explicit choice of representative for $\th$, $j_\xi$ and $Y_\xi$. 
In many cases of interest, including global symmetries and gauge transformations in Maxwell and Yang-Mills theories, $\d_\veps\th$ and $Y_\veps$ both vanish, and we immediately derive that the reference Noether current is the canonical generator. 

But how about the general case? 
Equation \eqref{Iepsom} shows that the Noether current is still a candidate Hamiltonian generator, but also that a symmetry vector field is not necessarily Hamiltonian! This property requires that  $\d\d_\veps\th=0$. 
To see when this happens, we use the definitions of symplectic potential and Lagrangian symmetry to derive
\be
d \d_\veps \th \eqons \d_\veps\d L = d \d Y_\veps \qquad \Rightarrow\qquad \d_\veps\th \eqons \d Y_\veps+dX_\veps,
\ee
for some $(2,1)$ spacetime/field-space form $X_\veps$ (we are still assuming as before that the spacetime cohomology is trivial).
Without further assumptions, $X_\veps$ may not be field-space closed, and thus the symmetry not Hamiltonian. The general result is thus
\be
-I_\veps\om \eqons  \d j_\veps - dX_\veps.
\ee
The specific form of $X_\veps$ depends on the theory and symmetry considered, but importantly, also on the  $Y$ and $\th$ representatives chosen, even at fixed $\om$. It is this quantity that determines whether the symmetry admits an Hamiltonian or not. Or in equivalent terms, whether the Noether current is `integrable' in field space.
From the general result we can already draw some important conclusions:
\bit
\item For a gauge symmetry, $j_\veps$ is exact by Noether theorem, hence 
\be
-I_\veps\om\eqons d(\d q_\veps-X_\veps),
\ee 
and has support only at the corners: bulk gauge transformations are degenerate directions of the symplectic 2-form, boundary ones generically are not.
Notice that this result contains also the canonical statement that the constraints generate bulk gauge transformations, as can be immediately seen going off-shell.
\item For a symmetry that corresponds to a Hamiltonian vector field, $\d X_\veps=0$. Assuming trivial cohomology in field space, we can write this as  $X_\veps=\d b_\veps$ for some $(2,0)$ spacetime/field-space form $b_\veps$. The Hamiltonian aspect generating the symmetry is thus
\be\label{hjb}
-I_\veps\om\eqons \d h_\veps, \qquad h_\veps\eqons j_\veps+db_\veps.
\ee 
\item For a gauge symmetry that corresponds to a Hamiltonian vector field, \eqref{hjb} holds with $j_\veps\eqons dq_\veps$ exact on-shell, that is
\be\label{hjg}
h_\veps\eqons d(q_\veps+b_\veps).
\ee 
Since the Noether current is only defined up to exact forms, it is possible to use this freedom to select an `improved Noether current' shifted by $db_\veps$ that acts as canonical generator. 
\eit

The analysis above gives the precise relation between the Noether currents and the canonical generators, and the necessary conditions for integrability. 
Specifically to gauge symmetries, we see that the fact that the Noether current is exact implies that bulk gauge transformations are degenerate directions of the symplectic 2-form, hence they don't affect the nature of the solution. 
On the other hand gauge transformations with support on the boundary can be non-trivial in general.\footnote{Sometimes the name \emph{large} is also used, but this term is also used for the completely unrelated notion of gauge transformations not connected to the identity. For this reason I prefer to avoid it and use instead boundary, or asymptotic gauge transformations.} The question is whether boundary conditions allow any gauge transformations at the boundary. 
A prominent example in the Maxwell case are the fall-off conditions at future null infinity, where a residual gauge symmetry of time-independent boundary $\l$'s is allowed, and gives rise to the conservation laws that have remarkably been related to the soft photon theorems of the quantum theory \cite{Strominger:2017zoo}.

We choose to work at the level of currents and  3-forms without committing to a specific pull-back on a given boundary, so that we can make statements that are general to the whole covariant phase space. Once a specific hypersurface $\Si$ is chosen and we integrate over it, we have the relation between the actual generator on the $\Si$ phase space and the Noether charge integrated over $\Si$.
The general analysis can be refined to more specific conclusions once we choose the theory, the boundary and its boundary conditions.
We can distinguish three possible outcomes:

{\bf Case 1: $b_\veps=0$}.
The Noether charge obtained from the reference symplectic potential is directly the canonical generator of the symmetry on a given phase space. 
This is the familiar case that occurs for instance in gauge theories. 

{\bf Case 2: $b_\veps\neq 0$}.
The canonical generator is shifted with respect to the reference Noether charge computed. 
This is the famous case that occurs for instance when deriving the gravitational ADM charges at spatial infinity \cite{Iyer:1994ys}.

{\bf Case 3: $X_\veps\neq \d b_\veps$}. 
The canonical generator is not integrable, and the symmetry is not a Hamiltonian vector field. 
In this case it is still possible to find a useful role for the Noether current identifying subsets of the field space or specific solutions for which integrability does occur, provided one deals with the ambiguities associated with the integrable/non-integrable split.
This is the situation with the BMS symmetries, where a unique prescription exists.

The relation between the Noether currents and the canonical generator gives us some useful insight on the ambiguities. 
Consider cases 1 and 2. The right-hand sides of \eqref{hjb} and \eqref{hjg} depend on the actual Noether current, and not on its equivalence class: adding an exact 3-form to the Noether current would spoil its relation to $I_\veps\om$, unless it is a field-space constant. This suggests that we can use the matching between Noether currents and canonical generators as a criterium to remove the ambiguity in the Noether current. However, we should not forget that $\om$ itself is also defined only up to exact 3-forms, under which
\be
-I_\veps\om'=-I_\veps\om+\d dI_\veps\vth \eqons \d d (q_\veps+b_\veps+I_\veps\vth).
\ee
 Therefore, matching the canonical generator removes the cohomology ambiguity of the Noether current (up to field-space constants) only if one has a prescription to fix the corner ambiguity of the symplectic 2-form. Otherwise, one is just repackaging the ambiguity of $j_\veps$ in the ambiguity $\vth$ of $\om$.
In case 3 the situation is more complicated. We still have the relation between the cohomology ambiguity of the Noether current and adding a corner term to the symplectic 2-form, but we also have the ambiguity in the split between integrable and non-integrable terms, and this can also be related to changing the symplectic potential hence the Noether currents,
\be
-I_\veps\om'=-I_\veps\om+\d dI_\veps\vth \eqons d\big(\d (q_\veps+b_\veps+I_\veps\vth) - X'_\veps\big),\qquad X'_\veps=X_\veps+\d b_\veps.
\ee 
Therefore one needs both a prescription for the corner term of $\om$ and for the integrable/non-integrable split in order to use this relation to remove the ambiguities of the Noether current. In practise, we will typically be able to resolve the issue of ambiguities in the Noether currents by a combination of arguments involving both the relation to the canonical generators and the specific properties of the flux-balance laws \eqref{NT2}.

The issues of ambiguities and non-integrability are typically very minor and easy to deal with in the case of gauge theories, but become central in the case of gravity, where they are furthermore often covered by an extra layer of complexity associated with ensuring general covariance of the prescription.
Let us present these two cases separately.

\subsection{Global vs. local U(1) gauge symmetry}

Consider a complex scalar field, with Lagrangian
\be
\cL = -  \p_\m\phi \p^\m\bar\phi-V(|\phi|).
\ee
From the variation one obtains the field equations and symplectic potential current,
\be\label{phi}
\p^2\phi-\p_{\bar \phi}V=0, \qquad
\th^\m = -\p^\m\bar\phi\d\phi -\p^\m\phi\d\bar\phi.
\ee
It is easy to see that the Lagrangian is invariant under the transformation $\phi\to e^{i\l}\phi$ with $\l\in\R$. This is a $U(1)$ symmetry, whose infinitesimal version is $\d_\l\phi=i\l\phi$. The invariance is exact without boundary term (that is, we have $Y_\l=0$), hence the associated Noether current is simply
\be\label{jphi}
j_\l^\m = I_\l\th^\m = i\l\bar\phi {\ala\p}{}^\m\phi.
\ee
Its on-shell conservation can be easily checked. The Noether charge over a space-like hypersurface $\Si$ is
\be\label{Qphi}
Q_\l[\Si] = \int_\Si j_\l^0 d^3x = -i\l \int_\Si (\bar\phi\dot\phi-\phi\dot{\bar\phi}) d^3x,
\ee
and satisfies \eqref{Qcons}. It is therefore constant in time if the fields satisfy conservative boundary conditions at the lateral boundary.
This is the example of a physical symmetry: different solutions can be distinguished by the value of the Noether charge.

Now let us `gauge' this symmetry, by coupling the complex scalar field to the Maxwell Lagrangian (`scalar electro-dynamics')
\be
\cL = 
-\f14 F^2 -  D_\m\phi \overline{D^\m\phi}-V(|\phi|),
\ee
where $D_\m \phi = (\p_\m+iA_\m)\phi $ is the covariant derivative.\footnote{In the mathematical literature, one often describes $\phi$ and $\bar\phi$ has duals in a complex line bundle, with covariant derivative $D_\m\phi = (\p_\m+A_\m)\phi $ and $D_\m \bar\phi = (\p_\m-A_\m)\bar\phi $.
Both conventions give the same results.}
From the variation one obtains the field equations
\begin{align}
& \p_\m F^{\m\n} = -J^\n,\qquad J^\m =  i\bar \phi \ala{D}^\m \phi= i \bar\phi \ala{\p}^\m\phi - 2A^\m|\phi|^2,\\
&D^2\phi-\p_{\bar\phi} V=(\p^2+2iA^\m\p_\m+i\p_\m A^\m-A^2)\phi-\p_{\bar\phi} V= 0, \qquad \p_\m J^\m\eqons 0,
\end{align}
and symplectic potential
\be\label{phiMth}
\th^\m = -F^{\m\n}\d A_\n -\overline{D^\m\phi}\d\phi -D^\m\phi\d\bar\phi.
\ee
The Lagrangian is invariant under
\be\label{Agt}
\d_\l\phi=i\l\phi, \qquad \d_\l A_\m = -\p_\m\l,  \qquad \d_\l \cL=0,
\ee
where $\l=\l(x)$ is now a real field. We still have $Y_\l=0$, and
the Noether current is
\be\label{jU1}
j^\m_\l = I_\l\th^\m=F^{\m\n}\p_\n\l + \l J^\m = \p_\n(\l F^{\m\n}) + \l(\p_\n F^{\n\m}+J^\m) \eqons \p_\n(\l F^{\m\n}). 
\ee
Or more elegantly using differential forms, 
\be
j_\l = \star F\w d\l + \l J \eqons d(\l \star F).
\ee
It is straightforward to verify that $j_\l$ is conserved on-shell, namely the first statement of Noether's theorem; and the third equality in \eqref{jU1} shows the second statement explicitly: we have a conserved current on-shell that is itself vanishing, up to a corner term. This is the essence of Noether's theorem for gauge symmetries.
It implies that the bulk Noether charge of a gauge symmetry vanishes on any solution. It can thus not be used to distinguish physical solutions. 
The situation changes if there are boundaries, and if the boundary conditions allow all or some gauge transformations at the boundary. In this case, we obtain a non-trivial Noether charge, given by a surface integral.

We can now distinguish two cases, depending on whether we take  $\l$ to be a constant or not. In the first case the gauge symmetry is \emph{global}, and it is an `isometry' in the sense that the gauge field is left invariant. To discuss this case, we can just set $\l=1$. Now the current is
\be
j^\m = J^\m \eqons \p_\n(F^{\m\n}),
\ee
and integrating it over a space-like hypersurface of constant time we find
\be\label{QU1g}
Q[\Si]=\int_\Si j^0 d^3x = -\int_\Si J^0d^3x \eqons \oint_{\p\Si} E^adS_a,
\ee
where we used that $F^{0a}=E^a$ is the electric field. We recognize this Noether charge as the total electric charge. It is conserved in time (consequence of the first statement of Noether's theorem, with conservative boundary conditions), and related to the electric flux by Gauss' theorem (here derived as a direct consequence of the second statement of Noether's theorem). 

Compare now \eqref{QU1g} to \eqref{Qphi}: if we ignore the electromagnetic field, the two $\Si$ integrals match. Thus the Noether charge for the global gauge symmetry can be related to the (non-gauge) symmetry of the matter alone.
This is the essence of the addendum to Noether's second statement we described earlier.
There is nonetheless a profound physical difference between the two cases, due to the last equality of \eqref{QU1g}.
In the case of matter alone, the Noether charge   \eqref{Qphi} is a 3d integral, non-vanishing even in the absence of boundaries. In the global gauge case, the 3d integral  \eqref{QU1g} is on-shell equivalent to a surface integral, and it vanishes in the absence of boundaries. The two pictures thus describe different physical situations:
a global U(1) symmetry of a complex scalar field is not necessarily related to any gauge field and needs not vanish on a spatially compact manifold;
conversely, a global U(1) gauge symmetry means that the total electric charge is a manifestation of the electromagnetic field, described by a gauge theory, and it vanishes on a spatially compact manifold even in the presence of charged particles.

In the second, more general case, $\l$ is a local function. The Noether charge is now
\be\label{U1FB}
Q_\l[\Si] 
= \int_\Si (F^{0\n}\p_\n\l - \l J^0 )d^3x \eqons \oint_{\p\Si}\l E^adS_a=Q_\l[\p \Si]. 
\ee
The 3d integral after the first equality has two contributions, one from the gauge field and one from the matter. 
These can be referred to as `soft' and `hard', in reference respectively to their photonic and matter origins.
These names particularly used in applications at null infinity \cite{Strominger:2017zoo}.\footnote{In this context, the residual gauge symmetries allowed by the boundary conditions are time-independent. The physical distinguishability of these asymptotic symmetries is realized via dephasing of test charges at different locations of the celestial sphere, see e.g. \cite{Susskind:2015hpa,Campoleoni:2019ptc}. This is similar to how a super-translation in the gravitational case can be understood as a desynchronization of test observes at different locations of the celestial sphere, see Fig.\ref{FigCuts} below.}
The second equality shows the relation between the Noether charge and the surface charges. If we use a $\Si$ with a single boundary $S$, its Noether charge coincides with the surface charge: $Q_\l[\Si]=Q_\l[S]$. If we use a $\Si$ with two disconnected boundaries $S_1$ and $S_2$, its Noether charge is the difference of two surface charges, namely their flux, which we denote $F$: $F_\l[\Si]=Q_\l[\Si]=Q_\l[S_2]-Q_\l[S_1]$. 
With two boundaries thus, Noether's current gives a flux-balance law that captures the dynamical content the Gauss constraint. Notice in fact that the relevant field equation in the second equality is the time component of Maxwell's equations, namely the Gauss constraint.

The application of Gauss's law with two disconnected boundaries is familiar from textbooks with applications to determining the total electric charge in an `annulus' region of a space-like hypersurface $\Si$, or to prove that the electric flux does not depend on the shape of the outer boundary if it lays entirely in a vacuum region, see left panel of Fig.~\ref{FigU1}.
It is also very useful in the case of dissipative boundary conditions. If these permit residual gauge transformations, then we can apply \eqref{U1FB} to a lateral boundary $\cal B$, and derive a flux-balance law for each allowed $\l$ that tells us how the surface charge changes under the dissipation, see the right panel of Fig.~\ref{FigU1}.

Let us now discuss the cohomology ambiguity. The reference choice of Noether current leads to useful flux-balance laws. Are there viable alternatives that can be obtained adding an exact 3-form to the Noether current? Let us explore a few possibilities to convince ourselves that this is not the case. 
As a first rule, we should require that the exact 3-form is gauge-invariant, otherwise the flux-balance laws we derive won't be. Then, we could consider exact terms constructed from $F_{\m\n}F^{\m\n}$ or $J^\m A_\m$. But adding such contributions to both sides of \eqref{U1FB} won't expose any new dynamical property, so there is no use in doing it. We can systematically rule them out requiring that we should preserve the order in the fields and their derivatives of the reference choice. 
A remaining possibility would be  $d\l\w dA$. This boils down to rewriting the Gauss law adding to it its `dual' the Bianchi identity $\vec\p\cdot\vec B=0$, and it will not provide new physical insights. Even though the symplectic potential and Noether current are ambiguous, the reference choices provide perfectly useful representatives, and can be singled out insisting on gauge-invariance and order of derivatives, removing the issue of ambiguities.

Finally, let us look at the canonical generators. We have $Y_\l=0$, and $\d_\l\th^\m=F^{\m\n}\p_\n\d\l=0$ for field-independent gauge transformations, hence by the general analysis above we know that the Noether current will be the canonical generator. To confirm this, we start from the symplectic 2-form current, which is given by
\be
\om^\m=-\d F^{\m\n}\cw\d A_\n,
\ee
and compute
\begin{align}
-I_\l\om=\d_\l F^{\m\n}\d A_\n -\d F^{\m\n}\d_\l A_\n = \d F^{\m\n}\p_\n\l = \d(F^{\m\n}\p_\n\l) = \d j_\l \eqons d\d q_\l.
\end{align}
We conclude that the Noether current is the aspect for the Hamiltonian generator of gauge transformations. We are thus in Case 1 of the general analysis of Sec.~\ref{SecHamGen}.
One can proceed similarly if matter sources are present, such as the complex scalar field of our previous example. The intermediate steps are longer, but using \eqref{phiMth} and \eqref{jU1} we arrive at the same end result.\footnote{Notice that we used explicitly $\d\l=0$: integrability requires the gauge parameter to be field-independent. There is no generator for a field-dependent gauge transformation: 
\begin{align}\nn
-I_\l\om=\d_\l F^{\m\n}\d A_\n -\d F^{\m\n}\d_\l A_\n = \d F^{\m\n}\p_\n\l = \d(F^{\m\n}\p_\n\l)-F^{\m\n}\p_\n\d\l = \d j_\l -j_{\d\l}\eqons d(\d q_\l-q_{\d\l}).
\end{align}
This is in agreement with the fact that the Poisson bracket of a field-dependent smearing of the constraints does not generate the correspondent field-dependent gauge transformation.}

\subsection{Chern-Simons}

The previous example extends to non-abelian Yang-Mills theory with minor differences. 
In particular, the Lagrangian is still exactly invariant. A situation with a non-trivial $Y_\l$ occurs for instance in Chern-Simons theory, where we have 
\be
Y_\l=-\f{k}{2\pi}\tr(A\w d\l).
\ee
The reference Noether current is 
\be
j_\l = I_\l\th-Y_\l=\f{k}{2\pi}\tr\big(\l F-d(\l A)\big)\eqons \f{k}{2\pi}d \tr(\l A),
\ee
and the canonical generator is
\be
-I_\l\om = \d j_\l\eqons \f{k}{2\pi}d \tr(\l \d A).
\ee
The reference Noether current is the canonical generator and there is no need of a `corner shift'; we are still in Case 1 of the general analysis of Sec.~\ref{SecHamGen}.
However an interesting difference is that the surface charge and canonical generators are \emph{not} gauge-invariant. This fact is usually interpreted in the light of the fact that allowed boundary gauge transformations are not redundancies but physical symmetries anyways, hence one should not require gauge-invariance of the boundary observables. The new degrees of freedom associated to the possibility of distinguishing gauge configurations at the boundary
lead to a rich relation to Wess-Zumino-Witten models, see e.g. \cite{Witten:1988hf,Carlip:1995qv,Gawedzki:1999bq}.

\subsection{Global vs. local diffeomorphisms}

Consider a matter Lagrangian $L_{\rm m}(\phi,g)$ that depends on dynamical matter fields $\phi$ and a background,  non-dynamical spacetime metric $g$. 
Under an infinitesimal diffeomorphism seen as an active transformation, the metric is untouched since it is a background field, and the dynamical fields transform as the Lie derivative,
\be\label{gdiffeo}
\d_\xi \phi=\pounds_\xi \phi. 
\ee
We then have
\begin{align}\label{dxiLm}
\d_\xi L_{\rm m} &= \d_\phi L_{\rm m}\, \d_\xi \phi =  \d_\phi L_{\rm m}\, \pounds_\xi \phi =  
\d_\phi L_{\rm m}\, \pounds_\xi \phi +  \d_g L_{\rm m}\, \pounds_\xi g - \d_g L_{\rm m}\, \pounds_\xi g \\
\nn &=\pounds_\xi L_{\rm m}  -\d_g L_{\rm m}\, \pounds_\xi g = d i_\xi L_{\rm m} -\d_g L_{\rm m}\, \pounds_\xi g, 
\end{align}
where in the last equality we used $dL_{\rm m}=0$, and
the last term can also be recognized as the background-dependence `anomaly' operator, see \eqref{Dxi2}. Since the right-hand side is not a total derivative, we conclude that a generic diffeomorphism is not a symmetry of a theory of matter fields coupled to a general, but non-dynamical, metric.
The only exception occurs if the metric has isometries, namely if there exist Killing vectors $\xi$ such that
\be\label{Keq}
\pounds_\xi g_{\m\n}=2\na_{(\m}\xi_{\n)}=0.
\ee
In this case the right-hand side of \eqref{dxiLm} is a total derivative, hence diffeomorphisms generated by Killing vectors are symmetries of the theory. 
To write the Noether current, we choose the representative $Y_\xi=i_\xi L_{\rm m}$, and then
\be
j_\xi = I_\xi\th_{\rm m}-i_\xi L_{\rm m}, \qquad j^\m_\xi = I_\xi\th^\m_{\rm m} - \xi^\m {\cal L}_{\rm m}.
\ee
The typical example is Poincar\`e transformations in flat spacetime, with $\xi$ given by \eqref{xiP4} below, and a standard calculation shows that the associated Noether charges are the energy-momentum and angular momentum tensors.
For example we can take the relativistic, real scalar field \eqref{Lphi}, then
\be\label{jximatter}
j^\m_\xi = I_\xi \th^\m-\xi^\m{\cal L} = -\xi^\n (\p^\m\phi \p_\n\phi + \d_\n^\m{\cal L})=-T^{\m\n}\xi_\n.
\ee
Notice that the energy-momentum tensor obtained in this way from Noether's theorem  generically matches the one obtained from the metric variation, and defined as
\be\label{Tdef}
T_{\m\n}:=-\f2{\sqrt{-g}}\f{\d{\tl{\cal L}}_{\rm m}}{\d g^{\m\n}}.
\ee 
To prove this, we include the metric variation in the Lagrangian:
\be
\d L_{\rm m}= \f{\d L_{\rm m}}{\d\phi} \d\phi -\f12 T_{\m\n}\d g^{\m\n} \eps + d\th_{\rm m}.
\ee 
Hooking this field-space 1-form with a diffeomorphisms and using \eqref{Tdef}, we find 
\begin{align}
\d_\xi L &= \f{\d L_{\rm m}}{\d\phi} \xi^\m\p_\m \phi + T_{\m\n}\na^\m\xi^\n \eps + dI_\xi\th_{\rm m}
= \left(\f{\d{\cal L}_{\rm m}}{\d\phi} \xi^\m\p_\m \phi - \xi_\m\na_\n  T^{\m\n} +\na_\m(T^{\m\n}\xi_\n +I_\xi\th^\m_{\rm m})\right)\eps.
\end{align}
It follows that
\be
\f{\d{\cal L}_{\rm m}}{\d\phi} \xi^\m\p_\m \phi - \xi_\m\na_\n  T^{\m\n} +\na_\m(T^{\m\n}\xi_\n +I_\xi\th^\m_{\rm m}-\xi^\m{\cal L}_{\rm m})=0.
\ee
The bulk part of this equation is the familiar statement that the matter field equations are equivalent to conservation of the energy-momentum tensor. The boundary part is what we are interested here, and it proves \eqref{jximatter} up to a total derivative. For the scalar field, we can take this total derivative to be simply zero, and recover \eqref{jximatter}. For electromagnetism, this total derivative is the standard `improvement' that makes Noether's original formula symmetric and gauge-invariant. The fact that the improved energy-momentum tensor has a geometric origin tells a long story about the connectivity of our physical theories.

Taking the divergence of the Noether current we find 
\be\label{najm}
\na_\m j_\xi^\m = - \na_\m T^{\m\n} \xi_\n-T^{\m\n}\na_\m \xi_\n\eqons -T^{\m\n}\na_\m \xi_\n
\ee
which vanishes for a Killing vector, in agreement with Noether's theorem.
The Noether charge over a space-like hypersurface $\Si$ is
\be\label{Qxi1}
Q_\xi[\Si] = \int_\Si j_\xi^0 d^3x = \int_\Si T_{0\m}\xi^\m d^3x,
\ee
and satisfies \eqref{Qcons}. Plugging the decomposition \eqref{xiP4} of the Poincar\'e generators we recover the familiar expressions for the total energy-momentum and relativistic angular momentum.

We now `gauge' the Killing symmetry, by treating the metric as a dynamical field, and adding a Lagrangian $L_{\rm g}$ for it. 
Writing $L=L_{\rm g}+L_{\rm m}$ for the total Lagrangian, and including the metric variation $\d g$ with $\d_\xi g=\pounds_\xi g$, we get
\be
\d_\xi L = \d_g L\, \d_\xi g+\d_\phi L\, \d_\xi \phi = \d_g L\, \pounds_\xi g+\d_\phi L\, \pounds_\xi \phi = \pounds_\xi L = d i_\xi L.
\ee
Now \emph{any} diffeomorphism is a symmetry. 
This is a simple and elegant way to state one version of Einstein's principle of general covariance: if every field in the Lagrangian is dynamical, including the metric, diffeomorphisms are a symmetry.
Choosing the representative $Y_\xi=i_\xi L$, the Noether current is 
\be\label{jxi}
j_\xi=I_\xi\th-i_\xi L \eqons dq_\xi, 
\ee
and it is conserved for any $\xi$, unlike \eqref{najm}.

Let us specialize to the Einstein-Hilbert Lagrangian coupled to matter fields $\phi$, and let's assume for simplicity that the matter Lagrangian $L_{\sscr m}$ couples to the metric but not its derivatives. Then an explicit calculation starting from \eqref{thEH} gives
\be\label{IxithEH}
I_\xi\th_{\sscr EH}^\m =  \f1{8\pi}\Big( R^\m{}_\n\xi^\n - \na_\n \na^{[\m} \xi^{\n]}\Big),
\ee
and
\be\label{jxiEH0}
j^\m_\xi =  \f1{8\pi}\Big( E^\m{}_\n\xi^\n - \na_\n \na^{[\m} \xi^{\n]}\Big),
\ee
where
\be
E^\m{}_\n=G^\m{}_\n+\L \d^{\m}_{\n}-8\pi T^\m{}_\n, 
\ee
and $T_{\m\n}$ is the energy-momentum tensor of the matter Lagrangian  given by \eqref{Tdef}.
The first statement of Noether's theorem can be verified taking the divergence of \eqref{jxi} and using the field equations of both metric and matter:
\be
\na_\m j^\m_\xi = \f{1}{8\pi } E^{\m\n}\na_\m\xi_\n- \na_\m T^{\m\n}\xi_\n \eqons 0.
\ee
The second statement of Noether's theorem is already manifest in the way we wrote \eqref{jxi}: the first term  vanishes on shell, and the second is a total derivative. It can be written using forms as in \eqref{jxi}, 
where
\be\label{jxiEH}
q_\xi := -\f1{32\pi}\eps_{\m\n\r\s} \na^{\r} \xi^{\s} dx^\m\w dx^\n, \qquad j_\xi\eqons dq_\xi
\ee
is known as Komar 2-form, and its integral as Komar charge.

We can investigate the content of the flux-balance laws associated with \eqref{jxiEH} using the identity
\be\label{Komarflux}
\na_\n \na^{[\m} \xi^{\n]} = \f12(R^\m{}_\n\xi^\n - \square \xi^\m+\na^\m \na_\n\xi^\n),
\ee
which follows from the definition of the Riemann tensor as the commutator of two covariant derivatives.
We can distinguish two cases, depending on whether $\xi$ is Killing or not. This is the analogue of the distinction between global and local gauge transformation in the electromagnetic example, with global here referring to isometries. In the first case, $\xi$ Killing vector, there is an important simplification: the right-hand side of \eqref{Komarflux} reduces to a single $R^{\m}{}_\n\xi^\n$ term. Integrating both sides of \eqref{Komarflux} over an hypersurface $\Si$ delimited by two boundaries $S_1$ and $S_2$, and using Stokes' theorem,  we find
\begin{align}\label{Komar}
& \int_\Si j_\xi \eqons \oint^{S_2}_{S_1}q_\xi = -\f1{8\pi}\int_\Si R^\m{}_\n\xi^\n d\Si_\m \eqons - \int_\Si \left(T^{\m\n}\xi_\n+\left(\f{\L}{8\pi}-\f T2\right)\xi^{\m}\right) dV_\m.
\end{align}
Notice the appearance on the right-hand side of the matter current \eqref{jximatter},  in agreement with the addendum discussed earlier.
If the right-hand side of \eqref{Komar} vanishes, the Komar charge is conserved in the sense that it has the same value independently of the surface $S$ used.
If the right-hand side does not vanish, the Noether charge varies by an amount determined by the matter energy-momentum in the enclosed region.
It could thus be taken as the analogue of the electro-magnetic flux-balance law \eqref{QU1g} associated with Gauss' law, with energy-momentum replacing electric charge. There are however some shortcomings. In the Kerr spacetime, which possesses Killing vectors corresponding to stationarity and axial symmetry, 
choosing $\xi=\p_t+\Om_{\sscr H}\p_\phi$ the horizon generator  in \eqref{Komar}, one reproduces the Smarr formula 
\be
\f M2-\Om_{\sscr H}J=\f{\k A}{8\pi}.
\ee
However, while the individual Komar integrals on an arbitrary 2-sphere $S$ encompassing the horizon correctly reproduce the mass and angular momentum, they have wrong relative factors, an issue which is known as the `factor of 2 problem' of the Komar mass. 
Another perplexing aspect of the flux-balance law \eqref{Komar} is that we have the trace-reversed energy-momentum appearing as source, as opposed to the expected one. These shortcomings already raise the question of whether it could be possible to find a improved Noether charges with better flux-balance laws, sieving the cohomology equivalence class.

In the second, more general case, $\xi$ is an arbitrary diffeomorphism, and \eqref{Komar} is replaced by
\begin{align}\label{Komar2}
\oint^{S_2}_{S_1}\k_\xi \eqons -\f12 \int_\Si \left(T^{\m\n}\xi_\n+\left(\f{\L}{8\pi}-\f T2\right)\xi^{\m}- \square \xi^\m+\na^\m \na_\n\xi^\n\right) dV_\m.
\end{align}
This would be the gravitational version of \eqref{U1FB} in the electromagnetic case, and one may hope that the new matter-independent terms are capturing the fact that  the gravitational field itself contributes to the charges. Unfortunately this hope is not fulfilled, as this flux-balance law for generic diffeomorphisms turns out not to have clear physical applications: we still have the trace-reversed energy-momentum tensor, the extra terms in $\xi$ don't have a geometric, coordinate-independent characterization, and if we apply it for instance to boundary symmetries of future null infinity, 
we would get a non-zero flux also in the absence of radiation.
The latter shortcoming of the reference Noether current can be traced back to the fact that we cannot interpret the flux on the right-hand side of \eqref{Komar2} as a $p\d_\xi q$ form, so we cannot control conservation of the charges using the principles of conservative or stationary boundary conditions evoked earlier.
Notice also that in many cases of interest the boundary diffeomorphisms are tangent to the boundary, then the $i_\xi L$ term vanishes, and it would be enough to identify  \eqref{IxithEH} as a $p\d_\xi q$ structure.

The idea to solve the shortcomings of the reference Noether current and its Komar charges is to use the cohomology ambiguity  to look for a different Noether current whose flux-balance laws have a wider range of applicability. In other words, the ambiguities should not be seen as a problem, but rather as a useful freedom. This is already familiar from the example of the electro-magnetic stress tensor. Noether's `reference' formula produces a result that is neither symmetric nor gauge-invariant, and can be `improved' using the corner ambiguity to a form that is symmetric and gauge-invariant.

Before doing so, let us complete the discussion with the analysis of the canonical generators. This will also be of help in addressing the selection of Noether charges.
A famous result by Iyer and Wald \cite{Iyer:1994ys} shows that 
\begin{align}\label{IW}
-I_\xi\om&=-\d_\xi\th +\d I_\xi\th = -\pounds_\xi \th+ \d (j_\xi + i_\xi L) \eqons \d j_\xi -d i_\xi\th \eqons d(\d q_\xi -i_\xi\th).
\end{align}
We used the key property that $\d_\xi\th=\pounds_\xi\th$, which follows from the fact that $\th$ only depends on dynamical fields, and assumed $\d\xi=0$.
With respect to the general analysis of Sec.~\ref{SecHamGen}, all three cases are now possible, depending on $\th$, on the boundary, the boundary conditions, and $\xi$. And in fact, all three cases occur in general relativity.
If $i_\xi\th$ vanishes, we are in Case 1 and the reference Noether current is also the canonical generator. An example of this situation is angular momentum for axial Killing vectors, and more in general tangential diffeomorphisms to the corner. So the fact that the Komar charge gives the right angular momentum for an axial Killing vector is nicely in synergy with the fact that it is the canonical generator with the standard symplectic 2-form.\footnote{The case of Killing vectors however introduces a subtlety, through the field-dependence that Killing vectors introduce (since their existence depend on choosing a specific solution). For a precise treatment of the case with field-dependent diffeomorphisms, the formula \eqref{IW} has to be replaced by
\begin{align}\label{IW2}
-I_\xi\om& \eqons d(\d q_\xi -q_{\d\xi}-i_\xi\th).
\end{align}
The extra term can be relevant when the symmetry itself is field-dependent, e.g. in the case of Killing vectors, or when field-dependence is introduced by gauge-fixing.} 
If it does not vanish but is a total variation, we are in Case 2, and matching the canonical generator suggests a specific corner improvement. An example of this occurs at spatial infinity, with standard ADM fall-off conditions. These boundary conditions allow residual diffeomorphisms which are isometries of the asymptotic flat metric and are given by the Poincar\'e group.
They further guarantee that the limit ${\th}\to \d b$ is exact, so we have a canonical generator for the boundary symmetries, with integrable charge $q'_\xi = q_\xi +i_\xi b$. 
In this way one can reconstruct the ADM charges from covariant phase space methods, and in particular the $b$ shift solves the famous issue of the missing factor of 2 in the Komar mass \cite{Iyer:1994ys}.
There are also situations in which $\th$ is not exact at the boundary, notably in the case of null infinity. We are in Case 3 and integrability fails. If $\th$ is not exact at the boundary, it follows that $\om$ does not vanish there, hence we are in the `dissipative' category of Table~\ref{TableTwoCases}. Failure of integrability is thus related to the use of dissipative boundary conditions. Notice also from the specific form of \eqref{IW} that if it fails, it will fail only for those vector fields that move the corner, namely those sensitive to dissipation.\footnote{Non-linearities also play an important role, for instance in the free theory super-translations move the corner but we have integrable charges since the non-integrable term is higher order, see e.g. \cite{Campoleoni:2017qot}.}

\subsection{Improved Noether charges  and Wald-Zoupas covariance}  \label{genWZsection}

To span the available Noether currents, and identify one that gives better flux-balance laws than \eqref{Komar2}, it is useful to tie its cohomology ambiguity to the covariant phase space ambiguities. We can do this if we start from the choice of representative \eqref{defjN}, and then use \eqref{defth'} to compute
\be
j_\xi'=I_\xi\th' - i_\xi L' = I_\xi \th - i_\xi L +\d_\xi \ell - i_\xi d\ell -d I_\xi\vth =j_\xi +d(i_\xi\ell-I_\xi\vth).
\ee
Using \eqref{NT2}, we conclude that
\begin{align}\label{jiN}
j_\xi' &= I_\xi\th'-i_\xi L' \eqons dq'_\xi,\qquad q'_\xi = q_\xi +i_\xi \ell-I_\xi\vth,
\end{align}
up to a closed 2-form, which as discussed earlier will be discarded. 
Proceeding in the same way we obtain 
\begin{align}\label{Ixiom2}
d(\d q'_\xi -i_\xi\th')\eqons -I_\xi\om' = -I_\xi\om +d\d I_\xi\vth\eqons d(\d q_\xi -i_\xi\th+\d I_\xi\vth).
\end{align}

These equations have appeared in various recent papers, e.g. 
\cite{Harlow:2019yfa,Freidel:2020xyx,Freidel:2021cjp,Odak:2021axr,Chandrasekaran:2021vyu,Odak:2022ndm}. They describe the combined set of $(b)$ and $(c)$ ambiguities, which in particular include corner changes of the symplectic 2-form.
If one has a fixed $\om$ to work with then we can restrict attention to $(b)$, and still have the possibility of changing the Noether current and the integrable/non-integrable split.
A subtlety that can sometimes be relevant is that for given $\th$ and $\th'$, there is a residual ambiguity $(\ell,\vth)\to(\ell+dc,\vth+\d c)$ that \emph{can affect} the charge \cite{Chandrasekaran:2021vyu,Odak:2022ndm}. 
Thus in general, $q'_\xi$ is not only determined by the choice of $\th'$ but also from the specific choice of $\ell$.

The relation \eqref{jiN} is the general formula for the `improved Noether charges' that we will use to seek better options for the flux-balance laws.
The general principle that we will use is:
\begin{center}\emph{
Fix the ambiguities so that the charges are conserved, and they are Hamiltonian generators, \\ in the subset of the phase space that satisfies the conservative or stationary conditions of interest \\ at the given boundary. }\end{center}

We see from \eqref{jxi} and \eqref{IW} that a sufficient condition to achieve this property is $\pbi{\th}=0$ and tangent diffeomorphisms.
This should happen for the boundary conditions specified at the given boundary as in Table~\ref{TableTwoCases}. If it does not happen for the reference solution, then we can start playing around with the freedom \eqref{pbth'} to change symplectic structures at the boundary, and associated with it change the Noether representatives to \eqref{jiN} and \eqref{Ixiom2}, to try to satisfy the requirements. 

A split like \eqref{pbth'} plays a prominent role in the analysis of gravitational radiation: a boundary can introduce crucial tools to identify radiation in a gauge-independent way also at the non-perturbative level, and allow on to split the phase space to dissipative and conservative solutions.
For instance, in the standard treatment of isolated gravitational systems, one introduces the idealized notions of spacetime asymptotes such as spatial and null infinity, where a background flat spacetime can be assigned. This background spacetime in turns introduces a notion of inertial observers, and can be used to identify radiative and non-radiative degrees of freedom.
On the other hand, the description of the boundary, and the pull-back of the symplectic potential on it, may introduce non-dynamical, background fields, and one has to make sure that such structures don't affect physical statements. This is the reason why for instance in the original formulation of BMS charges and fluxes one has to carefully check conformal invariance and foliation independence \cite{Geroch:1977jn}. 

A useful tool to test background-independence in the action of a boundary symmetry group is the `anomaly operator', defined by 
\be
\D_\xi F = (\d_\xi - \pounds_\xi)F,
\ee
see \cite{Hopfmuller:2018fni,Chandrasekaran:2020wwn,Freidel:2021cjp,Rignon-Bret:2024gcx}.\footnote{For field-dependent gauge transformations one has to also include a term $I_{\d\xi}$ in the definition of the anomaly operator. In that case however the definition of covariance should be kept as the matching of the field-space and spacetime Lie derivative, and not the vanishing of the anomalies, see discussions in \cite{Carrozza:2022xut,Odak:2023pga}.}
 If we denote generically $\phi$ the dynamical fields, and $\eta$ the background and non-dynamical ones, then
\be\label{Dxi2}
\D_\xi F = -\d_\eta F\pounds_\xi \eta.
\ee
The anomaly operator is therefore a probe of the background-dependence of a functional $F$ through a 
symmetry action. If a functional is background-independent, or in different words general covariant, it will have zero anomaly. Conversely, if it has zero anomaly, it will be background-independent at least in so far as symmetry group transformations are concerned. 
The anomaly operator will play an important role to determine that the preferred choice of symplectic potential is background-independent.
It is also very useful to understand the physical meaning of a given transformation, for instance, it allows to compute the field space transformation at null infinity locally on $\scri$ from geometric considerations alone, without knowing nothing about asymptotic expansion or postulated fall-off conditions \cite{Rignon-Bret:2024gcx}.
A functional that satisfy $\d_\xi F=\pounds_\xi F$ will be denoted \emph{Wald-Zoupas covariant}. It is a necessary condition for assuring background-independence, and a sufficient one if the symmetry group is large enough.

Starting from a general covariant Lagrangian, a symplectic potential can break covariance either through the boundary Lagrangian $\ell$, or the corner potential $\vth$. If one picks anomalous choices, then the formulas \eqref{jiN} and \eqref{Ixiom2} are no longer valid, and are replaced by \cite{Chandrasekaran:2020wwn,Freidel:2021cjp,Chandrasekaran:2021vyu}
\be\label{jxia}
j_\xi = I_\xi\th-i_\xi L- a_\xi,
\ee
and
\be\label{IxiomA}
-I_\xi \om \eqons d(\d q_\xi -q_{\d\xi}-i_\xi\th-A_\xi),
\ee
where
\be
a_\xi :=\D_\xi\ell, \qquad A_\xi :=\D_\xi\vth,
\ee
and we restored the possibility of field-dependent diffeos for full generality.
One has to make sure that the split preserves general covariance, by not introducing such anomalies.

This leads us to the \emph{(generalized) Wald-Zoupas prescription}, which we first motivate, and then define.
In general, characterizing the preferred symplectic potential requires the use of a background, and of background structures that can be associated with the boundary.
For instance in general relativity we would like to select preferred charges that are conserved in the absence of gravitational radiation. But this is hard to characterize in a diffeomorphism-invariant way. Just like it is hard to imagine that there is a preferred symplectic potential that applies to all situations. 
It seems instead easier to take advantage of the boundary to introduce some reference background structure that can be used to distinguish radiation from the other modes in the field. One has thus to first specify the system by specifying a boundary and the boundary conditions, and then look for the preferred symplectic potential at that boundary, possibly with variations restricted to preserving the boundary conditions. In other words, we look only for a preferred symplectic potential after pull-back on the boundary, in the equivalence class \eqref{pbth'}. This introduces the caveat discussed earlier that one has to make sure that the split chosen does not introduce dependence on the background structures of the boundary. In other words, the preferred potential should be general covariant. 
It may feel ironical that one should worry about general covariance in general relativity, but it is a consequence of introducing boundaries.

Accordingly, we give the following criteria for selecting the preferred symplectic potential \cite{Odak:2022ndm}:
\bit
\item[1.] Covariance: $\d_\xi \th=\pounds_\xi\th$, namely background-independence under symmetry action
\item[2.] Stationarity: $\th=p\d q$ according to the chosen boundary conditions of Table~\ref{TableTwoCases}
\eit
In the case of asymptotic symmetries, one has to make sure also that the preferred symplectic potential is well-defined, in other words any divergences should also be removed using the equivalence class freedom. The notion of stationarity should be prescribed based on the physical problem at hand. For instance, it could be all solutions without radiation, all solutions without dissipation, all solutions with a time-translation Killing vector, etc. These criteria are based on the seminal Wald-Zoupas paper \cite{Wald:1999wa}, where the prescription was applied at $\scri$, with the stationarity condition defined by the vanishing of the news function, and as we argued, can be applied systematically to more general situations.
The general analysis requires the freedom of allowing field-dependent diffeomorphisms and anomalous transformations, and the inclusion of corner terms in the equivalence class \eqref{pbth'}. The latter are important in situations where it is otherwise not possible to implement the two criteria, or because one wants to introduce corner degrees of freedom in the phase space. I have for instance seen an example of the former when full-filling this construction for the extended BMS symmetry in \cite{Rignon-Bret:2024gcx}, and an example of the latter when constructing a purely hard flux for BMS transformations in \cite{Rignon-Bret:2024mef}.

\subsection{Removal of field-dependent cocycles}

A very important consequence of our two criteria is that they guarantee that the Noether currents realize the symmetry algebra in the covariant phase space in terms of the Barnich-Troessaert bracket \cite{Barnich:2011mi}, without field-dependent cocycles \cite{Rignon-Bret:2024wlu}.
To prove this, 
consider a covariant split, and assume for simplicity field-independent diffeos.
From \eqref{IW}, we read off
\be
h_\xi = I_\xi\th=j_\xi+i_\xi L, \qquad
{\cal F}_\xi=\d_\xi \th\eqons di_\xi \th +i_\xi\d L.
\ee
Exploiting the ambiguities, we can equivalently take
\be
h_\xi = j_\xi, \qquad {\cal F}_\xi=di_\xi \th.
\ee
It follows that
\begin{align}
& \d_\chi {\cal F}_\xi = \d_\chi di_\xi \th \eqons  \d_\chi \pounds_\xi\th - \d_\chi i_\xi\d L \eqons \pounds_\xi \pounds_\chi\th - \d  i_\xi\pounds_\chi L, 
\end{align}
and
\be
 {\cal F}_{[\chi,\xi]} = di_{[\chi,\xi]}\th \eqons \pounds_{[\chi,\xi]}\th - \d i_{[\chi,\xi]}L.
\ee
Using these formulas, and the identity $(i_\xi\pounds_\chi-i_\chi\pounds_\xi)L=i_{[\xi,\chi]}L-di_\xi i_\chi L$, we can check that the condition \eqref{cocycle} holds, with $K_{(\xi,\chi)}=di_\xi i_\chi L$. 
Therefore, \emph{the Barnich-Troessaert bracket is well-defined}, and from its general definition \eqref{BTbracketgen} we find\footnote{In the second and third equality, we use
\be\nn
I_\xi{\cal F}_\chi = di_\chi (j_\xi+i_\xi L) \eqons \pounds_\chi j_\xi + d i_\chi i_\xi L.
\ee
}
\begin{align}\label{BTgr}
\{j_\xi,j_\chi\}_*&:=- I_\chi I_\xi\om +I_\chi{\cal F}_\xi -I_\xi{\cal F}_\chi =- I_\chi I_\xi\om + d(i_\xi j_\chi - i_\chi j_\xi) +2di_\xi i_\chi L
\\\nn& = (\d_\chi -\pounds_\chi )j_\xi+di_\xi i_\chi L = j_{[\xi,\chi]}+di_\xi i_\chi L,
\end{align}
up to a \emph{field-independent}, central extension. The residual field-dependent cocycle $i_\xi i_\chi L$ is absent if the allowed diffeomorphisms are tangent to the boundary.  

We also read from \eqref{BTgr} that a covariant Noether current provides a realization of the algebra via
\be\label{anojN}
(\d_\chi -\pounds_\chi )j_\xi = j_{[\xi,\chi]}.
\ee
The appearance of the operator $\d_\chi-\pounds_\chi$ in \eqref{BTgr} has a deep physical meaning: it coincides in fact with the anomaly operator acting on the Noether current. 
If the symmetry vector fields are universal hence field-independent (this is the most common situation), $\d\xi=0$ and $\D_\chi\xi=-\pounds_\chi\xi=-[\chi,\xi]$, hence absence of field-dependent cocycles is equivalent to the statement that \emph{the only background field allowed in the Noether current is the symmetry vector field.}
Furthermore, using the second statement of Noether's theorem, we can derive from \eqref{anojN} the following transformation property of the surface charges,
\be
\d_\chi q_\xi -i_\chi j_\xi \eqons q_{[\xi,\chi]},
\ee
up to a closed 2-form. This shows that the subtraction needed for the realization of the algebra by the surface charges is precisely the flux causing their non-conservation, whence the canonical algebra realization is recovered whenever the dissipation is turned off.

Had we instead started from a non-covariant split, then $a_\xi$ and $A_\xi$ entering 
\eqref{jxia} and \eqref{IxiomA}
will also contribute to a field-dependent cocycle, thus spoiling the realization of the algebra. Specifically, ${\cal F}_\xi=d(i_\xi \th+A_\xi)$, and
\begin{align}\label{BTgrAno}
\{j_\xi,j_\chi\}_*&:=- I_\chi I_\xi\om +I_\chi{\cal F}_\xi -I_\xi{\cal F}_\chi =
 j_{[\xi,\chi]}+d(i_\xi a_\chi-i_\chi a_\xi+i_\xi i_\chi L),
\end{align}
For the details of this derivation, as well as the complete general case including field-dependent diffeomorphisms, we refer the reader to \cite{Freidel:2021cjp}.
The field-dependent cocycle $K_{(\xi,\chi)} = d(i_\xi a_\chi-i_\chi a_\xi)$ is present also for diffeomorphisms tangent to the boundary. This is precisely the origin of the cocycle found in \cite{Barnich:2011mi}, as shown in \cite{Freidel:2021yqe,Rignon-Bret:2024gcx}.

\section{Gravitational charges at finite boundaries}\label{SecGRbc} 

Let us see some examples of polarizations in the gravitational case. 
To characterize the problem in geometric terms, we treat separately the case of null and not null boundaries.

\subsection{Space-like and time-like boundaries}

Consider a hypersurface $\Si$ located at $\Phi=0$, with $n_\m$ its unit normal, and boundary $\p\Si=S$, with $u_\m$ its unit normal within $T^*\Si$, so that $u_\m n^\m=0$. The corresponding volume forms are $\eps_\Si=i_n\eps$ and $\eps_S=i_u\eps_\Si$. The normal is not necessarily geodetic: in general,
$k=2\pounds_n\ln N$ and $a^\perp_\m=-q_\m^\n\p_\n\ln N$. The extrinsic geometry is automatically symmetric thanks to the normalization of $n$.

\begin{center}\begin{tcolorbox}[title=Geometric elements of a (non-null) hypersurface, colback=gray!10, colframe=gray, width=0.85\textwidth,
    boxsep=0mm,     
    left=1mm, right=1mm, top=1mm, bottom=4mm,  
    ]
\begin{center}
\begin{tabular}{lll}
\rule{0pt}{15pt} Boundary normal & $\Phi=0$, \quad $n_\m = s N\p_\m \Phi$, \quad $N=(sg^{\Phi\Phi})^{-1/2}$ \\
\rule{0pt}{15pt}  & $n^2=s$,\quad\quad $s=\pm1$, \quad\quad $n^\n\na_\n n_\m = k n_\m+a_\m^\perp$ \\ 
\rule{0pt}{15pt} Induced geometry & $q_{ab}=\pbi{g}_{ab}$,\quad $\det q=-s$, \quad $\eps_\Si=i_n\eps$ \\ 
\rule{0pt}{15pt} Projector & $q_{\m\n}:=g_{\m\n} -s n_\m n_\n$ \\
\rule{0pt}{15pt} Extrinsic geometry & $K_\m{}^\n= \pbi{\na}_\m n^\n = q^\r_\m \na_\r n^\n$ 
\end{tabular} \end{center}\end{tcolorbox} \end{center}

Taking the pull-back of \eqref{thEH} one finds (see e.g. \cite{Brown:2000dz,Lehner:2016vdi,Oliveri:2019gvm})
\begin{align}\label{thpb1}
\pbi{\th}^{\sscr EH} &= \f{s}{16\pi} \left( K_{\m\n} \d q^{\m\n}-2\d K \right) \eps_\Si + d\vth^{\sscr EH}, \qquad
\vth^{\sscr EH}= - \f1{16\pi}u_\m  \d n^\m \eps_S = \f1{16\pi}u^\m n^\n \d g_{\m\n}  \eps_S.
\end{align}
Let us compare different choices of \eqref{pbth'} and their corresponding polarizations. First, we introduce the gravitational momentum
\be
\tl\Pi^{\m\n}:=\sqrt{q}(K^{\m\n}-q^{\m\n}K),\qquad \tl\Pi:=g_{\m\n}\tl\Pi^{\m\n} = -2\sqrt q K,
\ee
familiar from the ADM analysis, here written as a spacetime tensor. It is then easy to see that
\begin{align}\label{thpb2}
\pbi{\th}^{\sscr EH} &= \f s{16\pi} \tl\Pi_{\m\n} \d q^{\m\n}  d^3x- \d \ell^{\sscr GHY}+ d\vth^{\sscr EH}
= \f s{16\pi} q_{\m\n}\d\tl\Pi^{\m\n} d^3x +d\vth^{\sscr EH},
\end{align}
where
\be\label{ellc}
\ell^{\sscr GHY}:= \f s{8\pi}K \eps_\Si
\ee
is the Gibbons-Hawking-York boundary Lagrangian. We see from the second equality in \eqref{pbth'} that the Einstein-Hilbert Lagrangian has a well-posed variational principle with Neumann boundary conditions (as could have been anticipated observing that it contains second derivatives of the fundamental field, the metric), and that to switch to Dirichlet boundary conditions we need to add a boundary term, given by \eqref{ellc}. 

Another interesting choice of polarization proposed by York  \cite{York:1986lje} is given by mixed boundary conditions where one uses the conformal equivalence class of boundary metrics, and the trace of the extrinsic curvature. The corresponding symplectic potential is obtained via \cite{Odak:2021axr}
\begin{align}\label{thpb3}
\pbi{\th}^{\sscr EH} &= -\f s{16\pi}\left(\tl P^{\m\n}\d \hat q_{\m\n} + \tl P_K \d K \right)d^3x -\d\ell^{\sscr Y}+d\vth^{\sscr EH},
\end{align}
where
\be
\tl P^{\m\n} := q^{1/3}(\tl\Pi^{\m\n}-\f13q^{\m\n}\tl\Pi),
\qquad
\tl P_K=\frac{4}{3}\sqrt{q},
\ee
and
\be
\ell^{\sscr C} = \f s{24\pi} K\eps_\Si
\ee
is the conformal, or York, boundary Lagrangian. The conformal case plays an important role in the study of the well-posedness of the initial boundary value problem \cite{An:2021fcq,An:2025rlw,Anninos:2023epi,Anninos:2024xhc,Galante:2025tnt,Liu:2025xij,Galante:2025emz}. 

More in general, we can consider a one-parameter family of polarizations \cite{Odak:2021axr,Liu:2025xij}
\be\label{thb}
\th^{b}=\pbi\th+\d\ell^b-d\vth^{\sscr EH},
\ee
where the corner term is always the same given by \eqref{thpb1}, and 
\be
\ell^{\sscr b} =\f{sb}{16\pi} K\eps_\Si.
\ee
For more details, and in particular the case of codimension-2 corner Lagrangians required 
for non-orthogonal corners, see \cite{Harlow:2019yfa,Odak:2021axr}. 
The three main examples are reported in the table~\ref{Tableb} below.
\begin{table}[H] \begin{center}
  \begin{tabular}{|l|l|c|l|}\hline  
\emph{polarization} & \emph{symplectic potential}  & \emph{b} & \emph{cons. boundary conditions}\\ \hline
Dirichlet & \ $\tl\Pi_{\m\n}\d q^{\m\n}$ & $2$ & \ $\d q_{\m\n}=0$  \\
Conformal & \ $-\tl P^{\m\n}\d \hat q_{\m\n} - \tl P_K \d K$ & $\f23$ & \ $\d \hat q_{\m\n}=\d K=0$ \\
Neumann& \ $q_{\m\n}\d\tl\Pi^{\m\n}$ & 0  & \ $\d\tl\Pi^{\m\n}=0$\\ \hline
\end{tabular} \end{center}   
\caption{\label{Tableb} \emph{\small{Different polarizations for a time-like boundary, and their corresponding conservative boundary conditions.}}}\end{table}

The sign $s$ in all these formulas depends on the signature of the boundary. Boundary conditions on the time-like and space-like boundaries have different meanings: the former determine the nature of the system, whereas the latter determines how one is specifying the initial conditions. 
The different polarizations possible in the time-like boundary case offer an explicit context to understand their physical meaning.
As we will see below, the stress-tensor with conformal or Neumann boundary conditions is different than the Brown-York stress-tensor \cite{Brown:1992br} associated with Dirichlet boundary conditions \cite{Odak:2021axr}. This modification can be understood in a similar way to the change between internal energy and free energy when changing ensembles in thermodynamics.
The boundary conditions at the time-like boundary are also relevant to determine the allowed boundary symmetries. For instance, these can be chosen to preserve the boundary, hence tangential diffeomorphisms only, and to preserve the boundary conditions.

\subsubsection{Covariance}

Before using the 1-parameter family of polarizations \eqref{thb} for physical applications, one has to make sure that it satisfies the covariance property $\d_\xi \th=\pounds_\xi\th$, which in turns guarantees the background-independence of Noether charges. To discuss the potential anomalies, consider first the case of arbitrary diffeomorphisms, and no boundary conditions.
The only background field is the scalar field $\Phi$ that identifies the boundary: $\d\Phi=0$. It follows that $\D_\xi\Phi=-\pounds_\xi\Phi = -\xi^\Phi$. 
This vanishes on the hypersurface if we consider only diffeomorphisms that are tangent to $\cN$, then $\xi^\Phi=\Phi\bar\xi^\Phi+O(\Phi^2)$ for some $\Phi$-independent $\bar\xi^\Phi$, and any functional of the metric and $\Phi$ will be anomaly-free. The extrinsic curvature and symplectic potential however depend on the normal 1-form, and this contains derivatives of $\Phi$. For these, $\D_\xi \p_\m\Phi=-\p_\m \xi^\Phi=-(s\bar\xi^\Phi/N)n_\m$,
hence a generic normal will carry an anomaly proportional to the first-order extension of the symmetry vector field off the hypersurface. The only case in which the anomaly vanishes for any tangential $\xi$ is if we use a \emph{unit-norm} 1-form
\cite{Odak:2021axr}. Therefore if we restrict to diffeomorphisms tangent to the boundary, \emph{and} we use a unit-norm,
\be
\D_\xi \th^{\rm b}=\D_\xi \ell^{\rm b}=0.
\ee
Under these conditions, any choice of symplectic potential in the $b$ family satisfies the covariance requirement.
It follows that any choice leads to Noether's second theorem in the form \eqref{jxi}:
\be\label{jNGRb}
j^b_\xi:=I_\xi\th^b-i_\xi L^b\eqons dq_\xi^b.
\ee
The surface charges will thus be automatically conserved when the boundary conditions associated with the chosen $b$ hold. 
The discussion also shows the precise relation between anomalies and background-dependence: anomalies occur if the residual; gauge transformations don't preserve the boundary, or if they do but the geometry of the boundary is described in a way that depends on its embedding.

\subsubsection{Conservative boundary conditions}

Open boundary conditions are complicated to study on a time-like boundary. There are two unrelated difficulties. The first, already present for a scalar field, is 
that it is difficult to disentangle incoming and outgoing waves (the usual no-incoming or no-outgoing conditions are global, and cannot be imposed on a local boundary), hence one cannot study a purely dissipative or purely absorbing system. The second, specific to gauge theories and gravity, is the presence of gauge modes. Some of the variables varying on the boundary are pure gauge, hence their variations give spurious contributions to the dynamics, and should be restricted. Restricting the variations of the pure gauge modes may look like conservative boundary conditions for these variables, but all it means is including gauge fixing conditions. Conversely, identifying the pure gauge modes means also being able to isolate local physical degrees of freedom. Implementing this procedure on a time-like boundary is difficult. Both problems are solved in the case of a null boundary, which we will treat below. 
For the rest of this Section, we focus on conservative boundary conditions only.

We can consider different choices of conservative boundary conditions, for instance the three listed in Table~\ref{Tableb}.
In all cases, we should restrict the boundary diffeomorphisms to preserve these boundary conditions. This will give us the notion of \emph{boundary symmetries}:
the residual diffeomorphisms compatible with the boundary conditions. They can be understood as `isometries of the boundary conditions', and are not gauge because as we have seen they are non-degenerate directions of the symplectic 2-form on a space-like hypersurface intersecting the boundary, where they identify a non-trivial Noether charge. This can be computed applying the improved Noether charge formula \eqref{jiN} to the family of polarizations \eqref{thb}, which gives 
\cite{Odak:2021axr}
\begin{align}\label{Qb}
Q_\xi^b[S]&=\oint_S \k_{\xi}+i_\xi\ell^b-I_\xi\vth^{\sscr EH} =-\f1{8\pi} \oint_S n^\m \xi^\n (\KB_{\m\n} - \f b2 \qB_{\m\n}\KB )  \eps_S,
\end{align}
where $n$ is the normal to $\Si$, and the barred quantities referred to the time-like boundary.
The charges are conserved when the fields on the time-like boundary $\cal B$ satisfy the boundary conditions corresponding to the chosen $b$. This follows immediately from \eqref{jNGRb}, which can be rewritten as
$Q_{\xi}^b[S]\eqons Q^b_\xi[S']$, where $S$ and $S'$ are the boundaries of $\Si$ and $\Si'$ connected by $\cal B$.
For $b=2$, these are the Brown-York charges, derived from covariant phase space methods in \cite{Harlow:2019yfa} (see also \cite{Freidel:2020xyx}).
For the other values of $b$ we obtain new charges, that can be interpreted as different notions of energy for different gravitational ensembles \cite{Odak:2021axr}.
The notable cases $b=0$ and $b=2/3$ correspond to Neumann and conformal boundary conditions, with the latter playing a useful role for the initial boundary value problem \cite{An:2021fcq,An:2025rlw,Anninos:2023epi,Anninos:2024xhc,Galante:2025tnt,Liu:2025xij,Galante:2025emz}.

Since the improved Noether charge involves a non-trivial corner term, it is the canonical generator for a corner-improved symplectic 2-form, different from the reference Einstein-Hilbert one based on \eqref{thEH}:
\be\label{Ixiomb}
-I_\xi \Om^b_\Si = Q_\xi^b[S] = -I_\xi \Om_\Si +\oint_S I_\xi\vth^{\sscr EH}.
\ee
The corner improvement is relevant at finite distances, but vanishes in the limit to $i^0$. Thus at spatial infinity the different choices of $b$ correspond to different polarizations of the same symplectic 2-form. The limit to $i^0$ requires some care, because the charges actually diverge. The divergence can however be subtracted adding a counter-term \cite{Brown:1992br}. This `renormalization' procedure can be understood as the existence of a finite representative associated with a specific boundary Lagrangian \cite{Odak:2021axr}. This is another useful way in which ambiguities play a constructive role, allowing one to select specific finite representatives.

It is instructive to see how the polarization and corner improvements solve the problems of the flux \eqref{IxithEH} associated with the reference symplectic potential and Komar charges: using \eqref{thpb2} we can rewrite the pull-back of \eqref{IxithEH} on the time-like boundary as
\be
I_\xi\pbi{\th}=\f1{16\pi}q_{\m\n}\d_\xi \tl\Pi^{\m\n} d^3x + dI_\xi\vth^{\sscr EH}, \qquad I_\xi\vth^{\sscr EH}=\f1{16\pi}u^\m n^\n\na_{(\m}\xi_{\n)}\eps_S.
\ee
The corner term vanishes for Killing vectors but not otherwise, and it is what spoils the structure $p\d_\xi q$ hence conservation for any of the conservative boundary conditions we considered.
Using the corner ambiguity to absorb it in the charge aspect removes this problem, improves the Komar expression to the $b=0$ charges \eqref{Qb} of \cite{Odak:2021axr}, which are now conserved by Neumann boundary conditions. A further polarization shift by boundary Lagrangians allows us to obtain the charges for the other values of $b$ that are conserved for other boundary conditions than Neumann.

\subsection{Null boundaries}\label{SecNull}

The main difference of a null boundary is that its normal 1-form defines a vector that is \emph{tangent} to the hypersurface, and not orthogonal to it. And furthermore, there is no canonical normalization for the normal, the induced metric is degenerate (with null direction the null tangent vector itself), and there is no projector on the hypersurface, nor unique induced Levi-Civita connection.
A very convenient way to deal with a null boundary is to use the Newman-Penrose formalism. One introduces a doubly null tetrad $(l,n,m,\bar m)$, of which one real vector is tangent to the null hypersurface (say $l$), and the second real null vector (then $n$) acts as a `rigging vector', or its 1-form as `rigging 1-form'. It provides a 2d space-like projector  via
$2m_{(\m}\bar m_{\n)} = \g_{\m\n}:=g_{\m\n}+2l_{(\m}n_{\n)}$ and, in the case in which it is hypersurface orthogonal, a $2+1$ foliation of $\cal N$ determined by $n$ and to which $(m,\bar m)$ are tangent.

\begin{center}\begin{tcolorbox}[title=Geometric elements of a null hypersurface, colback=gray!10, colframe=gray, width=0.85\textwidth,
    boxsep=0mm,     
    left=1mm, right=1mm, top=1mm, bottom=4mm,  
    ]
\begin{center}
\begin{tabular}{lll}
\rule{0pt}{15pt} Boundary normal & $\Phi=0$, \quad $l_\m = -f\p_\m \Phi$ \\
\rule{0pt}{15pt}  &  $l^2=0$, \quad $l^\n\na_\n l_\m =k l_\m$ \\
\rule{0pt}{15pt} Induced geometry & $q_{ab}=\pbi{g}_{ab}$,\quad $\det q=0$, \quad $q_{ab}l^b=0$,\quad $\eps_\cN=i_n\eps$, \quad $\eps_S=i_l\eps_\cN$ \\ 
\rule{0pt}{15pt} 2d projector & $ \g_{\m\n}=g_{\m\n}+2l_{(\m}n_{\n)}=2m_{(\m}\bar m_{\n)}$ \\
\rule{0pt}{15pt} Extrinsic geometry & $W_{\m}{}^\n:=\na_{\pbi{\m}} l^\n = \om_{\pbi{\m}}l^\n + \g^{\n}_{\r}B_{\pbi{\m}}{}^{\r}$, \qquad $W=\th+k$ \\
\end{tabular} \end{center}\end{tcolorbox} \end{center}

Being null and hypersurface orthogonal, $l$ is automatically geodesic. It is however not necessarily affinely parameterized, and an explicit calculation shows that
 \be\label{kmoche}
 k=  \pounds_l \ln f -\f f2 \p_\Phi g^{\Phi\Phi}.
 \ee
While there is no extrinsic curvature in the usual sense, one can still define the Weingarten map $W_\m{}^\n$, and with the help of a choice of rigging vector, split it into a vertical and a horizontal component. The horizontal component is purely intrinsic, and features the deformation tensor  $B$ occurring in the standard analysis of null congruences, see e.g. \cite{Wald}. Its antisymmetric part vanishes because $l$ is hypersurface orthogonal, and the rest can be decomposed in terms of shear $\s$ and expansion $\th$:\footnote{Hopefully there should be no confusion between the scalar $\th$ used for the expansion, and the 3-form or vector $\th$ used for the symplectic potential current. When both occur in the same equation, we will put a label to distinguish them.}
\begin{align}\label{defB}
& B_{\m\n}:=\g_\m^\r\g_\n^\s \na_\r l_\s = \f12 \g_\m^\r\g_\n^\s \pounds_l \g_{\r\s} {\eqonN}  \s_{\m\n}+\f12\g_{\m\n}\th, \\
& \s_{\m\n}:=\g_{\la \m}^\r\g_{\n\ra}^\s \na_\r l_\s = -\bar m_\m\bar m_\n \s+cc,
\qquad\ \ \th:=2m^{(\m}\bar m^{\n)}\na_\m l_\n =-2\r.\label{defsigma}
\end{align}
The vertical part is extrinsic, since it depends on the first derivatives of the metric off the hypersurface, and can be conveniently decomposed as follows,
\be
\om_{{\m}}:=- \eta_{\m}-kn_\m, \qquad\quad \eta_{\m}:=\g_\m^\r n^\s\na_\r l_\s=-(\a+\bar\b)m_\m+cc, 
\qquad l^\m\om_\m=k=2\re(\eps). \label{defdot}
\ee
Here $\om$ is the rotational 1-form of isolated and non-expanding horizons \cite{Ashtekar:2000sz,Ashtekar:2000hw,Ashtekar:2004cn}, satisfying $\om\cdot l=k$; $\eta$ is 
the connection 1-form on the normal time-like planes spanned by $(l,n)$, sometimes called Hajicek 1-form \cite{hajivcek1973exact}, or twist.
In these formulas, the complex scalars $\a,\b,\eps,\r$ and $\s$ make reference to the NP formalism.\footnote{With mostly-plus signature, we use the conventions of \cite{Ashtekar:2000hw}. The twist should not be confused with the 2-sphere connection of the covariant derivative $\eth$ used in NP calculus, which is given by $\a-\bar\b$ \cite{NP62,Booth:2006bn}. Both $\om_\m$ and $\eta_\m$ are connections, the first on the bundle of null geodesics on $\cN$, and the second on the normal bundle to the 2d cross sections.} 

The lack of 3d projector means also that there is no canonical Levi-Civita connection on a null hypersurface. In fact, even the pull-back of the ambient connection does not define a connection on the hypersurface. To see this, we can take two tangent vectors $X$ and $Y$ and compute:
\be\label{nopbconnection}
l_\m X^\n\na_\n Y^\m = -X^\n Y^\m \na_\n l_\m = -X^\n Y^\m(\s_{\m\n}+\f\th2\g_{\m\n}).
\ee
For a general null hypersurface the right-hand side does not vanish, hence the pull-back of ambient covariant derivative takes them outside of the hypersurface. Only  special hypersurface that are shear-free and expansion-free admit a canonical connection, given by the pull-back of the ambient connection. These hypersurfaces play indeed an important role in the study of non-expanding horizons \cite{Ashtekar:2024bpi} and future null infinity, see below.
For more general hypersurfaces, there is no Levi-Civita connection, but one can take advantage of the rigging vector and introduce a family of \emph{rigging connections}, defined by 
\be\label{riggingD}
\Dr_\m v^\n:=\Pi^\r{}_\m\Pi^\n{}_\s \na_\r v^\s,
\ee
where $\Pi^\m{}_\n = \d^\m_\n +l^\m n_\n$ is a `half-projector'.  
The pull-back of \eqref{riggingD} gives a well-defined 3d connection acting on hypersurface tensors and forms. There is however no canonical choice, and we have a different connection for each choice of rigging.\footnote{One can reduce the freedom choosing for instance the rigging 1-form to be hypersurface orthogonal and Lie dragged by the null tangent vector, which leaves a super-translation residual freedom. There is also some gauge freedom, for instance changing the rigging by a global translation does not change the connection, so given a rigging connection, there is not a unique rigging vector associated to it.} 
If the hypersurface is shear and expansion free, all rigging connections become rigging-independent and match the canonical, induced connection.

From the point of view of holography, it is useful to consider a purely intrinsic description of a null surface, that makes no reference to its embedding. An elegant approach to this problem is to exploit the natural fibration of $\cN$ by null geodesics, and replace the (pull-back of the) rigging 1-form with a choice of \emph{Ehresmann connection} on this fibre bundle. The intrinsic description of null hypersurfaces is the subject of \emph{Carollian geometry}, which you will also learn about in this school.
The rigging connections considered here are related to the Carollian connections. See e.g. \cite{Ciambelli:2019lap,Herfray:2021qmp,Fiorucci:2025twa,Ciambelli:2025unn}.

Taking the pull-back of \eqref{thEH} one finds (see e.g. \cite{Parattu:2015gga,Lehner:2016vdi,Hopfmuller:2016scf,Oliveri:2019gvm}), in units $16\pi G=1$,
\begin{align}\label{ThN2}
\pbi{\th}^{\sscr EH} = \big[ \s^{\m\n} \d \g_{\m\n}  + \pi_\m \d l^\m +2\d(\th+k)
\big]\eps_\cN +\th\d\eps_\cN+ d\vartheta^{\sscr EH}, 
\end{align}
where
\be
\pi_\m:=-2\left(\om_\m+\f\th2n_\m\right) = 2 \left(\eta_{\m}+\left(k-\f\th2\right)n_\m\right),
\ee
and
\be\label{vthEH}
\vth^{\sscr EH}=n^\m\d l_\m \eps_S - i_{\d l}\eps_\cN =
\left(n^\m\d l_\m + n_\m\d l^\m \right)\epsilon_S - n\w i_{\d l}\eps_S.
\end{equation}
Notice that since $\s_{\m\n}$ is traceless, only the conformal class of $\g_{\m\n}$ enters the first pair of variables in the symplectic potential.
This expression is the $p\d q$ form of the reference Einstein-Hilbert symplectic potential. 

We now explore possible changes of polarizations and their boundary Lagrangians. 
Flipping the spin-2 pair only produces a trivial minus sign, since $\g^{\m\n}\s_{\m\n}=0$, and no boundary Lagrangian,
in agreement with the fact that the configuration and momentum variables of this pair capture twice the same information.\footnote{The shear is the Lie derivative of the induced metric, see \eqref{defsigma}. The dependence of momentum on position is a general property of null hypersurfaces, occurring also in the canonical formalism, and due to the presence of second class constraints, see e.g. \cite{Alexandrov:2014rta}.} Flipping the spin-1 pair is not useful, because $\pi_\m$ is determined on-shell by the induced metric, which already enters the spin-2 pair. It remains the spin-0 sector, where one can consider changes of polarization in both inaffinity $k$ and expansion $\th$. This leads to the 2-parameter family of polarizations
\cite{Odak:2023pga,Chandrasekaran:2023vzb}
\begin{align}\label{ThN3}
\pbi{\th}^{\sscr EH} &= \th^{(b,c)} -\d\ell^{(b,c)}+ d\vartheta^{\sscr EH}, \qquad \ell^{(b,c)}=-(bk +c\th)\eps_\cN,
\\\label{Thbc}
& \th^{(b,c)} = \big[ \s^{\m\n} \d \g_{\m\n}  + \pi_\m \d l^\m +(2-b)\d k+(2-c)\d\th \big]\eps_\cN 
-(bk+(c-1)\th)\d\eps_\cN.
\end{align}
They correspond to different choices of conservative and dissipative boundary conditions.
Before discussing them, let us keep arbitrary variations, to talk about anomalies and see Noether's theorem in action.

\subsubsection{Covariance and flux-balance laws}

The issue of covariance is more delicate on a null boundary, because there are non-dynamical, background fields that necessarily enter the description.
One is the embedding field $\Phi$, and at fixed embedding, the choice of rescaling factor $f$ of $l^\m$; another one is the auxiliary vector field $n^\m$. 
Thus while $\th^{\sscr EH}$ is general covariant by construction, there is no guarantee that every $b,c$ choice in \eqref{Thbc} is also covariant, and one has to explicitly check this.
This background dependence can be conveniently studied using the Newman-Penrose formalism, because changes in the background fields (rescaling the normal and changing the rigging vector) can be generated using internal Lorentz transformations that preserve the direction of the null vector $l$ (called class-III and class-I in the nomenclature of \cite{Chandra}). The anomaly of quantities built from $l$ and $n$ can thus be equivalently stated as their gauge-dependence under class I and III transformations. This was studied in \cite{Odak:2023pga,Chandrasekaran:2023vzb} (see also \cite{Ciambelli:2023mir} for another approach to this question).
If we restrict attention to diffeomorphisms that are tangent to $\cN$, the anomalies are
\be\label{defw}
\D_\xi l_\m \eqonN - w_\xi l_\m, \qquad \D_\xi l^\m \eqonN - w_\xi l^\m, \qquad w_\xi := (\pounds_\xi - \d_\xi )\ln f+\bar\xi^\Phi
\ee
and
\begin{align}
\D_\xi\eps_\cN=w_\xi\eps_\cN, \qquad \D_\xi\th=-w_\xi\th, \qquad \D_\xi\eps_S=0,
\qquad \label{anok}
\D_\xi k= -(\pounds_l+k)w_\xi. 
\end{align}
Using these one can compute 
\be\label{anoell}
\D_\xi \ell^{\sscr (b,c)} = b\pounds_l w_\xi \, \eps_\cN, \qquad \D_\xi \vth^{\sscr EH} = -\d\D_\xi\ln f\eps_S
\ee
We see that a sufficient condition for anomaly-freeness of tangent diffeomorphisms is constant $f$, and $b=0$. 
For any non-zero value of $b$, we have a non-covariant boundary Lagrangian, hence we should expect a modification to Noether's flux formula in the form \eqref{jxia}.

Since the boundary variations are arbitrary, we have for the moment no restrictions on the boundary diffeomorphisms, other than taking them tangent to $\cN$ so to preserve the fact that it is null.
Then, using the Raychaudhuri and Damour equations, 
it is possible to show that \cite{Odak:2023pga} (see also \cite{Hopfmuller:2018fni})
\begin{align}\label{IxiNull}
I_\xi\th^{\sscr (b,c)}&= \big[ \s^{\m\n} \d_\xi \g_{\m\n}  + \pi_\m \d_\xi l^\m +(2-b)\d_\xi k+(2-c)\d_\xi\th \big]\eps_\cN -(bk+(c-1)\th)\d_\xi\eps_\cN \\
&=-2R_{\m\n}(l^\m l^\n+l^\m Y^\n)\eps_\cN -2d\left[\left(Y^\m\eta_\m+\f c2f\th+\f{b-2}2fk+w_\xi\right)\eps_S\right]+b\pounds_lw_\xi\eps_\cN,
\end{align}
up to terms that are total derivatives on the cross-sections and thus do not contribute to the surface charges for compact surfaces. 
The first line is the flux in $p\d_\xi q$ form. The field space variations $\d_\xi q$ are in general not simply Lie derivatives because of the presence of anomalies, and should be computed using $\d_\xi=\pounds_\xi+\D_\xi$ and the anomalies derived from \eqref{defw} etc.
The last term of the second line can be recognised as the boundary Lagrangian anomaly \eqref{anoell}. 
We conclude that 
\begin{align}\label{genflux}
dq{}^{\sscr (b,c)}_\xi &\eqons I_\xi\th^{\sscr (b,c)}-\D_\xi\ell^{\sscr (b,c)}, 
\end{align}
where we dropped the $i_\xi L$ term since we restrict to tangent diffeomorphisms, and 
\be\label{qbc1}
q_\xi^{\sscr (b,c)} = -2\left(\f c2f\th + w_\xi +\f{b-2}2fk+Y^\m\eta_\m\right) \eps_S.
\ee
The same result \eqref{qbc1} can be obtained using directly the improved Noether charge formula \eqref{jiN} \cite{Odak:2023pga}.
We thus have two alternative ways to get the charges: `integrating the fluxes', namely manoeuvring the terms of $p\d_\xi q$ as to isolate the field equations and a corner term, as we did in \eqref{IxiNull}, or starting from the reference Noether charge and computing the improvement shift following \eqref{jiN}. 

The equation \eqref{genflux} describes the flux-balance laws of general relativity on an arbitrary null hypersurface, for an arbitrary tangent diffeomorphisms.
A priori, every boundary diffeomorphism can be a non-degenerate direction of the symplectic 2-form. The question is which ones are compatible with the boundary conditions. Therefore to talk about boundary symmetries, we first have to decide which boundary conditions to use. These can be distinguished in the two categories of Table~\ref{TableTwoCases}. Once this choice is made, one specializes \eqref{genflux} to the corresponding diffeos and $(b,c)$ parameters, and deduces mechanical laws for null hypersurfaces and horizons, and can investigate their thermodynamical interpretation.

The general charges \eqref{qbc1} have many important applications for the dynamics of horizons and their thermodynamical interpretation, and once specialized to the boundary conditions and values of $(b,c)$, reproduce many notable cases. The Weyl charge at the bifurcation surface $\l=0$ is the famous Wald's entropy \cite{Wald:1993nt}, and more in general at an arbitrary cross-section it provides a notion of dynamical entropy \cite{Chandrasekaran:2018aop,Chandrasekaran:2019ewn,Rignon-Bret:2023fjq,Odak:2023pga,Chandrasekaran:2023vzb,Ciambelli:2023mir,Hollands:2024vbe,Rignon-Bret:2024zhj}.
The angular momentum charge provides the multipole moments of isolated horizons \cite{Ashtekar:2004gp,Ashtekar:2021wld}, where it is  commonly rewritten in terms of $\im(\Psi_2)$ using a Newman-Penrose tetrad.
Finally, the flux of the super-translation charge is related to Wall's entropy \cite{Wall:2015raa}.

\subsubsection{Conservative boundary conditions}

To discuss the conservative boundary conditions, recall that $\d\g_{\m\n}=0$ implies that also $\d\s_{\m\n}$, $\d\th$ and $\d\eps_\cN$ vanish.
On the other hand, $\d l^\m$ and $\d k$ are independent variations. As we have seen from the previous discussion, there is a clear sense in which these variations are `pure gauge', in the sense that one can always find coordinates on $\cN$ and its neighbourhood to achieve any desired value of $l^\m$ and $k$, for any given metric.
If we impose only $\d\g_{\m\n}=\d l^\m=0$, we see from \eqref{Thbc} that we need the polarization $b=2$. This was considered in \cite{Lehner:2016vdi}, and the problem there exposed is that the required boundary Lagrangian is not covariant, since it depends on the inaffinity $k$ which is an arbitrary, coordinate-dependent quantity. In our language, we would say that the required boundary Lagrangian is anomalous. 
A natural way to fix this gauge freedom is to choose affine coordinates, namely $k=0$. 
Other choices have also been considered in the literature. For instance, the `dressing time' approach of \cite{Ciambelli:2023mir} replaces the affine gauge-fixing with the (field-dependent) condition $2k+\th=0$.
A covariant resolution to the problem is to include the $\d k=0$ condition, as was suggested in \cite{Chandrasekaran:2018aop}. This option allows us to set $b=0$, and there is then no need of any anomalous boundary Lagrangian. 
Since any $b=0$ is anomaly-free, we have a 1-parameter family of covariant polarizations. The choice $c=2$ was considered in \cite{Chandrasekaran:2018aop}.
The choice $c=1$ was proposed in \cite{Odak:2023pga}, where it was called conformal or York, because it is reminiscent of the conformal boundary conditions in the time-like case. These choices are summarized in Table~\ref{TablebcCons}.
\begin{table}[H]
\begin{center}
  \begin{tabular}{l|l|c|l}  
\emph{polarization} & \emph{symplectic potential} & $(b,c)$ & \emph{cons. boundary conditions} \\ \hline
Neumann & $(\s^{\m\n} \d \g_{\m\n}  + \pi_\m \d l^\m )\eps_\cN +2\d(k+\th)\eps_\cN +\th\d\eps_\cN$ & (0,0) & $\d\g_{\m\n}=\d l^\m=0$ \\
Dirichlet & $(\s^{\m\n} \d \g_{\m\n}  + \pi_\m \d l^\m )\eps_\cN -(2k+\th)\d\eps_\cN$ & (2,2) & $\d\g_{\m\n}=\d l^\m=0$ \\
CFP & $(\s^{\m\n} \d \g_{\m\n}  + \pi_\m \d l^\m +2\d k )\eps_\cN - \th\d\eps_\cN$ & (0,2) & $\d\g_{\m\n}=\d l^\m=\d k=0$ \\
ORBS & $(\s^{\m\n}\d\g_{\m\n}  + \pi_\m \d l^\m +\d (2k+\th )\eps_\cN $ & (0,1)& $\d\s_{\m\n}=\d l^\m=\d(2 k+\th)=0$
\end{tabular}
\end{center}   
\caption{\label{TablebcCons} \emph{\small{Different polarizations for a null boundary, and their corresponding conservative boundary conditions. The names Neumann and Dirichlet are chosen by analogy with the time-like case, since the first coincides with the original Eistein-Hilbert polarization -- up to the corner term --, and the second with the null equivalent of trace $K$ boundary Lagrangian, namely trace $W$. CFP refers to \cite{Chandrasekaran:2018aop}, and ORBS to \cite{Odak:2023pga}.}}}
\end{table}

Once the conservative boundary conditions are chosen, with the associated polarization of the symplectic potential, one can study the boundary symmetries as the residual diffeomorphisms that preserve the boundary conditions. Since conservative boundary conditions include the induced metric, one would need $\d_\xi \g_{\m\n}=0$, which typically restricts the phase space to very special solutions admitting boundary isometries. A much richer set of boundary symmetries occurs with dissipative boundary conditions, which we discuss next.

\subsubsection{Dissipative boundary conditions} 

Assuming that the null hypersurface is incoming, any flux through it would correspond to dissipative boundary conditions. Conversely if the null hypersurface is outgoing, flux through it corresponds to absorbing boundary conditions. Let fix ideas taking the former example.
We now compare \eqref{Thbc} with the dissipative case of Table~\ref{TableTwoCases}, namely ask what solutions correspond to the stationarity condition $p=0$.
For any value of the parameters, the stationarity condition will include vanishing shear.
A natural choice of stationary solutions are non-expanding horizons \cite{Ashtekar:2000sz,Ashtekar:2021wld}. There is no gravitational radiation across them, and can be considered to be in equilibrium. In vacuum, they are characterized by vanishing shear and expansion: $\s_{\m\n}=\th=0$. On the other hand, $\eta_\m\neq 0$ unless we restrict to spherical symmetry, and $k\neq 0$ unless we restrict to representatives with affine normal, a choice known as \emph{weakly isolated horizons} in the literature. These two additional restrictions have thus no relation to the physical notion of stationarity in the sense of lack of gravitational radiation, and it makes sense not to impose them. Looking at \eqref{Thbc}, we can select this notion of stationarity and preserve covariance if we take $b=0$ and $c=2$, and boundary conditions $\d l^\m=\d k=0$ which as argued previously can be understood as gauge fixing. This is the CFP choice \cite{Chandrasekaran:2018aop} also used in \cite{Ashtekar:2021kqj}.

A drawback of this choice of stationarity, and its corresponding polarization and boundary conditions, is that it does not include a null cone in Minkowski: this is obviously  an equilibrium and non-radiative configuration, but is expanding. An alternative choice that includes flat null cones as stationary is to take $c=1$ and add $\d\th=0$ to  the ORBS boundary conditions \cite{Odak:2023pga}. In the ORBS proposal, the stationarity condition $\pbi{\th}=0$ is relaxed to include every hypersurface of  vanishing shear. This is achieved adding $\d\th=0$ as a boundary condition. Non-expanding horizons and flat null cones are still stationary but in the weaker sense that $I_\xi\pbi{\th}=0$. In other words the symplectic flux can be non-zero, but in a way that does not affect the flux-balance laws. 
The ORBS choice gives rise to a very elegant description of dynamical entropy \cite{Rignon-Bret:2023fjq} that improves upon the one based on the CFP polarization \cite{Rignon-Bret:2023fjq,Hollands:2024vbe,Visser:2024pwz}.
Relaxing the boundary conditions further has been explored in the literature, but leads to notions of stationarity and dissipation whose physical meaning is unclear. For instance $\d k\neq 0$ and $b\neq 2$ leads to a non vanishing flux in Minkowski, and $\d l^\m\neq 0$ means that a vanishing flux would require $\pi_\m=0$, namely spherical symmetry. 

Some of these possibilities are summarized in Table~\ref{TablebcDiss}, and showcase neatly the viewpoint we take with the generalized Wald-Zoupas prescription: adapt your choice to different physical problems and the boundary conditions they define. Always insist on covariance, but allow different notions of stationarity to be relevant.
\begin{table}[H]
\begin{center}
  \begin{tabular}{l|l|c|l|l}  
\emph{polarization} & \emph{symplectic potential} & $(b,c)$ & \emph{diss. boundary cond.} & \emph{stationarity cond.} \\ \hline
Neumann & $\s^{\m\n} \d \g_{\m\n} \eps_\cN + \th\d\eps_\cN$ & (0,0) & $\d l^\m=\d(k+\th)=0$ & $\s_{\m\n}=\th=0$\\
Dirichlet & $\s^{\m\n} \d \g_{\m\n} \eps_\cN -(2k+\th)\d\eps_\cN$ & (2,2) & $\d l^\m=0$ & $\s_{\m\n}=2k+\th=0$\\
CFP & $\s^{\m\n} \d \g_{\m\n} \eps_\cN - \th\d\eps_\cN$ & (0,2) & $\d l^\m=\d k=0$ & $\s_{\m\n}=\th=0$ \\
ORBS & $\s^{\m\n}\d\g_{\m\n} $ & (0,1)& $\d l^\m=\d k=\d\th=0$ & $\s_{\m\n}=0$
\end{tabular}
\end{center}   
\caption{\label{TablebcDiss} \emph{\small{Different polarizations for a null boundary with dissipative boundary conditions, and their  corresponding stationarity condition.}}}
\end{table}

The boundary symmetries associated with these boundary conditions were studied e.g. in \cite{Donnay:2015abr,Donnay:2016ejv,Chandrasekaran:2018aop,Adami:2020amw,Adami:2021kvx,Ashtekar:2021wld,Ashtekar:2021kqj,Odak:2023pga,Chandrasekaran:2023vzb}, and are summarized in Table~\ref{TableGnull}. They can be equivalently characterized in terms of universal structures, as opposed to boundary conditions \cite{Chandrasekaran:2018aop,Odak:2023pga}.
\begin{table}[H]
\begin{center}\begin{tabular}{l|l|l}
\emph{symmetry group} & \emph{boundary conditions} & \emph{universal structure} \\\hline
$\Diff(\cN)$  & $\d l_\m=0$ & $\{l_\m\sim\om l_\m\}$ \\
$\Diff(\cN)_l = \Diff(S)\ltimes\Diff(\R)^S$ & $\d l_\m=\d l^\m=0$ & $\{(l_\m,l^\m)\sim(\om l_\m,\om l^\m)\}$ \\
CFP= $(\Diff(S)\ltimes\R^S_W)\ltimes\R^S_T$ & $\d l_\m=\d l^\m=\d k=0$ & $\{(l_\m,l^\m,k)\sim(\om l_\m,\om l^\m,\om(k+\pounds_l\ln\om))\}$ \\
AKKL= $(\SL\ltimes\R)\ltimes\R^S_T$ & $\th=\s_{\m\n}=0$ & $\{([\os{l}],\os{q}_{\m\n})\sim(\om^{-1}[\os{l}],\om^2\os{q}_{\m\n})\}$
\end{tabular}\end{center}
\caption{\label{TableGnull} \emph{\small{Boundary symmetries for different dissipative boundary conditions on  a null hypersurface, and their corresponing universal structures as equivalence class of conformal transformations. }}}
\end{table}
These symmetry groups also appear in the study of null infinity, which can be realized as a non-expanding horizon in the conformally compactified spacetime, and more precisely a weakly isolated horizon \cite{Ashtekar:2024mme,Ashtekar:2024bpi}. In particular, the `little group' $\Diff(\cN)_l$ preserving a null geodesic congruence $l$ is the same as the Newman-Unit group, the CFP group is the same as the BMSW group. There is no analogy with the gBMS group. The AKKL group \cite{Ashtekar:2021wld} is the same as the BMS group plus a dilation generator, and arises if we restrict the boundary conditions to allow only non-expanding horizons, which come with a universal equivalence class of round metrics and normals.

\section{BMS symmetry and gravitational charges at null infinity}

\subsection{Future null infinity}

To study the asymptotic symmetries of gravitational waves, we are interested in the behaviour of asymptotically flat spacetimes along null directions. To gain some intuition about these asymptotics, let us first consider the case of flat spacetime.
If we use spherical coordinates and retarded time $u:=t-r$, the metric reads
\be\label{etaB}
 ds^2 = -du^2-2dudr +r^2q_{AB}dx^Adx^B,
\ee
where $q_{AB}$ is the standard round sphere metric, and $x^A=(\th,\phi)$. Hypersurfaces of constant $u$ describe outgoing null cones, ruled by null geodesics, and $r$ is an affine parameter along them. Taking the limit $r\to\infty$ at constant $u$ is thus a way to reach future null infinity. A difficulty with this limit is that the metric becomes ill-defined, since $r^2$ diverges, and $dr$ is no longer defined. We can improve the mathematical control using Penrose's idea of conformal compactification. 
To do that, we change the radial coordinate $r$ to
\be\label{BondiOm}
\Om= \f1r.
\ee
We have
\be
d\Om = -r^{-2}dr, \qquad \p_\Om = -r^2\p_r, \qquad dr = -\Om^{-2}d\Om, \qquad \p_r= -\Om^2\p_\Om.
\ee
It follows that vector and form components change as follows,
\be
v^\Om = -\Om^2v^r, \qquad v_\Om = -\Om^{-2}v_r.
\ee
This has a strong impact into the study of limits. Something that looks divergent in $r$ coordinates may not be so, once a well-defined coordinate system is used.
For instance, the vector field $r\p_r=-\Om\p_\Om$ may look divergent as $r\to\infty$, but it is in fact well-defined, and actually vanishing, if we use the good coordinate $\Om$ at $\Om\to 0$. This opens up the possibility of adding points corresponding to $\Om=0$ to the spacetime manifold.
We thus obtain a new manifold $\hat M$, which we refer to as the conformally completed manifold.
The hypersurface $\Om=0$ is the boundary of $\hat M$.

Using coordinates $x^\m=(u,\Om,x^A)$ the Minkowski metric reads
\be
{d s}^2 = -du^2+\Om^{-2}( 2dud\Om+q_{AB}dx^Adx^B).
\ee
It still blows up at the boundary of $\hat M$, but in a more uniform way, and that can now be controlled. 
Penrose's key idea is to introduce a conformally rescaled metric (aka `unphysical metric') $\hat \eta_{\mu\nu}=\Om^{2}\eta_{\mu\nu}$,
so that
\be\label{hetaBOm}
{d\hat s}^2 =  \Om^2ds^2 = 2dud\Om+q_{AB}dx^Adx^B - \Om^2du^2\eqonS 2dud\Om+q_{AB}dx^Adx^B.
\ee
Or in matrix form,
\be\label{hatetaBondiOm}
\hat \eta_{\mu\nu}= \left(\begin{matrix} -\Om^2 & 1 & 0 \\& 0 & 0 \\&& {q}_{AB}\end{matrix}\right), \qquad
\hat\eta^{\mu\nu}= \left(\begin{matrix} 0 & 1 & 0 \\& \Om^2 & 0 \\&& {q}^{AB}\end{matrix}\right).
\ee

The unphysical metric at $\Om=0$ is given by the last equality in \eqref{hetaBOm}: it is well-defined and invertible.
The pair $(\hat M,\hat g)$ is the conformally rescaled spacetime, and its boundary $\Om=0$ is the hypersurface we refer to as future null infinity $\scri^+$, 
or $\scri$ in short, since we will talk mostly about future null infinity alone. 
While $\scri$ does not `exist' in the physical spacetime, it is the boundary of the conformally completed spacetime.
Since $\scri$ is a hypersurface of $\hat M$, its properties can be studied using local differential geometry. 
In particular, we can take as normal to $\scri$
\be\label{nG}
\nG:=d\Om,
\ee
and observe that its norm $\nG^2=\hat \eta^{\Om\Om}$ vanishes at $O(\Om^2$), hence $\scri$ is a null hypersurface. Furthermore as a vector,
\be
\nG = \p_u +\Om^2\p_\Om \eqonS \p_u.
\ee
The retarded time vector $\p_u$ is  time-like everywhere in the bulk, and becomes null at $\scri$.
Finally, the induced, 3d metric $q_{ab}$ at $\scri$ is degenerate, with null vector precisely $\p_u$, which provides a an affinely parameterized tangent vector to the null geodesics of $\scri$. In the $(u,x^A)$ coordinates induced from the bulk coordinates, 
\be
q_{ab}= \left(\begin{matrix}  0 & 0 \\& {q}_{AB}\end{matrix}\right).
\ee

Remark: We noticed in Section~\ref{SecNull} that on a null hypersurfaces there is no canonical choice of normal. However in the case of $\scri$ the situation is special, because the conformal compactification provides a preferred choice, given by \eqref{nG}. The existence of this choice tying up the normal to the conformal factor is ultimately responsible for the extra generator of the symmetry group of physical non-expanding horizons with respect to the BMS group at $\scri$ \cite{Ashtekar:2021wld,Ashtekar:2024bpi}.

\subsection{Global and asymptotic symmetries}

A metric possesses isometries if there are non-trivial solutions to the Killing equation \eqref{Keq}.
Minkowski spacetime is maximally symmetric and admits ten Killing vectors. They form an algebra under Lie bracket which is isomorphic to the Poincar\'e algebra.
The description of the Killing vectors is simplest in Cartesian coordinates, where the metric is constant everywhere, and we get
\be\label{xiP4}
\xi = a^{\mu}+b^{\mu}{}_{\nu}x^{\nu}, 
\ee
where $a$ and $b$ are constants, and $b_{(\m\n)}=0$.
Changing coordinates to $x^\m=(u,r,x^A)$, the same vectors read
\begin{align}\label{Keta}
\xi =  f\p_u +Y^A\p_A - r\dot f\p_r - \f1r\Dd^A f\p_A +\f12\Dd^2 f \p_r,
\end{align}
where $\Dd_A$ is the covariant derivative on the 2-sphere, and $f=T+\f u2 \Dd_A Y^A$. See Appendix~\ref{AppA} for details.
The function $T=T(x^A)$ is a linear combination of the lowest harmonics $l=0,1$ and encodes the translation parameters via
$T=a^0-\vec a\cdot\vec n$.
The vectors $Y^A$ are conformal Killing vectors of the sphere, which span the Lorentz group and encode the rotation $r^a=-\f12\eps^{abc}b_{bc}$ and boost $b^a=b^{0a}$ parameters via
$Y^A=\eps^{AB}\p_B(\vec r\cdot\vec n)+\p^A(\vec b\cdot \vec n)$.
Using the inverse radius coordinate $\Om=1/r$ makes it manifest that the Killing vector are tangent to $\scri$:
\begin{align}\label{KetaOm}
\xi&= f\p_u + Y^A\p_A +\Om(\dot f\p_\Om - \Dd^A f \p_A) - \f12\Om^2 \Dd^2 f \p_\Om.
\end{align}
These expressions are exact to all orders in $r$, or $\Om$: these are the global Killing vectors. 
Notice that the Killing vectors are also \emph{conformal} Killing vectors of the unphysical metric, since they satisfy
\be\label{xihateta}
\pounds_\xi \hat \eta_{\m\n} = 2\pounds_\xi\ln\Om \, \hat\eta_{\m\n} +\Om^2\pounds_\xi\eta
= 2\a_\xi \hat\eta_{\m\n},\qquad  \a_\xi :=\pounds_\xi\ln\Om= \f{\nG\cdot\xi}{\Om}. 
\ee

Now let's look at \eqref{hetaBOm}. We can ask for a weaker condition than global Killing vectors, namely \emph{asymptotic} Killing vectors that preserve only the leading order at $\scri$ of \eqref{hetaBOm}: 
\be\label{xihateta1}
\pounds_\xi \hat \eta_{\m\n} \eqonS 2\a_\xi \hat\eta_{\m\n}.
\ee
Requiring $\xi$ to solve \eqref{xihateta1} as opposed to \eqref{xihateta} has two effects on the global solution \eqref{Keta}: first, only the $O(\Om)$ is determined, all higher orders are left free.\footnote{ 
The reason why the $O(\Om)$ is fixed is because we are requiring \eqref{xihateta1} for the spacetime metric. If we restrict the requirement to hold only for the pull-back, then only the tangent part of the vector field is determined.}
Second, $T$ no longer needs to be in the lowest two harmonics, it can be an arbitrary function on the sphere. 
Again, see Appendix~\ref{AppA} for details.
We write the result as 
\be\label{xiBMS}
\xi=  f\p_u +Y^A\p_A + \Om\left( f\p_\Om - \p^A f\p_A\right)+O(\Om^2)
\ee
with the understanding that now $T$ inside $f$ is an arbitrary function on the sphere. 
The sub-group of global translation is characterized by the $l=0,1$ modes of $T$, namely by solutions of the equation
\be\label{Tgl}
\Dd_{\la A}\Dd_{B\ra} T=0.
\ee
The $Y$'s are on the other hand still CKVs as in the global case.
The higher orders can be fixed requiring preservation of bulk coordinate choices, for instance.

\begin{figure}[H] 
\begin{center}  \includegraphics[width=3cm]{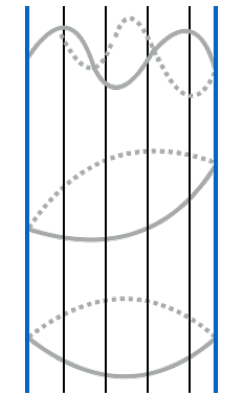}\hspace{2cm}
\includegraphics[width=8cm]{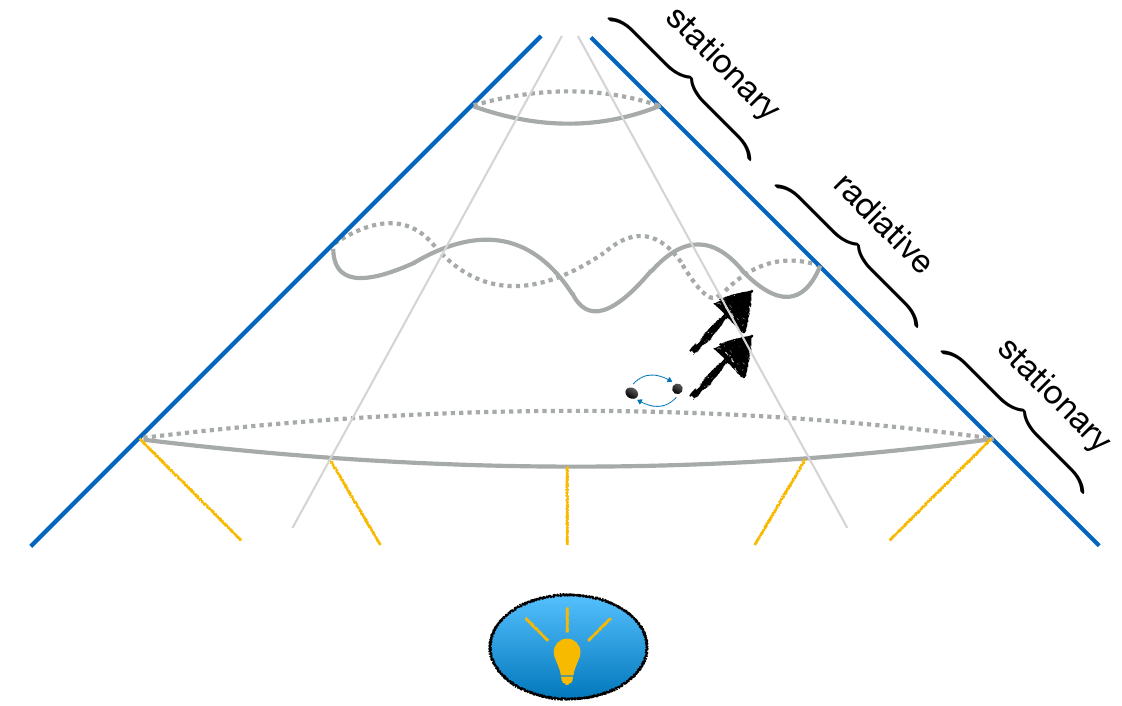} \end{center}
\caption{\label{FigCuts}    {\small Left panel: {\emph{Different cuts of $\scri$, from bottom up: an initial good cut, a translated one and super-translated one. The translated one is still a good cut, its ray back-tracing identifies a point translated from the origin. The super-translated one is now a bad cut, its ray back-tracing forms caustics and does not identify an point in the bulk of flat spacetime. } Right panel: \emph{The physical meaning of super-translations: in the initial stationary epoch, we calibrate the clocks of the asymptotic family of observers using light rays from a point source; during the radiative epoch, angle-dependent redshift produced by gravitational waves desynchronizes the clocks; in the final stationary epoch, a super-translation can be used to resynchronize the clocks.}} }}
\end{figure}
The new allowed diffeomorphisms are arbitrary angle-dependent time translations,
\be\label{u'T}
u'\eqonS u+T(x^A),
\ee
and are called \emph{super-translations}. Pictorially, global translations are like `ellipsoidal' deformation of a constant $u$ cut, and super-translations are `wriggly' deformations, see Fig.\ref{FigCuts}.
To gain more intuition about super-translations, let us fix the extension to all orders, requiring it to satisfy $\pounds_\xi \eta_{r\m}=0$ in retarded time coordinates, or better for later purposes, Bondi coordinates.
We can then compute the action of an asymptotic Killing vector on the Minkowski metric \eqref{etaB}, and observe that it adds new terms to it, constructed from $\Dd_{\la A}\Dd_{B\ra} T$ \cite{Compere:2018ylh}: 
\begin{align}\label{etaorbit}
& ds^2=-du^2-2dud\r -2\Dd^B\os{\s}_{AB}^vdudx^A+\left(\Big(\r^2+\f12\os{\s}_{CD}\os{\s}^{CD}\Big)q_{AB}-2\r \os{\s}_{AB}^v\right)dx^Adx^B, 
\end{align}
where
\be
\r=\sqrt{r^2+\f12\os{\s}_{AB}\os{\s}^{AB}}
\ee
is an affine parameter for the null geodesic congruences of constant $u$, and the `vacuum shear'
\be
\os{\s}_{AB}=\Dd_{\la A}\Dd_{B\ra}T, \qquad \Dd^B\os{\s}_{AB}= \f12\Dd_A(\Dd^2+2)T
\ee
is generated by a super-translation, and vanishes for global translations.

The consequence is that while the codimension-2 leaves of constant $(u,r)$ are round spheres, the new codimension-2 leaves of constant $(u',r')$ are non-round spheres. It further follows that while for $r\to 0$ the constant $u$ ingoing geodesics of different angles $x^A$ all converge to a point, the constant $u'$ ingoing geodesics of different angles start individually crossing before and don't focus to a point. From the point of view of $\scri$, there is no difference between $u$ and $u'$, both are equally good coordinates, corresponding to different choice of  foliating $\scri$. But from the bulk perspective, some cuts come from light emitted from a point, and are called good cuts. The rest are the bad cuts.

The vector fields \eqref{xiBMS} form a closed sub-algebra of the diffeomorphism algebra at $\scri$, given by
\be\label{BMSalgebra}
[\xi,\chi]=(T_{\xi}\dot f_\chi +Y_\xi[f_\chi] - (\xi\leftrightarrow \chi))\p_u + [Y_\xi,Y_\chi]^A\p_A.
\ee
This algebra exponentiates to a finite group action, a subgroup of the full diffeomorphism group of $\scri$ that we call BMS group:
\be
G^{\sscr BMS}=\SL\ltimes\R^S.
\ee
Here $\SL$ is the (covering) group of CKVs on the sphere, $\R^S$ the arbitrary function $T(x^A)$, and the semi-direct product structure follows from the action of $Y$ on $T$. 
To further discuss the properties of this group, we can consider various special cases of \eqref{BMSalgebra}. First, the commutator of any $\xi$ with a super-translation is still a super-translation, hence super-translations from an ideal sub-algebra, and a normal sub-group.
Any two super-translations commute, $[\xi_{T},\xi_{T'}]=0$, but a super-translation does not commute with a Lorentz transformation,
\begin{align}\label{anoY}
& [\xi_Y,\chi_T] = \xi_{T'}, \qquad  T'= Y^A\p_AT -\dot f T.
\end{align}
Notice that if $T$ is $l=0,1$, so is $T'$. Therefore global translations are also an ideal. The algebra of global $T$ and $Y$ reduces to the Poincar\'e algebra in retarded time coordinates, in particular $T'=\vec a\times \vec r\cdot \vec n $ for rotations. However, while there is a global notion of $T$, there is no global notion of $Y$. This can be anticipated from the discussion of null hypersurfaces where we saw that the vertical part of a tangent vector is canonical, but the horizontal part is ambiguous, and can be changed arbitrarily changing Ehresmann connection, or foliation. The same happens here: the $Y$ vector makes explicit reference to the foliation of $\scri$ by constant $u$ cross-sections. If we change the foliation by a super-translation as in \eqref{u'T} we find a \emph{different} Lorentz subgroup. A 4-parameter family of these Lorentz subgroups are the analogue of the different Lorentz subgroups of the Poincar\'e groups obtained by translations, but all the others are new.

We conclude that there is no unique subgroup corresponding to the Lorentz group, the notion of Lorentz subgroup of the BMS group is super-translation dependent. 
The situation is similar to what happens for the Poincar\'e group, whose Lorentz subgroup is not unique but depends on a choice of origin, and the freedom in doing so is spanned by the finite-dimensional group of translations, here we have to pick a cut, and this is an infinite-dimensional freedom unless we have access to the bulk and we can restrict attention to good cuts only. 

The interplay between super-translations and Lorentz transformations can also be read the other way around. If we act with a rotation, all that happens is that the different $m$ modes of the super-translation parameter are mixed, and $l$ stays invariant. But if we act with a boost, all $l$ modes of $T$ are mixed up.
This can be understood geometrically also because a boost is a conformal transformation of a round sphere, operation that leaves the curvature invariant but changes the metric, and in particular the two sets of spherical harmonics associated with the initial and the transformed metric get mixed up. Only the global $l=0,1$ sector is `pure', in the sense that these harmonics mix amongst themselves, without contribution from higher harmonics. But higher harmonics with all harmonics, including the global ones. This means is that while global translations are characterized by \eqref{Tgl} in any frame, non-global-super-translations are not: We first pick a frame, then we can talk about the $l\geq 2$ modes in that frame. But changing frame, these modes will mix, and get contributions from the $l=0,1$ modes as well.

Summarizing, the global Killing vectors preserve all of \ref{hetaBOm}, including the $O(\Om^{-2})$; the asymptotic Killing instead only the lowest order of \eqref{hetaBOm}. 
These define the BMS group, which can thus be identified as the asymptotic symmetry group of Minkowski spacetime.

\subsection{Asymptotically flat spacetimes} 

In the literature one can find two different approaches to asymptotically flat spacetimes at null infinity. The Bondi-Sachs approach \cite{Bondi:1960jsa,Sachs:1962wk}, based on a bulk gauge fixing and asymptotic expansion of the metric, and the Penrose-Geroch-Ashtekar approach \cite{Penrose:1964ge,Geroch:1977jn,Ashtekar:1978zz}, based on the idea of conformal compactification used above for Minkowski.
They are complementary, and it is both useful to know both.

In the Bondi-Sachs approach, we start from a coordinate patch $(u,r,x^A)$, where $A=1,2$ are coordinates on topological 2-spheres, and require that the level sets of $u$ are null, and that $x^A$ are Lie dragged along the null geodesics at constant $u$. This implies that 
\be
g^{uu}=g^{uA}=0 \quad \Leftrightarrow \quad g_{rr}=g_{rA}=0.
\ee
These 3 gauge-fixing conditions can be referred to as partial Bondi gauge (e.g. \cite{DePaoli:2017sar,Geiller:2022vto,Geiller:2024amx}).
The remaining gauge freedom can be used to fix $r$ to be the area radius of the 2-spheres, as in the Bondi (aka Bondi-Sachs) coordinates:
\be\label{Bg3}
\p_r(\det{}^{\sscr(2)} g_{AB}/r^4)=0;
\ee 
or an affine parameter for the null geodesics at constant $u$, as in the Newman-Unti coordinates, where one fixes $g^{ur}=-1$. 
Then, let us parametrize the gauge-fixed metric as
\begin{align}\label{gBondi}
& g_{\mu\nu}= \left(\begin{matrix} -V e^{2\exb}+\g_{AB} V^A V^B & -e^{2\exb} & -\g_{AB}V^B \\ & 0 & 0 \\
&& \g_{AB}\end{matrix}\right), \qquad 
 g^{\mu\nu}= \left(\begin{matrix} 0 & -e^{-2\exb} & 0 \\ &V e^{-2\exb} & -e^{-2\exb}V^A \\
&& \g^{AB}\end{matrix}\right),
\end{align}
with determinant
\be
\sqrt{-g}=e^{2\exb}\sqrt\g.
\ee
Switching to the conformal picture with $\Om=1/r$ and $\hat g_{\m\n}=\Om^2 g_{\m\n}$, requiring the the unphysical metric is smooth at $\Om=0$ gives
\begin{subequations}\label{falloffs}\begin{align}
& V=\Om^{-2}V^{\sscr (-2)} + \Om^{-1}V^{\sscr (-1)} + V^{\sscr (0)} +\Om V^{\sscr (1)}+O(\Om^2), \\
& \exb= {\exb}^{\sscr (0)}+\Om \exb^{\sscr (1)} + \Om^2 \exb^{\sscr (2)}+O(\Om^3), \\
& V^A = {V}^{{\sscr (0)}A} +\Om V^{{\sscr (1)}A} +\Om^2 V^{{\sscr (2)}A}, \\
& \hat q_{AB} = q_{AB} +\Om C_{AB} + \Om^2D_{AB}+O(\Om^3).
\end{align}\end{subequations}
The determinant condition \eqref{Bg3} imposes
\be
q^{AB}C_{AB}=0, \qquad q^{AB }D_{AB}=\f12C^{AB}C_{AB},
\ee
and similar conditions on the lower orders of the expansion.\footnote{The traceless piece $D_{\la AB\ra}$ vanishes on-shell if one insists on perfect smoothness of $\scri$, but it can be included in more general descriptions with only partial smoothness, e.g. \cite{Bieri:2023cyn,Geiller:2024ryw}.}
We can define as in the flat case the normal to $\scri$ via \eqref{nG}, then
\be
\nG^\m:=\hat g^{\m\n}\nG_\n = e^{-2\exb}(1,\Om^2V,V^A).
\ee
We can use the freedom of choosing coordinates on $\scri$ to fix $n\eqonS\p_u$ namely ${\exb}^{\sscr (0)}=1$ and $V^{\sscr (0)}{}^A =0$. 
The asymptotic Einstein's equations then give (see e.g. \cite{Barnich:2010eb,Geiller:2024amx,McNees:2024iyu})
\begin{align}\label{AEE1}
& \dot q_{AB}\eqons q_{AB} \p_u\ln\sqrt q, \qquad V^{\sscr (-2)} \eqons 0, \qquad V^{\sscr (-1)} \eqons \p_u\ln\sqrt q,\qquad
 V^{\sscr (0)} \eqons \f\cR2, \\ & {\exb}^{\sscr (1)}\eqons 0, \qquad {\exb}^{\sscr (2)}\eqons\bb:= -\f1{32}C_{AB}C^{AB},
\qquad V^{\sscr (1)}{}^A \eqons 0, \qquad U^A := V^{\sscr (2)}{}^A \eqons -\f12\Dd^BC_{AB},
 \end{align}
where $\cR$ is the curvature of $q_{AB}$, and $\cR=2$ if $q_{AB}$ is round.
The conformal picture makes it clear that the explicit form of $q_{AB}$ is completely irrelevant, because it can be changed arbitrarily changing conformal factor, without changing the \emph{geometric} (i.e., diffeomorphism-invariant) properties of being asymptotically flat. So in particular it is always possible to restrict attention to $\dot q=0$, know as \emph{Bondi condition} or \emph{divergence-free condition} in the conformal picture, and within that to round spheres, known as \emph{Bondi frames} in the conformal picture.\footnote{While this is always possible, it may not always be the most convenient option. For instance, the Robinson-Trautman solution can be naturally written in BS coordinates with $\dot q\neq 0$, and changing radial coordinate so to have a round sphere makes it bulk expression much more complicated \cite{GlennAli}.}

An important role is also played by the equations for $V^{\sscr (1)}$ and $V^{{\sscr (2)}A}$. To write them, it is convenient to parameterize
\begin{align}
V^{\sscr (1)} = -2M, \qquad & V^{{\sscr (2)}A} = - \frac{2}{ 3} (J^A+\p^A\bb+ C^{AB}U_C  ).
\end{align}
Then,
\begin{align}\label{Mdot}
\dot M & = -\f18 \dot C^2+\f14 \Dd\Dd \dot C+\f18\Dd^2\cR, \\\nn
 \dot J_A &= \f14 \dot C^{BC}\Dd_B C_{AC} +\f12 C_{AB}\Dd_C\dot C^{BC}-\f14\dot C_{AB}\Dd_C C^{BC}-\f18\p_A(C_{BC}\dot C^{BC}) +\f14C_{AB}\p^B \cR\\
 &\qquad +\p_AM+\f12 \Dd^B\Dd_{[A}\Dd_C C_{B]}{}^C.
\end{align}
The reason for this parameterization is that $J_A$ is chosen to match Dray-Streubel's Lorentz charge aspect. It corresponds to the choice $(1,1)$ in the parametrization of \cite{Compere:2019gft}, and it is related to the common Barnich-Troessaert (BT) \cite{Barnich:2011mi} and Flanagan-Nichols (FN) \cite{Flanagan:2015pxa} choices by
\be
J_A = N^{\sscr BT}_A - \p_A\b= N_A^{\sscr FN}+2\p_A\b+\f12C_{AB}U^B.
\ee
After these choices, the general fall-off on an arbitrary frame satisfying the Bondi condition is
 \begin{subequations}\label{gexp}\begin{align}
g_{uu} &= -\f\cR 2+\f{2M}r +  O(r^{-2}), \qquad
g_{ur} = -1-\f{2\bb}{r^2}+O(r^{-3}), \\
g_{uA} &=-U_A+\f2{3r}(J_A+\p_A\bb-\f12 C_{AB}U^B), \label{guAexp} \qquad 
g_{AB} = r^2 q_{AB}+r C_{AB}+ O(1).
\end{align}\end{subequations}
Further restricting the background $q_{AB}$ to be a round sphere we have the same expression but now $\cR=2$, and the gravitational waves, or news function, can be identified with $\dot C_{AB}$. 

We can derive the symmetry group of asymptotically flat metrics in two different ways. 
In the Bondi-Sachs approach, we define the asymptotic symmetries as the residual diffeomorphisms preserving the bulk coordinates and the boundary conditions.
That is, 
\begin{subequations}\label{BMSxicond}\be
\pounds_\xi g_{rr}=0, \qquad \pounds_\xi g_{rA}=0, \qquad \p_r(g^{AB}\pounds_\xi g_{AB})=0,
\ee
and
\begin{align}
& \pounds_\xi g_{ur}=O(r^{-2}), \qquad \pounds_\xi g_{uA}=O(1) \qquad \pounds_\xi g_{AB}=O(r), \qquad \pounds_\xi g_{uu}=O(r^{-1}).
\end{align}\end{subequations}
Solving these equations give
\begin{align}\label{xi}
\xi &=  f\p_u +Y^A\p_A - \f r2 \Dd^{A} Y_A\p_r - \f1r\Dd^A f\p_A +\f12\Dd^2 f \p_r+ \f1{2r^2} C^{AB} \Dd_B f\p_A + O(r^{-3}) \\ 
&= f\p_u + Y^A\p_A +\Om(\dot f\p_\Om - \Dd^A f\p_A) - \f12\Om^2(\Dd^2 f \p_\Om - C^{AB} \Dd_B f \p_A) + O(\Om^3),
\end{align}
where $f$ and $Y$ satisfy 
\be\label{BMS1}
f=T(x^A)+\f u2\Dd_A Y^A(x^B), \qquad \Dd_{\la A}Y_{B\ra}=0.
\ee
This coincides with the BMS symmetry that we found studying the asymptotic Killing vectors of Minkowski's spacetime.
The only novelty is that the bulk extension is now fixed to preserve the Bondi gauge of an arbitrary metric. In the flat case this reduces to \eqref{etaorbit},
and the $C_{AB}$ in the extension is then the vacuum shear of the flat metric.

In the conformal approach, we  rewrite \eqref{BMSxicond} in terms of the unphysical metric, which gives
\be\label{xihateta2}
\pounds_\xi \hat g_{\m\n} \eqonS 2\a_\xi \hat g_{\m\n},
\ee
and since the metric on $\scri$ is flat, we obtain the same result as before in \eqref{KetaOm}. 
Notice also that 
\be
\pounds_\xi \nG^\m = \pounds_\xi \hat g^{\m\n}\p_\m\Om+\hat g^{\m\n}\p_\n\pounds_\xi \Om = (\xi^\Om-2\a_\xi)\nG^{\m} =-\a_\xi\nG^{\m}
\ee
where in the last step we used \eqref{xihateta}. Since $\nG^\m$ is tangent to $\scri$, we can use hypersurface indices and write the equations that define the symmetry generators in terms of intrinsic quantities only, as
\be\label{xiUS}
\pounds_\xi q_{ab}=2\a_\xi q_{ab}, \qquad \pounds_\xi \nG^a=-\a_\xi \nG^{a}.
\ee
These equations allows us to interpret the BMS symmetries as diffeomorphism preserving the equivalence class under conformal transformations
\be
(q_{ab}, \nG^a)\sim (\om^2 q_{ab}, \om^{-1}\nG^a).
\ee
This is the \emph{universal structure} of asymptotically flat metrics.

The upshot is that we have two equivalent ways of thinking about BMS asymptotic symmetries:
as boundary diffeomorphisms that preserve the boundary conditions \eqref{BMSxicond}, or equivalently as isometries of the universal structure \eqref{xiUS} determined by the boundary conditions.
The result is the same, and it can also be identified simply looking at the asymptotic symmetries of Minkowski spacetime alone, which is sufficient since any asymptotic symmetry satisfies \eqref{xihateta2} which is equivalent to \eqref{xihateta1} under the boundary conditions. The analysis of boundary symmetries can be extended to weaker fall-off conditions, leading to larger symmetries than BMS, but we do not have time to cover these cases, nor the inclusion of logarithmic terms and non-smoothness also discussed in the literature.

The BMS transformations on the fields can also be derived in two different ways. In the Bondi-Sachs approach, 
we compute the transformations of the asymptotic phase space variables $\Phi=(M,J_A,C_{AB})$ through the definition
\be
\pounds_\xi g_{\m\n}[\Phi] \equiv g_{\m\n}[\Phi+\d_\xi\Phi] - g_{\m\n}[\Phi].
\ee
The result can be written as 
\be\label{dBMS2}
\d_\xi \Phi = \pounds_\xi \Phi +\D_\xi\Phi,
\ee
where the second term is the anomaly discussed before, namely $\D_\xi \Phi=-\p_\eta\Phi\pounds_\xi\eta$. It captures the dependence of $\Phi$ on the background fields $\eta$, which in this case are the conformal factor, and the foliation $u$ used to define the different metric components being identified as mass, angular momentum and shear. Knowing the dependence of $\Phi$ on the background fields, and computing their Lie derivative under BMS vector fields, we can derive \eqref{dBMS2} without going through the asymptotic expansion. This gives a more geometric and intuitive way of understanding the transformation laws, and which furthermore can be derived intrinsically at $\scri$ without the need of any bulk expansion \cite{Rignon-Bret:2024gcx}.

\subsection{Fluxes and charges for the BMS symmetry}

Charges and fluxes for the BMS symmetry in the radiative phase space were first identified in
\cite{Geroch:1977jn,Ashtekar:1981hw,Ashtekar:1981bq,Dray:1984rfa,Dray:1984gz}. 
Using the Bondi-Sachs parametrization \eqref{gexp} and a Bondi frame with $\cR=2$, for any BMS symmetry $\xi$ we can write the flux between \emph{any} two cuts of $\scri$ as
\be\label{jNBMS}
F_\xi = \int_\cN j_\xi = -\f1{32\pi}\int_\cN \dot C_{AB}\d_\xi C^{AB} \eps_\scri\eqons Q_\xi[S_2]-Q_\xi[S_1],
\ee
where
\be
\d_{\xi} \,C_{A B}= (f\p_u+\pounds_Y-\dot{f}) C_{A B}-2 {\Dd}_{\langle A} \Dd_{B\rangle} f,
\ee
and
\begin{align}\label{QBMS}
Q_\xi[S] = \f1{8\pi}\oint_S(2f M+ Y^AJ_A)\eps_S.
\end{align}
Specializing \eqref{jNBMS} to a super-translation we have
\begin{align}\label{STflux}
\D Q_T = -\f1{32\pi}\int (T\dot C_{AB} \dot C_{AB} - 2C^{AB} \Dd_{A} \Dd_{B} T)\eps_\scri.
\end{align}
The super-translation flux has two terms, the one quadratic in the news is called `hard', and the one linear in the news, which vanishes for global translations, and is called `soft'. The soft term is the famous displacement memory effect \cite{Christodoulou:1991cr,Thorne:1992sdb}, that is expected to be observed in future detections, see e.g. \cite{Grant:2022bla}. Isolating the memory, we see that it has two contributions, $\D Q_T$ which contains the linear memory (namely what would be present also in the linearized theory), and a non-linear term, the `hard' flux. The super-translation charge $Q_T$ is known as Bondi super-momentum, also known as Geroch super-momentum for its coordinate-independent description in  \cite{Geroch:1977jn}. The Lorentz charge $Q_Y$ was first identified by Dray-Streubel \cite{Dray:1984rfa}, and among its useful properties, it reduces to the Komar integral for radiative but axially symmetric spacetimes \cite{Chen:2022fbu,Ashtekar:2023zul}.

The charges can be written in a coordinate-independent way using the Newman-Penrose formalism, 
\be\label{DS}
M = -\left( \psi_2+\s\dot{\bar\s} +\f12\left(\eth^2 \bar\s - cc\right)\right) = -\re(\psi_2 + \s \dot{\bar\s}), 
\qquad m^AJ_A = -\left( \psi_1 + \s\eth \bar\s+\f12\eth(\s\bar\s)\right).
\ee
The relation to the Bondi expression is obtained choosing for Newman-Penrose tetrad $n\eqonS\p_u$ and $l\eqonS -du$, and then $\s=\f12m^Am^BC_{AB}$.

\begin{figure}[H] 
\begin{center}  \includegraphics[width=3cm]{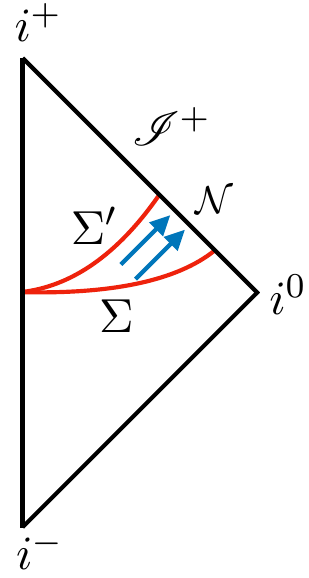} 
\end{center}
\caption{\label{FigPD}    {\small {\emph{Two space-like hypersurfaces $\Si$ and $\Si'$ intersecting $\scri$ and delimiting a portion $\cN$ of it. By the conservation law $d\om\eqons 0$, the canonical generator on the phase space defined on the portion of $\scri$ is equal to the difference of the two canonical generators on the space-like hypersurfaces.}} }}
\end{figure}

The fluxes can be understood as canonical generators for the radiative phase space at $\scri$ (or for a portion $\cN$ of it), and the surface charges as canonical generators for a hyperbolic space-like hypersurface $\Si$ intersecting $\scri$ at $S$, see Fig.~\ref{FigPD}, in the sense of the general principle given in Section~\ref{genWZsection}.\footnote{The existence of the generator requires the cross sections under consideration to be non-radiative, otherwise the associated vector fields are not Hamiltonian.}

The expressions \eqref{jNBMS} and \eqref{QBMS} can be derived in four different ways: the Ashtekar-Streubel approach, the Wald-Zoupas approach, the improved Noether charge approach, and the Barnich-Brandt approach.
We will now briefly describe the methods and their equivalence.
For simplicity we restrict attention to the case of Bondi frames, and refer to the literature for generalizations. We further restrict attention to the BMS case, for which there is no need of symplectic renormalization, nor of corner term improvements. We refer to the literature for the eBMS and gBMS cases where the situation is more complicated and these additional techniques are required.

\subsubsection{Ashtekar-Streubel approach}
\bit
\item\emph{Advantages:} Intrinsic at $\scri$; independent of bulk coordinates and bulk extensions of the symmetry vector fields; independent of symplectic potential ambiguities that don't affect the symplectic 2-form.

\item\emph{Caveats:} Integration of the angular momentum charge complicated; closed-form ambiguities to be resolved in a second stage; no relation to canonical generators on $\Si$.
\eit

In this approach one first evaluates the pull-back at $\scri$ of the standard Einstein-Hilbert symplectic 2-form.
The result can be described in terms of geometric quantities in an arbitrary coordinate system of $\scri$, however for simplicity
let us use the specific foliation induced by Bondi coordinates, and the parametrization \eqref{gexp}. This gives 
\be\label{OmAS}
\Om_\cN = -\f1{32\pi}\int \d \dot C_{AB}\cw \d C^{AB} \eps_\scri,
\ee
known as Ashtekar-Streubel symplectic form since it was first derived in \cite{Ashtekar:1981bq}. Here $\cN$ could be all of $\scri$, or a portion of it.
From this formula one can compute 
\be\label{IxiAS}
-I_\xi\Om_\cN = \d F_\xi + \f1{32\pi}\oint^{S_2}_{S_1} f \dot C_{AB} \d C^{AB} \eps_S,
\ee
where $F_\xi$ is the flux given in \eqref{jNBMS}, and $S_{1,2}$ two arbitrary cuts of $\scri$. The boundary terms make the  infinitesimal  generator not integrable, however the obstruction is measure zero with respect to the natural measure that one can introduce in the field space. This makes  it possible to define a  generator in the full radiative phase space starting from the integrable one by introducing a topology in the radiative phase space \cite{Ashtekar:1981bq,Ashtekar:2024stm}. 
This procedure identifies the flux \eqref{jNBMS} as the canonical generator of BMS symmetries. A simple example of this approach is reported in Appendix~\ref{AppB}. 

Next, in order to introduce the surface charges, one can use Einstein's equations to show that the flux is on-shell exact, the same procedure that we used in \eqref{IxiNull}. Hence if we evaluate it between two arbitrary cuts of $\scri$, it would provide the difference of two surface terms, each of which would coincide with the Noether current integrated on the hyperbolic $\Si$ intersecting $\scri$ at that cut, see Fig.~\ref{FigPD}.
It is particularly simple and illuminating to do this for the super-translation part of the symmetry. 
Writing the integrand of \eqref{STflux} as a 3-form, we have
\begin{align}
j_T^{\sscr BMS}&= -\f1{32\pi}\dot C_{AB} \d_T C^{AB}\eps_\scri = -\f1{32\pi}\dot{C}^{AB} ( T\dot C_{AB} - 2 \Dd_{A} \Dd_{B} T)\eps_\scri \nn\\
	&= \f1{32\pi}\Big[T (- \dot{C}^{AB} \dot{C}_{AB} + 2 \Dd_{A} \Dd_{B} \dot{C}_{AB} ) + 2\Dd_A( \dot{C}^{AB} \p_B T - T \Dd_B \dot{C}^{AB} )\Big]\eps_\scri \nn \\
	&\eqons  \f1{4\pi}\p_u\left( T M + \f14\Dd_A(C^{ AB}\Dd_B T-T\Dd_B C^{AB}) \right)\eps_\scri.\label{jTBMS}
\end{align}
The result can be written as
\be\label{DP}
j^{\sscr BMS}_T \eqons \f1{4\pi}D_aP^a_T\eps_\scri = \f1{4\pi}dP_T,
\ee
where
\be\label{GSM}
P_T^a
= \left(TM_\r, \ \f14 \big(T\Dd_B N^{AB}-  N^{AB}\Dd_B T\big) \right),
\ee
is the Geroch super-momentum  \cite{Geroch:1977jn}. 
Its Hodge dual defines the 2-form $P_T := \f12P^a_T \eps_{\scri abc}dx^b\w dx^c$, whose pull-back on the cross sections gives $P^u_T\eps_S=TM_\r\eps_S$,
hence we recover \eqref{QBMS}.
This calculation shows explicitly the Ashtekar-Streubel strategy of obtaining the surface charges `integrating the fluxes'.
The same procedure for the angular momentum was carried out in \cite{Dray:1984gz}, and one obtains the Dray-Streubel charge given in \eqref{QBMS}.

Deriving the charges in this way  leaves the ambiguity of adding closed forms, which at the level of surface integrals means 
time-independent terms, since
\be
\pounds_n c =i_ndc+di_nc=di_nc
\ee
if $c$ is closed, and integrates to zero.
 The procedure can then be completed showing that all time-independent terms that could be added would spoil covariance. 
This leads to the unique identification of the charges seen earlier.

%

\subsubsection{Wald-Zoupas approach}
\bit
\item\emph{Advantages:} Closed-form ambiguity fixed (for fixed symplectic 2-form); possibility of bootstrapping the charge from the Komar formula, 
simplifying in principle the calculation of the charges; relation to canonical generators on $\Si$;

\item\emph{Caveats:} Depends on choice of $\Si$, bulk coordinates and bulk extensions of the symmetry vector fields; subtlety with certain choices of field-dependent extensions due to extension-dependence of Komar formula; field-space constant ambiguity to be resolved in a second stage.
\eit

In the Wald-Zoupas approach, we first select a preferred symplectic potential for \eqref{OmAS}. The condition of stationarity is identified with the vanishing of the news function, which for the special case of Bondi frames (and this case only!) can be identified with $\dot C_{AB}$. This leads to the choice
\be\label{thBMS}
\th^{\sscr BMS} = -\f1{32\pi}\int \dot C_{AB}\d C^{AB} \eps_\scri,
\ee
for which one can check covariance, namely conformal invariance and foliation independence.
The Noether current is then immediately seen to match the Ashtekar-Streubel flux, and choosing \eqref{thBMS} for the split leads to \eqref{IxiAS}. In fact, the same symplectic potential had been already identified in \cite{Ashtekar84book}.

To extract the charges however, the Wald-Zoupas strategy is different. Instead of `integrating the flux', they propose to match the charge with the integrable part of the canonical generator at $\Si$ obtained subtracting the preferred symplectic potential:
\be
\d Q_\xi :=-I_\xi\Om_\Si +\oint_S i_\xi\th^{\sscr BMS}.
\ee 
The charge so obtained is automatically a potential for the flux over a portion of $\scri$ thanks to $d\om\eqons 0$.
This procedure has the advantage of fixing the ambiguity of adding time-independent terms of the previous procedure, thanks to the matching to the canonical generator on a given cut. There is instead the ambiguity of adding a field-space independent term. This can be eliminated requiring that the charges are all zero in a reference solution, say Minkowski.

Another potential advantage of this procedure is that one can bootstrap the calculation from the Komar 2-form, using the improved Noether charge formula. 
The explicit calculation can however be delicate, and there is an aspect of it which is left implicit in the original Wald-Zoupas paper, as well as in the references \cite{Flanagan:2015pxa,Grant:2021sxk}.
The reason is that the Komar 2-form depends on the second and third order expansion of the symmetry vector field off $\scri$. And these are not canonical. To make things worse, if one uses the common choice of extension \eqref{xi}, they are field dependent. In this case \eqref{IW} is no longer valid, and has to be replaced by
\eqref{IW2}.
The correction $q_{\d\xi}$ gets rid of the spurious contribution coming from the field dependence of the extension of \eqref{xi}, and only including it one recovers \eqref{QBMS}. 
Omitting the correction changes the flux by a soft term that spoils conservation in Minkowski for all BMS generators. 
For more details on this see \cite{Odak:2022ndm}, and the discussion in \cite{Rignon-Bret:2024gcx,Ashtekar:2024stm}.

\subsubsection{Improved Noether charge approach}
\bit
\item\emph{Advantages:} Explicit formula without computing variations in field space.

\item\emph{Caveats:} Need to carefully analyse the anomalies.
\eit

Even though the Wald-Zoupas paper never mentions that their procedure leads to charges that can be interpreted as improved Noether charges, it was proven in \cite{Odak:2022ndm,Chandrasekaran:2021vyu} that this is indeed the case. There is a caveat though. One could think that since
$\th^{\sscr BMS}=\pbi\th+\d b$ for a certain $b$, it is possible to use directly the formula $q_\xi^{\sscr BMS}\stackrel{?}=q_\xi+i_\xi b$.
The caveat has to do again with the extension-dependence of the Komar 2-form, that created the necessity of the term $\d q_\xi -q_{\d\xi}$ in the canonical generator. From the point of view of the Noether charges, it is possible to show that $q_{\d\xi} = \d s_\xi$,
and this can be generated adding a corner term to the boundary Lagrangian, 
\be
c=-\f1{8\pi}\b\eps_S.
\ee
With this corner term one can indeed recover \eqref{QBMS} starting from the limit of Komar $q_\xi$:
\be
q^{\sscr BMS}_\xi=q_\xi +i_\xi \ell^{\sscr BT}- I_\xi \d c.
\ee
This leads to a prescription for the BMS charges as Noether charges satisfying the stationary condition, and the covariance condition should be imposed on both the symplectic potential \emph{and} the boundary Lagrangian.

\subsubsection{Barnich-Brandt approach}
\bit
\item\emph{Advantages:} Explicit formula directly in terms of the metric; avoids the subtlety with field-dependent extensions.

\item\emph{Caveats:} Hides the role of the symplectic structure; needs to be supplemented by Wald-Zoupas prescription in order to identify a covariant and stationary split.
\eit

The Barnich-Brandt formula gives
\begin{align}\label{IxiomBB}
-I_\xi\om &\eqons d\sd k_\xi, \\
\sd k_\xi& = - \f1{32\pi} \eps_{\m\n\r\s}\big[ (\d\ln{\sqrt{-g}}) \na^\r\xi^\s + \d g^{\r\a}\na_\a\xi^\s 
+ \xi^\r \left(\na_\a \d g^{\a\s} + 2 \na^\s \d\ln\sqrt{-g}\right) \nn\\ &\hspace{3cm} - \xi_\a\na^\r\d g^{\s\a} \big] dx^\m\w dx^\n,
\end{align}
where we have subtracted off the extra contribution that comes from the different corner term in the symplectic structure, and which anyways plays no role in the BMS symmetry because it vanishes in the limit to $\scri$.
An immediate advantage is that this formula gives directly \eqref{IW2}, so any spurious field dependence introduced by the choice of extension \eqref{xi} is removed, and one gets  \cite{Barnich:2011mi} 
\be\label{BT11}
-I_\xi\Om_\Si \eqons \d Q_\xi - {F}_\xi,
\ee
with the candidate charges and fluxes given precisely by \eqref{jNBMS} and \eqref{QBMS}. So the only thing that remains to be done is to identify them in a canonical way, which can be done applying the Wald-Zoupas prescription. This leads to non-trivial insights, for instance the need for Geroch tensor correcting the formulas \eqref{jNBMS} and \eqref{QBMS} for frames which are not round spheres. This correction removes the cocycle found in \cite{Barnich:2011mi}, thus leading to a center-less realization of the BMS algebra \cite{Rignon-Bret:2024gcx}. The existence of a center-less realization is a remarkable consequence of insisting on covariance, and explains that the cocycle found in \cite{Barnich:2011mi}, which has exactly the structure of \eqref{BTgrAno}, was due to a lack of conformal invariance.

These results have motivated us also to apply the Wald-Zoupas prescription to larger symmetry groups. For eBMS, this is indeed possible, see \cite{Rignon-Bret:2024gcx} again, and the result provides a solid foundation for the symplectic structure used in \cite{Campiglia:2020qvc,Donnay:2022hkf}. For gBMS this is harder, and after succeeding for an intermediate construction that we called the group of rest-frame Bondi spheres (RBS), we now have a candidate and we hope to finish verifications soon.

\section{Conclusions and further reading}

The ambiguities of the covariant phase space and of the Noether charges, and the difficulties due to the non-integrability of the canonical generators, have hindered their  understanding in dynamical gravitational systems for a long time.
An important message of these lectures is that insisting on covariance and on the importance of identifying radiation (and its absence) through boundary conditions, allows one to identify typically a unique set of canonical fluxes and charges for the boundary symmetries, satisfying the crucial properties of being conserved in the absence of radiation and of providing a correct realization of the symmetry algebra in the phase space. 

In spite of its length, this material offers only a limited view on this large and important subject.
Regrettably, I have not discussed in much details how specific boundary conditions are chosen, and how this leads to the identification of the relevant boundary symmetries as residual gauge transformations at the boundary, with the exception of the BMS case.
I also did not cover the relation between the symmetries and the mechanical laws of horizons and quasi-local horizons. Nonetheless, 
I hope this introduction succeeded in giving you some initial perspective on these fascinating problems. I conclude reiterating that this was personal view, biased both in terms of selecting relevant questions, and identifying the most interesting answers. There is a large and beautiful literature on the subject, and I hope you will fill motivated to explore it and learn about further problems and their motivations. Here is a small selection to get you started.

\bit
\item BMS symmetries: \cite{BMS,Newman:1966ub,Geroch:1977jn,Barnich:2011mi,Barnich:2011ty,Ashtekar:2014zsa,Flanagan:2015pxa,Barnich:2017ubf,Grant:2021sxk,Barnich:2019vzx,Ashtekar:2019rpv,Barnich:2021dta,Barnich:2022bni,Prabhu:2019fsp,Wieland:2020gno,Prabhu:2021cgk,Herfray:2021qmp,Capone:2022gme,Wieland:2021eth,Prabhu:2022zcr,Mitman:2024uss,Ashtekar:2023zul,AndradeeSilva:2026mpa}
\item Relation to soft theorems: \cite{Strominger:2013lka,Strominger:2013jfa,He:2014laa,Kapec:2014opa,Kapec:2016jld,Campiglia:2014yka,Choi:2024ygx}
\item Additional symmetries at $\scri$: \cite{Barnich:2010eb,Campiglia:2014yka,Barnich:2016lyg,Compere:2018ylh,Compere:2020lrt,Campiglia:2020qvc,Campiglia:2021bap,Freidel:2021yqe,Barnich:2021dta,Geiller:2022vto,McNees:2024iyu,Satishchandran:2019pyc,Flanagan:2023jio,Geiller:2024bgf,Freidel:2021ytz,Donnay:2022hkf,Donnay:2022wvx,Capone:2021ouo,Kmec:2024nmu,Geiller:2024bgf,Pranzetti:2025flv,Geiller:2025dqe,Cresto:2024fhd,Cresto:2024fhd}
\item Dual gravitational charges and other extensions: \cite{Godazgar:2018qpq,Godazgar:2018vmm,Godazgar:2019dkh,Godazgar:2020kqd,Oliveri:2020xls,Godazgar:2022foc,Godazgar:2022pbx,Cerdeira:2025elp,Gomez-Fayren:2023qly,Ortin:2021ade,Ortin:2022uxa,Bandos:2023zbs,Mittal:2022ywl}
\item Symmetries and mechanical laws of horizons: \cite{Iyer:1994ys,Jacobson:1993vj,Ashtekar:2000hw,Ashtekar:2000sz,Gao:2003ys,Mars:2003ud,Ashtekar:2013qta,Donnay:2015abr,Donnay:2016ejv,Donnay:2018ckb,Jacobson:2018ahi,Ashtekar:2021wld,Ashtekar:2021kqj,Banihashemi:2022htw,Odak:2023pga,Chandrasekaran:2023vzb,Rignon-Bret:2023lyn,Rignon-Bret:2023fjq,Hollands:2024vbe,Visser:2024pwz,Wall:2024lbd}
\item Horizons and more general null boundary symmetries: \cite{Hajian:2015xlp,Wieland:2017zkf,Wieland:2018ymr,Wieland:2019hkz,Wieland:2021vef,Adami:2021nnf,Ciambelli:2018wre,Chandrasekaran:2018aop,Ciambelli:2018xat,Safari:2019zmc,Ciambelli:2019lap,Ciambelli:2019bzz,Ciambelli:2021vnn,Ciambelli:2022vot,Speranza:2022lxr,Chandrasekaran:2019ewn,Freidel:2021ytz,Ciambelli:2022cfr,Donnay:2022aba,Ciambelli:2023mir,Adami:2021kvx,Adami:2023wbe,Shajiee:2025cxl,Ruzziconi:2025fuy,Agrawal:2025fsv,Adami:2021nnf,Wieland:2024dop,Wieland:2025qgx,Parvizi:2025shq,Parvizi:2025wsg}

\item Tetrad variables: 
\cite{Ortin:2002qb,Ashtekar:2008jw,Corichi:2010ur,Corichi:2013zza,Jacobson:2015uqa,Prabhu:2015vua,Corichi:2015cqa,Corichi:2016zac,Barnich:2016rwk,DePaoli:2017sar,Frodden:2017qwh,DePaoli:2018erh,Oliveri:2019gvm,Margalef-Bentabol:2020teu,G:2021xvv,Kabel:2023jve}
\item BRST, BV and related methods:
\cite{Cattaneo:2015xca,Canepa:2020ujx,Hadfield:2020mnn,Simao:2021xgw,Riello:2022din,Cattaneo:2023hxv,Riello:2025ktk,Baulieu:2025itt,Baulieu:2024oql}
\item Some relations to the Post-Minkowskian approach:
\cite{Compere:2017wrj,Hosseinzadeh:2018dkh,Compere:2019gft,Blanchet:2020ngx,Seraj:2021qja,Seraj:2021rxd,Seraj:2022qyt}

\item Relation to flat holography: \cite{Donnay:2018neh,Compere:2020lrt,Donnay:2020guq,Donnay:2022wvx,Campoleoni:2022wmf,Campoleoni:2023fug,Fiorucci:2023lpb}
\item Higher spins:
\cite{Campoleoni:2017mbt,Campoleoni:2017qot,Francia:2018jtb,Campoleoni:2020ejn,Campoleoni:2023fug,Francia:2024hja,Campoleoni:2025bhn}

\eit

\subsubsection*{Acknowledgements}

I thank the Galileo Galilei Institute for Theoretical Physics for the hospitality and the INFN for partial support.
I would like to thank my fellow co-organizers of the workshop, Dario Francia, Andrea Campoleoni, Laura Donnay, Andrea Puhm and Sabrina Pasterski, and all the staff at GGI for their support and help.
It is a pleasure to thank various people that have contributed to my understanding of the subject over the years, among them:
Abhay Ashtekar, Glenn Barnich, Luca Ciambelli, Adrien Fiorucci, Eanna Flanagan, Marc Geiller, Yannick Herfray, Laurent Freidel, Thierry Masson, Gloria Odak, Roberto Oliveri, Pierre Piovesan, Antoine Rignon-Bret, Michele Schiavina, Antony Speranza, Bob Wald, Wolfgang Wieland, Celine Zwikel.
I am also grateful to the organizers of the Classical and Quantum Gravity Focus Issue on Finite Boundaries in Theories of Gravity Dionysios Anninos, Dami\'an Galante and Ana-Maria Raclariu for their invitation to contribute and encouragement to put this materila together.

\appendix

\section{Derivation of the BMS group in flat spacetime}\label{AppA}

In this Appendix we provide a pedagogical description of the BMS group. Unlike most reviews in the literature, this derivation is based only on flat spacetime, and allows one to appreciate the origin of the difference with the Poincar\'e group in the switch from preserving the metric globally, to only preserving it at $\scri$. 

\subsubsection*{Killing vectors in spherical coordinates}
We start from the expression of the Poincar\'e Killing vectors in Cartesian coordinates \eqref{xiP4}, which we decompose 
\be\label{xidecGen}
\xi=a^{\mu}P_{\mu}+b^{a}B_{a}+r^{a}R_{a}, \qquad b^a=b^0{}_a=b^a{}_0, \qquad r^a=-\f12\eps^a{}_{bc}b^b{}_c, \qquad b^a{}_b=-\eps^a{}_{bc}r^c
\ee
where
\be
P_\m=\p_\m, \qquad R_a=\eps_{ab}{}^cx^b\p_c, \qquad B_a=t\p_a+x_a\p_t. 
\ee
Let us first rewrite these vectors using spherical coordinates. The coordinate transformation is:
\begin{subequations}\label{CartToSpher}
\begin{align}
& \p_{t}=\p_{t}\\
& \p_{x}=\sin\th\cos\phi\p_r+\frac{1}{r}\left(\cos\th\cos\phi\p_{\theta}-\frac{\sin\phi}{\sin\th}\p_\phi\right) \\
& \p_{y}=\sin\th\sin\phi\p_r+\frac{1}{r}\left(\cos\th\sin\phi\p_{\theta}+\frac{\cos\phi}{\sin\th}\p_\phi\right) \\
& \p_{z}=\cos\th\p_r-\frac{1}{r}\sin\th\p_{\th}.
\end{align}\end{subequations}
We can write its inverse as
\be
\p_r=n^a\p_a, \qquad \p_\th=re_1^a\p_a, 
\qquad \p_\phi=re^a_2\p_a,
\ee
where we introduced the right-handed, orthonormal basis
\begin{align}
& n = (\sin\th\cos\phi,\sin\th\sin\phi,\cos\th),\\
\label{mbasis}
& e_1 = (\cos\th\cos\phi,\cos\th\sin\phi, -\sin\th)=\p_\th n, \qquad e_2= (-\sin\phi,\cos\phi,0)=\f1{\sin\th}\p_\phi n.
\end{align}
A generic 4-vector can thus be written as
\be
a^\m\p_\m=a^0\p_t + \vec a\cdot \vec n\,\p_r +\f1r q^{AB}\p_B (\vec a\cdot \vec n)\,\p_A.
\ee

For the rotations, we have
\be\label{GenRotSpher}
\begin{aligned}
R_{x}=y\p_{z}-z\p_{y}&=-\sin\phi\p_{\theta}-\frac{\cos\th}{\sin\th}\cos\phi\p_{\phi}, \\
R_{y}=z\p_{x}-x\p_{z}&=\cos\phi\p_{\theta}-\frac{\cos\th}{\sin\th}\sin\phi\p_{\phi}, \qquad
R_{z}=x\p_{y}-y\p_{x}=\p_{\phi}.
\end{aligned}
\ee
A generic rotation can thus be written as
\be\label{rR}
r^aR_a = r^ae_{2a}\p_\th+\f1{\sin\th}r^ae_{1a}\p_\phi =\eps^{AB}\p_B(\vec r\cdot \vec n)\p_A,
\ee
where we introduced the area 2-form
\be
\eps_{AB}=\sin\th\mat{0}{1}{-1}{0},\qquad\eps^{AB}=\f1{\sin\th}\mat{0}{1}{-1}{0}.
\ee

For the boosts,
\be\label{GenBooSpher}
\begin{aligned}
B_{x}=x\p_{t}+t\p_{x}&=\sin\th\cos\phi(r\p_{t}+t\p_r)+ \frac{t}{r}\left(\cos\th\cos\phi\p_{\th}-\f{\sin\phi}{\sin\th}\p_{\phi}\right), \\
B_{y}=y\p_{t}+t\p_{y}&=\sin\th\sin\phi(r\p_{t}+t\p_r)+ \frac{t}{r}\left(\cos\th\sin\phi\p_{\th}+\f{\cos\phi}{\sin\th}\p_{\phi}\right),  \\
B_{z}=z\p_{t}+t\p_{z}&=\cos\th(r\p_{t}+t\p_r)-\frac{t}{r}\sin\th\p_{\theta}.
\end{aligned}
\ee
A generic boost can thus be written as
\be
b^aB_a=b^an_a(r\p_{t}+t\p_r)+ \f tr\left(b^ae_{1a}\p_\th+\f1{\sin\th}b^ae_{2a}\p_\phi\right) 
=\vec b\cdot \vec n (r\p_{t}+t\p_r)+\f tr\p^A(\vec b\cdot\vec n)\p_A.
\ee

Putting everything together, a Poincar\'e Killing vector in spherical coordinates reads
\be\label{xiP4sph}
\xi  =  a^0\p_t +\vec a\cdot\vec n\p_r +\vec b\cdot \vec n (r\p_{t}+t\p_r)+\Big(\eps^{AB}\p_B(\vec r\cdot\vec n)+\f tr\p^A(\vec b\cdot\vec n)+\f1r \p^A (\vec a\cdot \vec n)\Big)\p_A.
\ee

\subsubsection*{Spherical harmonics}

Let us notice at this point that there is a convenient interpretation of these coefficients in terms of spherical harmonics.
For the time and radial parts can be written using
\begin{align}
& \sin\th\cos\varphi = - \sqrt{\f {2\pi}3} (Y_1^{+1} - Y^{-1}_1),\quad \sin\th\sin\varphi= i \sqrt{\f {2\pi}3} (Y_1^{+1} + Y^{-1}_1) , \quad \cos\th=2\sqrt{\f \pi3}Y^0_1,
\end{align}
where
\be
Y_0^0=\f1{2\sqrt{\pi}}, \quad Y_1^{\pm1}=\mp\f12\sqrt{\f 3{2\pi}}\sin\th e^{\pm i\varphi}, \quad Y_1^0=\f12\sqrt{\f 3{\pi}}\cos\th
\ee
are the $l=0,1$ modes of spherical harmonics.
It follows that
\be\label{SHsp}
\vec a\cdot \vec n = \sum_{m=\pm1,0}a_m Y_l^{m}, \qquad a_\pm=\mp\sqrt{\f{2\pi}3}(v^x\mp iv^y), \qquad a_0=\sqrt{\f{4\pi}3}v^z.
\ee
Let us also define for later convenience the function
\begin{align}
& T=a^0-\vec a\cdot \vec n=\sum T_{lm}Y_{lm}, \\
& T_{00}= 2\sqrt\pi a^0, \qquad T_{10}=-\sqrt{\f{4\pi}3}a^z, \qquad T_{1\pm1}=\pm\sqrt{\f{2\pi}3}(a^x\mp ia^y).
\end{align}
It can be explicitly checked that any linear combination of $l=0,1$ modes satisfies
\be
\Dd_{\la A}\Dd_{B\ra}T=0.
\ee

The angular part can be left as derivatives of spherical harmonics, or rewritten using
\begin{align}
& \cos\th \sin\phi=\sqrt{\f {4\pi}3}\im( Y_{1,11} + \bar Y_{1,1-1}), \qquad \cos\th \cos\phi=\sqrt{\f {4\pi}3}\re( Y_{1,11} + \bar Y_{1,1-1}), \\
& \sin\th =  \sqrt{\f{8\pi}{3}} Y_{1,10},   
\qquad  \cos\phi=\sqrt{\f {4\pi}3}\re(\bar Y_{1,11} - Y_{1,1-1}), \qquad  \sin\phi=\sqrt{\f {4\pi}3}\im(\bar Y_{1,11} - Y_{1,1-1}),
\end{align}
where
\begin{align}
Y_{1,1,0}=\sqrt{\f{3}{8\pi}}\sin\th, \qquad
Y_{1,1\pm1}=\pm\sqrt{\f{3}{16\pi}}(\cos\th\mp 1)e^{\pm i\phi}
\end{align}
are the spin-1 weighted spherical harmonics.

For the rotations and boost parts, we define
\be
Y^A_{\rm r}= \eps^{AB}\p_B\Psi, \qquad \Psi=\vec r\cdot \vec n, \qquad 
Y^A_{\rm b}= \p^A\Phi, \qquad \Phi=\vec b\cdot\vec n.
\ee
We can collect $Y^A_{\rm r}$ and $Y^A_{\rm b}$ into a unique 2-vector
\be
Y^A=Y^A_{\rm r}+Y^A_{\rm b}=\Dd^A\Phi +\eps^{AB}\p_B\Psi.
\ee
Since both the boost (aka electric) part $\Phi$ and the rotation (aka magnetic) part $\Psi$ are composed of $l=1$ modes only, it follows that
\be
\Dd Y = \Dd^2\Phi=-2\Phi, \qquad \eps^{AB}\Dd_AY_B = -\Dd^2\Psi=2\Psi, \qquad \Dd_{\la A}Y_{B\ra}=0.
\ee 
Namely, the $Y$'s are conformal Killing vectors of the sphere. 
Then,
\be\label{rR1}
r^aR_a =Y^A_{\rm r}\p_A,\qquad
b^aB_a=\vec b\cdot \vec n (r\p_{t}+t\p_r)+\f trY^A_{\rm b}\p_A.
\ee

Putting everything together, \eqref{xiP4sph} can be rewritten as
\be\label{xiP4sph2}
\xi  =  a^0\p_t+(a^0-T)\p_r-\f12\Dd Y (r\p_{t}+t\p_r)+(Y^A_{\rm r}+\f trY^A_{\rm b}+\f1r\p^A T)\p_A.
\ee

\subsubsection*{Killing vectors in Bondi coordinates}

Next, we consider the transformation from Cartesian to retarded time \eqref{etaB}:
\begin{subequations}\label{CartToBondi}
\begin{align}
& \p_{t}=\p_{u}\\
& \p_{x}=\sin\th\cos\phi(\p_r-\p_u)+\frac{1}{r}\left(\cos\th\cos\phi\p_{\theta}-\frac{\sin\phi}{\sin\th}\p_\phi\right) \\
& \p_{y}=\sin\th\sin\phi(\p_r-\p_u)+\frac{1}{r}\left(\cos\th\sin\phi\p_{\theta}+\frac{\cos\phi}{\sin\th}\p_\phi\right) \\
& \p_{z}=\cos\th(\p_r-\p_u)-\frac{1}{r}\sin\th\p_{\th},
\end{align}\end{subequations}
whose inverse is
\be
\p_u=\p_t,\qquad \p_r=\p_t+n^a\p_a, \qquad \p_\th=re_1^a\p_a, 
\qquad \p_\phi=re^a_2\p_a.
\ee
Therefore,
\be
a^\m\p_\m=(a^0-\vec a\cdot \vec n)\p_u + \vec a\cdot \vec n\,\p_r +\f1r \vec a\cdot ( q^{AB}\p_B \vec n)\,\p_A.
\ee
The rotations are the same as in spherical coordinates. The boosts can be
read from \eqref{GenBooSpher} replacing $r\p_{t}+t\p_r$ with $(u+r)\p_{r}-u\p_u$, and $t/r$ with $(u+r)/r$.
This leads to 
\be
b^aB_a=\vec b\cdot \vec n ((u+r)\p_{r}-u\p_u)+\f {u+r}rY^A_{\rm b}\p_A=
\f u2\Dd Y\p_u+ Y^A_{\rm b}\p_A -\f{u+r}2\Dd Y\p_{r}+ \f ur Y^A_{\rm b}\p_A. 
\ee

We now introduce
\be
f=T+\f u2\Dd Y,
\ee
and use
\begin{align}
& \f12\Dd^2 f=-\vec a\cdot \vec n-\f u2\Dd Y, \qquad
 \p^Af=q^{AB}\p_B(T+\f u2\Dd Y)= -  \vec a\cdot ( q^{AB}\p_B \vec n)-u Y_{\rm b}^A,
\end{align}
which follow immediately from the properties of $l=1$ modes.
Putting everything together, \eqref{xidecGen} reads
\be\label{xiP4Bondi}
\xi  =  f\p_u +Y^A\p_A - \f r2 \Dd Y\p_r - \f1r\p^A f\p_A +\f12\Dd^2 f \p_r.
\ee

We conclude that the global Killing vectors of Minkowski in retarded time have a finite expansion in $1/r$, and that $f$ only contains $l=0,1$ modes.

\subsubsection*{Global versus asymptotic symmetries} 

With the inverse radius coordinate $\Om:=1/r$, 
\be\label{xiP4Om}
\xi=  f\p_u +Y^A\p_A + \Om\left( \f12 \Dd Y\p_\Om - \p^A f\p_A\right) -\f{\Om^2}2 \Dd^2 f \p_\Om.
\ee
We now consider the unphysical metric, given by $\hat \eta_{\mu\nu}=\Om^{2}\eta_{\mu\nu}$.
Under a general diffeomorphism,
\be
\pounds_\xi\hat\eta_{\m\n}=\pounds_\xi(\Om^2\eta_{\m\n})=\Om^2\pounds_\xi\eta_{\m\n}+2\Om^{-1}\hat \eta_{\m\n}\pounds_\xi\Om.
\ee
For the global Killing vectors \eqref{xiP4Om} we have $\pounds_\xi \eta_{\m\n}=0$, hence
\be\label{GKVeq}
\pounds_\xi\hat\eta_{\m\n}=2\a_\xi\hat \eta_{\m\n}, 
\ee
where
\be
\a_\xi = \Om^{-1}\pounds_\xi\Om=\f12\Dd Y -\f{\Om}2\Dd^2 f. 
\ee
We see that the global Killing vectors are exact conformal Killing vectors of the unphysical metric, to all orders in $\Om$ or $1/r$.

We define the asymptotic symmetries as those diffeomorphisms that preserve only the leading order of the unphysical metric. That is,
\be\label{AKVeq}
\pounds_\xi\hat\eta_{\m\n}=2\a_\xi\hat \eta_{\m\n} +O(\Om).
\ee
To solve these equations, 
let us parametrize the generic expansion as follows,
\be\label{xiParamOm}
\xi=  f\p_u +Y^A\p_A + \Om\bar\xi^\m\p_\m +\Om^2\bar{\bar\xi}^\m\p_\m+\ldots
\ee
where $f$ and $Y^A$ are initially arbitrary functions of all 3 coordinates. Using \eqref{xiParamOm} and \eqref{hatetaBondiOm}, we can explicitly evaluate all components of $\pounds_\xi \hat\eta_{\m\n}$. Equating them to the global Killing condition \eqref{GKVeq}, or the asymptotic Killing condition \eqref{AKVeq}, we get respectively 
\begin{align}
& (uu) \qquad 2(\dot\xi^\Om-\Om\xi^\Om-\Om^2\dot\xi^u) && = -2\a_\xi \Om^2; && = O(\Om);\\
& (u\Om) \qquad \p_\Om\xi^\Om-\Om^2\p_\Om\xi^u+\dot\xi^u  && = 2\a_\xi; && = 2\a_\xi+O(\Om);\\
& (\Om\Om) \qquad \p_\Om\xi^u && = 0; && = O(\Om); \\
& (uA) \qquad \p_A\xi^\Om-\Om^2\p_A\xi^u+q_{AB}\dot\xi^B && = 0; && = O(\Om) ;\\
& (\Om A) \qquad \p_A\xi^u+q_{AB}\p_\Om\xi^B && = 0; && = O(\Om); \\
& (AB)\qquad 2\Dd_{(A}\xi_{B)}  && = 2\a_\xi q_{AB}; && = 2\a_\xi q_{AB}+O(\Om);
\end{align}
where $\xi_A:=q_{AB}\xi^B$. Let us consider the first column, corresponding to the global Killing vector condition. Recalling that $\a_\xi=\Om^{-1}\xi^\Om=\bar\xi^\Om+\Om\bar{\bar\xi}^\Om+...$, we get
\begin{align}
& (\Om\Om) \qquad \xi^u=f \\
& (\Om A) \qquad \bar\xi^A=-\p^Af, \qquad \bar{\bar\xi}^A=...=0\\
& (AB)\qquad \bar\xi^\Om=\f12\Dd Y, \qquad \bar{\bar\xi}^\Om= -\f12\Dd^2f, \qquad \bar{\bar{\bar\xi}}^\Om=...=0, \qquad 2\Dd_{(A}\xi_{B)}=0.
\end{align}
This means that $Y^A$ and $\bar\xi^A$ are CKVs of the sphere, and therefore also $\p^A f$ is; which in turns implies that $T$ only has $l=0,1$ modes.
It follows that 
\be\label{D2fid}
\bar{\bar\xi}^\Om= -\f12\Dd^2f = f.
\ee
The remaining conditions are
\begin{align}
& (u\Om) \qquad \bar\xi^\Om=\dot f,\\
& (uA) \qquad \dot Y^A=0, \qquad \p_A\bar\xi^\Om=-\p_u\bar\xi_A=\p_A\dot f, \\
& (uu) \qquad \p_u\bar\xi^\Om=\ddot f=0, \qquad \dot\xi^u=\p_u\bar{\bar\xi}^\Om,
\end{align}
Solving these we recover the previous result \eqref{xiP4Om}.

Consider now the second column, that imposes the asymptotic Killing condition. We now get
\begin{align}
& (\Om\Om) \qquad \xi^u=f+O(\Om), \\
& (\Om A) \qquad \xi^A=Y^A-\Om\p^Af+O(\Om^2), \\
& (AB)\qquad \bar\xi^\Om=\f12\Dd Y, \qquad 2\Dd_{(A}Y_{B)}=0.
\end{align}
Only $Y$ has to be a CKV, and $\bar\xi^A$ and $f$ can be arbitrary functions on the sphere. 
Notice also that we have lost the condition on $\bar{\bar\xi}^\Om$. This, as well as all other higher order conditions, can be restored if one fixes the extension requiring preservation of the bulk gauge. Then,
\begin{align}
& (u\Om) \qquad \bar\xi^\Om=\dot f, \\
& (uA) \qquad \dot Y^A=0, \\
& (uu) \qquad \p_u\bar\xi^\Om=0,
\end{align}
The result is 
\be\label{xiBMS1}
\xi=  f\p_u +Y^A\p_A + \Om\left( f\p_\Om - \p^A f\p_A\right)+O(\Om^2).
\ee

\section{Hamiltonian generator for a scalar field on a null hypersurface}\label{AppB}

In this Appendix we give a simple example of the construction of Hamiltonian generators for diffeomorphisms proposed in \cite{Ashtekar:2024stm},
and show how it compares with the prescription obtained with the generalized Wald-Zoupas critera of \cite{Odak:2022ndm}.
The example is a scalar field, and the boundary a general null hypersurface $\cN$.
For this example, we don't need to worry about the boundary conditions for the gravitational field. This can be fully arbitrary, and the allowed diffeos the whole of Diff$(\cN)$.

Up to numerical constants, the symplectic 2-form current of a scalar field pulled-back on a generic null hypersurface $\cN$ with (outgoing, future-pointing) normal $l$ gives
\be
\om= \d\dot\phi\cw\d\phi \, \eps_\cN,
\ee
where $\dot\phi:=\pounds_l\phi$ is just a shorthand notation (no preferred time coordinate needs to be chosen), and $\eps=-l\w\eps_\cN$ defines the induced volume form in terms of the spacetime volume form. 
Notice that we are not including metric variations; these will contribute to the gravitational part of the symplectic 2-form, which we neglect here to focus on the scalar field degrees of freedom only. To compute the flow associated with a boundary diffeomorphism, we first recall that 
the scalar field is dynamical, hence by definition its field-space transformation under the residual diffeos allowed by the boundary conditions is the Lie derivative, that is $\d_\xi \phi=\pounds_\xi\phi$. On the other hand, the normal contains a background dependence in its choice of scaling, since there is no canonical normalization for a null vector. This leads to an anomalous transformation, which affects also the induced volume form. In the notation of \cite{Chandrasekaran:2020wwn,Odak:2023pga},  $\D_\xi l^\m=-w_\xi l^\m$, and
\be
\d_\xi l^\m = \pounds_\xi l^\m - w_\xi l^\m, \qquad \d_\xi \eps_\cN = \pounds_\xi \eps_\cN+ w_\xi \eps_\cN.
\ee
It follows that
\be
(\d_\xi \dot\phi)\eps_\cN = (\pounds_{\d_\xi l}\phi +\pounds_l\pounds_\xi \phi)\eps_\cN
= (\pounds_{\d_\xi l}\phi +\pounds_\xi\dot \phi - \pounds_{\pounds_\xi l} \phi)\eps_\cN = \pounds_\xi(\dot\phi\eps_\cN).
\ee
We can now compute the flow, obtaining
\begin{align}\nn
-I_\xi\om &= (-\d_\xi\dot\phi\cw\d\phi+\d\dot\phi\cw\d_\xi\phi) \eps_\cN =-\pounds_\xi(\dot\phi \eps_\cN)\cw\d\phi+\d\dot\phi\cw\pounds_\xi\phi \, \eps_\cN
\\&=-\pounds_\xi (\dot\phi\d\phi\eps_\cN)+\d(\dot\phi\pounds_\xi)\eps_\cN = -d (\dot\phi\d\phi \, i_\xi \eps_\cN)+\d(\dot\phi\pounds_\xi\phi)\eps_\cN.\label{Dpol}
\end{align}

The two problems of integrability and ambiguities are now manifest. First, we have a corner term which is not $\d$-exact, hence the Hamiltonian generator does not exist in general. Second, we could change polarization, integrating by parts in field space, and rewrite this expression as
\begin{align}\label{Npol}
-I_\xi\om & = d (\phi\d\dot\phi \, i_\xi \eps_\cN) - \d(\phi\d_\xi\dot\phi)\eps_\cN.
\end{align}
This changes both the candidate integrable term, and integrability condition.
The strategy proposed in \cite{Ashtekar:1981bq,Ashtekar:2024stm} is to treat both issues at once, by introducing a topology in the field space. 
A way to do so is to choose a norm, which we take to be
\be\label{norm}
\norm{\phi}^2=\int_\cN (\dot\phi^2+q^{ab}\Dd_a\phi\Dd_b\phi+\phi^2)\eps_\cN,
\ee
where $\Dd_a$ is the 2d derivative on the space-like cross-sections, and $q^{ab}$ any choice of inverse.
In mathematical terms, we are now describing the boundary fields using the first Sobolev space. This specific choice of norm is motivated by analogy with the expression of the energy of a scalar field on a space-like hypersurface, where 
\be
E=\f12\int (\pi^2+D_a\phi D^a\phi + m^2\phi^2)\eps_\Si.
\ee 
However, no physical meaning should be given to \eqref{norm}: the comparison with the energy is only used to motivate the specific choice, and its only use is to introduce a topology on the field space, which we can now use to select and define a canonical generator. 
Using the norm \eqref{norm} in fact, we can see that the subspace with $\dot\phi|_{\p\cN}=0$ is dense, whereas $\phi|_{\p\cN}=0$ is not.\footnote{Being boundary configurations, both conditions are measure zero. Then, recall that a function with arbitrary boundary value can be obtained as a sequence of $L_2$ functions with vanishing boundary value, but not as a sequence of functions in the first Sobolev norm. Therefore only the first subspace is dense, and not the second. In loose language, $L_2$ does not see the boundary, whereas the Sobolev norm does.} This means that 
the integrable term in \eqref{Dpol} is densely defined, whereas the alternative option given by the integrable term in \eqref{Npol} is not.
Furthermore because it is densely defined and continuous, it admits a unique extension to the full space. Our choice of norm thus equips the field space with a unique Hamiltonian generator, well-defined everywhere.\footnote{A word of caution about this result: even though the Hamiltonian is well-defined everywhere, the existence of the non-integrable corner term still means that the action of the Hamiltonian cannot be exponentiated except for those $\xi$'s for which the corner term vanishes.}

We can now compare this resolution of the ambiguity and non-integrability issues with the generalized Wald-Zoupas prescription as introduced in \cite{Odak:2022ndm}. The bare symplectic potential is
\be\label{thscalar}
\th = \dot\phi\d\phi\,\eps_\cN,
\ee
and corresponds to Dirichlet boundary conditions in the conservative case, and to a notion of stationarity associated with homogeneous Neumann $\dot\phi=0$ in the open/dissipative case. 
The natural alternative is to integrate by parts in field space and switch to Neumann boundary conditions. Now the notion of stationarity for the open/dissipative case would be associated with homogeneous Dirichlet $\phi=0$. Both options are covariant and can be used: the choice depends on the physical problem at hand. 
Typically, no radiation going through the boundary is associated with homogeneous Neumann condition, and then \eqref{thscalar} is the preferred symplectic potential. In this case one obtains the same prescription as in the approach based on introducing the norm \eqref{norm}.


\providecommand{\href}[2]{#2}\begingroup\raggedright\endgroup

\end{document}